\documentstyle[12pt,a4,epsfig]{article}
\begin{document}
\newcommand{\be}{\begin{equation}}
\newcommand{\ee}{\end{equation}}
\newcommand{\w}{wavelet}
\newcommand{\an}{analysis}
\newcommand{\1}{one-dimensional}
\newcommand{\2}{two-dimensional}
\newcommand{\m}{two-microlocal}
\newcommand{\co}{coefficient}
\newcommand{\scl}{\varphi}
\newcommand{\mr}{multiresolution}
\begin{center}

{\bf WAVELETS AND THEIR USE}

\vspace{2mm}

I.M. Dremin, O.V. Ivanov, V.A. Nechitailo

\vspace{2mm}

{\it Lebedev Physical Institute, Moscow 117924, Russia}

\vspace{3mm}

{\bf Contents}

\end{center}

\noindent 1. Introduction       \\
2. Wavelets for beginners \\
3. Basic notions and Haar \w s       \\
4. Multiresolution analysis and Daubechies \w s\\
5. Fast wavelet transform and coiflets   \\
6. Choice of wavelets         \\
7. Multidimensional \w s         \\
8. The Fourier and wavelet transforms\\
9. Wavelets and operators  \\
10.Nonstandard matrix multiplication \\
11.Regularity and differentiability \\
12.Two-microlocal analysis  \\
13.Wavelets and fractals   \\
14.Discretization and stability           \\
15.Some applications                          \\
15.1. Physics\\
15.1.1. Solid state and molecules\\
15.1.2. Multiparticle production processes \\
15.2. Aviation (engines)\\
15.3. Medicine and biology  \\
15.3.1. Hearts     \\
15.3.2. DNA sequences  \\
15.3.3. Blood cells  \\
15.4. Data compression  \\
15.5. Microscope focusing \\
16.Conclusions \\
Appendix 1 \\
Appendix 2  \\
Appendix 3  \\
Acknowledgements \\
References\\

\newpage

\begin{abstract}

This review paper is intended to give a useful guide for those who want to
apply discrete wavelets in their practice. The notion of wavelets and their use
in practical computing and various applications are briefly described, but
rigorous proofs of mathematical statements are omitted, and the reader is
just referred to
corresponding literature. The multiresolution analysis and fast
wavelet transform became a standard procedure for dealing with discrete
wavelets. The proper choice of a wavelet and use of nonstandard matrix
multiplication are often crucial for achievement of a goal. Analysis of
various functions with the help of wavelets allows to reveal fractal structures,
singularities etc. Wavelet transform of operator expressions helps solve
some equations. In practical applications one deals often with the
discretized functions, and the problem of stability of wavelet transform
and corresponding numerical algorithms becomes important. After discussing
all these topics we turn to practical applications of the wavelet machinery.
They are so numerous that we have to limit ourselves by some examples only.
The authors would be grateful for any comments which improve this review paper
and move us closer to the goal proclaimed in the first phrase of the abstract.

\end{abstract}

\section{Introduction}
Wavelets became a necessary mathematical tool in many investigations. They
are used in those cases when the result of the analysis of a particular
{\it signal}\footnote{The notion of a signal is used here for any
ordered set of numerically recorded information about some processes, objects,
functions etc. The signal can be a function of some coordinates, would it be
the time, the space or any other (in general, $n$-dimensional) scale.} should
contain not only the list of its typical
frequencies (scales) but also the knowledge of the definite local coordinates
where these properties are important. Thus, \an\ and processing of different
classes of nonstationary (in time) or inhomogeneous (in space)
signals is the main field  of applications of \w\ \an\ .
The most general principle of the \w\ construction is to use dilations and
translations. Commonly used wavelets form a complete orthonormal
system of functions with a finite support constructed in such a way.
 That is why by changing a scale (dilations) they can distinct the local
characteristics of a signal at various scales, and by translations they
cover the whole region in which it is studied. Due to the completeness of the
system, they also allow for the inverse transformation to be done. In the \an\
of nonstationary signals, the locality property of \w s leads to their
substantial advantage over
Fourier transform which provides us only with the
knowledge of global frequencies (scales) of the object under investigation
because the system of the basic functions used (sine, cosine or imaginary
exponential functions) is
defined on the infinite interval\footnote{Comparison of the wavelet transform
with the so-called windowed (within the finite interval) Fourier transform will
be briefly discussed below.}. However, as we shall see, the more general
definitions and, correspondingly, a variety of forms of \w s are used which
admit a wider class of functions to be considered. According to Y. Meyer
\cite{ymey}, "the wavelet bases are universally applicable: "everything that
comes to hand", whether function or distribution, is the sum of a wavelet
series and, contrary to what happens with Fourier series, the coefficients of
the wavelet series translate the properties of
the function or distribution simply, precisely and faithfully."

The literature devoted to wavelets is very voluminous, and one can easily get a
lot of references by sending the corresponding request to Internet web sites.
Mathematical problems are treated in many monographs in detail (e.g., see
\cite{ymey, daub, mcoi, meye}). The introductory courses on \w s can be found in
the books \cite{chui, hwei, kais, koor}. The nice review paper adapted for beginners
and practical users with demonstration of wavelet transform of some signals was
published in this journal about four years ago \cite{asta} and attracted much
attention. However continuous wavelets are mostly considered there whereas
discrete wavelets are just briefly mentioned. This choice was dictated by the
fact that the continuous \w s admit somewhat more visual and picturesque
presentation of the results of the \an\ of a signal
in terms of local maxima and sceleton graphs of \w\ coefficients
with continuous variables.

At the same time, the main bulk of papers dealing
with practical applications of wavelet analysis uses discrete wavelets which
will be our main concern here. This preference of discrete wavelets is related
to the fact that widely used continuous wavelets are not, strictly speaking,
orthonormal because they are both infinitely differentiable and
exponentially decreasing at infinity what violates orthonormalization property
whereas there is no such problem for discrete wavelets. That is why the
discrete \w s allow often for more accurate transform and presentation of the
analyzed signal,
and, especially, for its inverse transform after the compression procedure.
Moreover, they are better adapted to communication theory and practice.
These comments do not imply that we insist on use of only discrete \w s
for signal \an\ . On the contrary, the continuous \w s may provide sometimes
more transparent and analytical results in modelling the signal \an\ than the
discrete \w s.

The choice of a specific \w\ , would it be discrete or
continuous one, depends on the analyzed signal and on the problem to be solved.
Some functions are best analyzed using one method or another, and it depends
on the relative simplicity of the decomposition achieved. The researcher's
intuition helps resolve this "miracle". As an analogy, the example with number
systems is often considered. It is a matter of tradition and convenience to
choose the systems with the base 10, 2 or $e$. However, the Roman number system
is completely excluded if one tries to use it for multiplication. At the same
time, different problems can ask for more or less efforts both in their solution
and in graphical presentation depending on the system chosen, and our
intuition is important there.

The programs exploiting the wavelet transform are
widely used now not only for scientific research but for commercial purposes as
well. Some of them have been even described in books (e.g., see \cite{chto}).
At the same time, the direct transition from pure mathematics to computer
programming and applications is non-trivial and asks often for the individual approach
to the problem under investigation and for a specific choice of wavelets used.
Our main objective here is to describe in a suitable way the bridge that
relates mathematical wavelet constructions to practical signal processing.

The discrete wavelets look as strangers to all those accustomed to analytical
calculations because they can not be represented by analytical expressions
(except of a simplest one) or by solutions of some differential equations,
and instead are given numerically as solutions of definite functional equations
containing rescaling and translations. Moreover, in practical calculations their
direct form is not required even, and the numerical values of the coefficients
of the functional equation are only used. Thus wavelets are defined by the
iterative algorithm of the dilation and translation of a single function. This
leads to a very important procedure called multiresolution analysis which gives
rise to the multiscale local analysis of the signal and fast numerical
algorithms. Each scale contains an independent non-overlapping set of
information about the signal in form of \w\ \co s, which are determined from
an iterative procedure called fast \w\ transform. In combination, they provide
its complete {\it analysis} and the {\it diagnosis} of the underlying processes.

After such an analysis was done, one can {\it compress} (if necessary) the
resulting data by
omitting some inessential part of the {\it encoded} information. This is done with
the help of the so-called {\it quantization} procedure which commonly allocates
different weights to various \w\ \co s obtained. In particular, it helps erase
some statistical fluctuations and, therefore, increase the role of dynamical
features of a signal. It can however falsify the diagnostic if the compression
is done inappropriately. Usually, the accurate compression gives rise
to a substantial reduction of the required computer {\it storage} memory and
{\it transmission} facilities, and, consequently, to a lower expenditure. The
number of vanishing moments of \w s is important at this stage. Unfortunately,
the compression introduces unavoidable systematic errors. The mistakes
one has made will consist of multiples of the deleted \w\ \co s, and, therefore,
the regularity properties of a signal play an essential role.
{\it Reconstruction} after such compression schemes is then not perfect any
more. These two objectives are clearly antagonistic. Nevertheless, when
one tries to reconstruct the initial signal, the inverse transformation
({\it synthesis}) happens to be rather stable and reproduces its most important
characteristics if proper methods are applied. The regularity properties of \w s
used become also crucial at the reconstruction stage. The distortions of the
reconstucted signal due to quantization can be kept small, although significant
compression ratios are attained. Since the part of the signal which is not
reconstructed is often called noise, in essence, what we are doing is
denoising signals. Namely at this stage the
superiority of the discrete \w s becomes especially clear.

Thus, the objectives of signal processing consist in the accurate \an\ ,
effective coding, fast transmission and, finally, careful reconstruction
(at the transmission destination point) of the initial signal. Sometimes
the first stage of signal \an\ and diagnosis is enough for the problem to be
solved and anticipated goals to be achieved.

It has been proven that any function can be written as a superposition of
admissible \w s, and there exists a numerically stable algorithm to compute
the \co s for such an expansion. Moreover, these \co s completely characterize
the function, and it is possible to reconstruct it in a numerically stable way
by knowing these \co s.
Because of their unique properties, wavelets were used in functional analysis
in mathematics, in studies of (multi)fractal properties, singularities
and local oscillations of
functions, for solving some differential equations, for investigation of
inhomogeneous processes involving widely different scales of interacting
perturbations, for pattern recognition, for image and sound compression,
for digital geometry processing,
for solving many problems of physics, biology, medicine, technique etc
(see the recently published books \cite{chto, aapi, mall, ehja, auns}).
This list is by no means exhaustive.

One should however stress that, even though this method is very powerful,
the goals of \w\ \an\ are rather modest. It helps us describe and reveal some
features, otherwise hidden in a signal, but it does not pretend to explain the
underlying dynamics and physics origin although it may give some crucial hints
to it.
Wavelets present a new stage in optimization of this description providing
the best known representation of a signal. With the help of \w s, we merely
see things a little more clearly. To understand the dynamics, standard
approaches introduce models assumed to be driving the mechanisms generating
the observations. To define the optimality of the algorithms of \w\ \an\ , some
(still debatable!) energy and entropy criteria have been developed. They are
internal to the algorithm itself. However, the choice of the best algorithm
is also tied to an objective goal of its practical use, i.e., to some
external criteria. That is why in practical applications one should submit
the performance of a "theoretically optimal algorithm" to the judgements of
experts and users to estimate its advantage over the previously developed ones.

Despite a very active
research and impressive results, the versatility of \w\ \an\ and existence
of many different inversion formulas for \w\ transform (redundancy) imply
that these studies are presumably not in their final form yet. We shall try
to describe the situation in its {\it status nascendi}.

The main part of this paper (Sections 2--14) is devoted to description of
general properties of wavelets and use of the wavelet transform in computer
calculations. Some applications to different fields are briefly described
in Section 15.

\section{Wavelets for beginners}

Each signal can be characterized by its averaged (over some intervals) values
(trend) and by its variations around this trend. Let us call these variations
as fluctuations independently of their nature, would it be dynamic, stochastic,
psychological, physiological or any other origin. When processing a signal, one
is interested in its fluctuations at various scales because therefrom one can
learn about their origin. The goal of the \w\ \an\ is to provide tools for such
processing.

Actually, physicists dealing with experimental histograms analyze their data at
different scales when averaging over different size intervals. This is a
particular example of a simplified \w\ \an\ treated in this Section. To be more
definite, let us consider the situation when an experimentalist measures some
function $f(x)$ within the interval $0\leq x\leq 1$, and the best resolution
obtained with the measuring device is limited by 1/16th of the whole interval.
Thus the result consists of 16 numbers representing the mean values of $f(x)$
in each of these bins and can be plotted as a 16-bin histogram shown in the
upper part of Fig. 1.
It can be represented by the following formula
\be
f(x)=\sum _{k=0}^{15}s_{4,k}\scl _{4,k}(x),     \label{fscl}
\ee
where $s_{4,k}=f(k/16)/4$, and $\scl _{4,k}$ is defined as a step-like block of
the unit norm (i.e. of a height 4 and a widths 1/16) different from zero only
within the $k$-th bin. For an arbitrary $j$, one imposes the condition
$\int dx\vert \scl _{j,k}\vert ^2=1$, where the integral is taken over the
intervals of the lengths $\Delta x_j=1/2^j$ and, therefore, $\scl _{j,k}$ have
the following form $\scl _{j,k}=2^{j/2}\scl (2^jx-k)$ with $\scl $ denoting a
step-like function of the unit height in such an interval. The label 4 is
related to the total
number of such intervals in our example. At the next coarser level the average
over the two neighboring bins is taken as is depicted in the histogram just
below the initial one in Fig. 1. Up to the normalization factor, we will
 denote it as $s_{3,k}$ and the difference between the two levels shown to the
right of this histogram as $d_{3,k}$. To be more explicit, let us write down
the normalized sums and differences for an arbitrary level $j$ as
\be
s_{j-1,k}=\frac {1}{\sqrt 2}[s_{j,2k}+s_{j,2k+1}];\;\;\;      
d_{j-1,k}=\frac {1}{\sqrt 2}[s_{j,2k}-s_{j,2k+1}],      \label{dssr}
\ee
or for the backward transform (synthesis)
\be
s_{j,2k}=\frac {1}{\sqrt 2}(s_{j-1,k}+d_{j-1,k});\;\;\;
s_{j,2k+1}=\frac {1}{\sqrt 2}(s_{j-1,k}-d_{j-1,k}).   \label{s2s1}
\ee
Since, for the dyadic partition considered, this difference has
opposite signs in the neighboring bins of the previous fine level, we introduce
the function $\psi $ which is 1 and -1, correspondingly, in these bins and the
normalized functions $\psi _{j,k}=2^{j/2}\psi (2^jx-k)$. It
allows us to represent the same function $f(x)$ as
\be
f(x)=\sum _{k=0}^{7}s_{3,k}\scl _{3,k}(x)+\sum _{k=0}^{7}d_{3,k}\psi _{3,k}(x).
   \label{fsc3}
\ee
One proceeds further in the same manner to the sparser levels 2, 1 and 0 with
averaging done over the interval lengths 1/4, 1/2 and 1, correspondingly.
It is shown in the subsequent drawings  in Fig. 1.
The most sparse level with the mean value of $f$ over the whole interval
denoted as $s_{0,0}$ provides
\begin{eqnarray}
f(x)=s_{0,0}\scl _{0,0}(x)+d_{0,0}(x)\psi _{0,0}(x)+
\sum _{k=0}^1d_{1,k}\psi _{1,k}(x) \nonumber \\+
\sum _{k=0}^{3}d_{2,k}\psi _{2,k}(x)+\sum _{k=0}^{7}d_{3,k}\psi _{3,k}(x).
   \label{fsc0}
\end{eqnarray}
The functions $\scl _{0,0}(x)$ and $\psi _{0,0}(x)$ are shown in Fig. 2. The
functions $\scl _{j,k}(x)$ and $\psi _{j,k}(x)$ are the normalized by the
conservation of the norm, dilated and translated versions of them. In the next
Section we will give explicit formulae for them in a particular case of
Haar scaling functions and \w s. In practical signal processing, these
functions (and their
more sophisticated versions) are often called as the low and high-path
filters, correspondingly, because they filter the large and small scale
components of a signal. The subsequent terms in Eq. (\ref{fsc0}) show the
fluctuations (differences $d_{j,k}$) at finer and finer levels with larger $j$.
In all the cases (\ref{fscl})--(\ref{fsc0}) one needs exactly 16 \co s to
represent the function.
In general, there are $2^j$ \co s $s_{j,k}$ and $2^{j_n}-2^j$
\co s $d_{j,k}$, where $j_n$ denotes the finest resolution level (in the above
example, $j_n=4$).

All the above representations of the function $f(x)$
 (Eqs.~(\ref{fscl})-(\ref{fsc0})) are mathematically
equivalent. However, the latter one representing the \w\ analyzed function
directly reveals the fluctuation structure
of the signal at different scales $j$ and various locations $k$ present in a
set of \co s $d_{j,k}$ whereas the original form (\ref{fscl}) hides the
fluctuation patterns in the background of a general trend. The final form
(\ref{fsc0}) contains the overall average of the signal depicted by $s_{0,0}$
and all its fluctuations with their scales and positions well labelled by
15 normalized \co s $d_{j,k}$ while the initial histogram shows only the
normalized average values $s_{j,k}$ in the 16 bins studied. Moreover, in
practical applications the latter \w\ representation is preferred because for
rather smooth functions, strongly varying only at some discrete values of their
arguments, many of the high-resolution $d$-\co s in relations similar to
Eq. (\ref{fsc0}) are close to zero and can be discarded. Bands of zeros
(or close to zero values) indicate those regions where the function is smooth
enough.

At first sight, this simplified example looks somewhat trivial. However, for
more complicated  functions and more data points with some elaborate forms of
\w s it leads to a detailed \an\ of a signal and to possible strong compression
with subsequent good quality restoration.
This example provides also the illustration for the very important feature of
the whole approach with successive coarser and coarser approximations to $f$
called the \mr\ \an\ and discussed in more detail below.

\section{Basic notions and Haar \w s}

To analyze any signal, one should, first of all, choose the corresponding
basis, i.e., the set of functions to be considered as "functional coordinates".
In most cases we will deal with signals represented by the square integrable
functions defined on the real axis (or by the square summable sequences of
complex numbers). They form the infinite-dimensional Hilbert space $L^2(R)$
($l^2(Z)$). The scalar product of these functions is defined as
\be
\langle f,g\rangle =\int _{-\infty }^{\infty }dxf(x)\bar g(x), \label{lfgr}
\ee
where the bar stands for the complex conjugate.
For finite limits in the sum it becomes a finite-dimensional
Hilbert space. Hilbert spaces always have orthonormal bases, i.e., families
of vectors (or functions) $e_n$ such that
\be
\langle e_n,e_m\rangle =\delta _{nm},    \label{enem}
\ee
\be
\vert \vert f\vert \vert ^2 =\int dx\vert f(x)\vert ^2=
\sum _n\vert \langle f,e_n\rangle \vert ^2.
\label{vvfv}
\ee

In the Hilbert space, there exist some more general families of linear
independent basis vectors
called a Riesz basis which generalize the equation (\ref{vvfv}) to
\be
\alpha \vert \vert f\vert \vert ^2 \leq \sum _n\vert \langle f,e_n\rangle \vert ^2
\leq \beta \vert \vert f\vert \vert ^2      \label{avvf}
\ee
with $\alpha >0, \;\; \beta <\infty $. It is the unconditional basis where the
order in which the basis vectors are considered does not matter. Any bounded
operator with a bounded inverse maps an orthonormal basis into a Riesz basis.

Sometimes we will consider the spaces $L^p(R) (1\leq p<\infty ;\;\; p\neq 2)$,
where the norm is defined as
\be
\vert \vert f\vert \vert _{L^p}=[\int dx\vert f(x)\vert ^p]^{1/p}, \label{lpno}
\ee
as well as some other Banach spaces\footnote{Those are linear spaces equipped
with norm which generally is not derived from a scalar product and complete
with respect to this norm ($L^p(R) (1\leq p<\infty ;\;\; p\neq 2)$ is a
particular example of it.) } (see Sections 11, 12).

The Fourier transform with its trigonometric basis is well suited for the \an\
of stationary signals. Then the norm $\vert \vert f\vert \vert $ is often called
energy. For nonstationary signals, e.g., the location of that moment when the
frequency characteristics has abruptly been changed
is crucial. Therefore the basis should have a compact support. The \w s are
just such functions which span the whole space by translation of the dilated
versions of a definite function. That is why every signal can be decomposed
in the \w\ series (or integral). Each frequency component is studied with
a resolution matched to its scale. The above procedure of normalization of
functions $\scl _{j,k}$ is directly connected with the requirement of
conservation of the norm of a signal at its decompositions.

The choice of the analyzing \w\ is, however, not unique. One should choose it
in accordance with the problem to be solved. The simplicity of operations
(computing, in particular) and of representation (minimum parameters used)
plays also a crucial role. Bad choice may even prevent from getting any answer
as in the above example with Roman numbers. There are several methods of
estimating how well the chosen function is applicable to the solution of
a particular problem (see Section 6).

Let us try to construct functions satisfying the above criteria. "An educated
guess" would be to relate a function $\scl (x)$ to its dilated and translated
version. The simplest linear relation with $2M$ coefficients is
\be
\scl (x)=\sqrt 2 \sum _{k=0}^{2M-1}h_k\scl (2x-k)    \label{sclx}
\ee
with the dyadic dilation 2 and integer translation $k$. At first sight, the
chosen normalization of the \co s $h_k$ with the factor $\sqrt 2$ looks
somewhat arbitrary. Actually, it is defined {\it a'posteriori} by the
traditional form of fast algorithms for their calculation (see Eqs. (\ref{sshs})
and (\ref{dsgs}) below) and normalization of functions
$\scl _{j,k}(x), \psi _{j,k}(x)$. It is used in all the books cited above.
However, sometimes (see \cite{daub}, Chapter 7) it is replaced by
$c_k=\sqrt 2h_k$.

For discrete values
of the dilation and translation parameters one gets discrete \w s.
The value of the dilation factor determines the size of cells in the lattice
chosen. The integer $M$ defines the number of \co s and the length
of the \w\ support. They are interrelated because from the definition of $h_k$
for orthonormal bases
\be
h_k=\sqrt 2 \int dx\scl (x)\bar {\scl }(2x-k)  \label{hkde}
\ee
it follows that only finitely many $h_k$ are nonzero if $\scl $ has a finite
support. The normalization condition is chosen as
\be
\int _{-\infty }^{\infty }dx\scl (x)=1.  \label{norm}
\ee
The function $\scl (x)$ obtained from the solution of this equation is called
a scaling function\footnote{It is often called also a "father \w\ " but we will
not use this term.}. If the scaling function is known, one can form a
"mother \w\ " (or a basic \w\ ) $\psi (x)$ according to
\be
\psi (x)=\sqrt 2 \sum _{k=0}^{2M-1}g_k\scl (2x-k),  \label{psix}
\ee
where
\be
g_k=(-1)^{k}h_{2M-k-1}.   \label{gkhk}
\ee

The simplest example would be for $M=1$ with two non-zero \co s $h_k$ equal to
$1/\sqrt 2$, i.e., the equation leading to the Haar scaling function $\scl _H(x)$:
\be
\scl _H(x)=\scl _H(2x)+\scl _H(2x-1).    \label{sclh}
\ee
One easily gets the solution of this functional equation
\be
\scl _H(x)=\theta (x)\theta (1-x),    \label{thth}
\ee
where $\theta (x)$ is the Heaviside step-function equal to 1 at positive
arguments and 0 at negative ones.
The additional boundary condition is $\scl _H(0)=1,\; \scl _H(1)=0$.
This condition is important for the simplicity of the whole procedure of
computing the \w\ \co s when two neighboring intervals are considered.

The "mother \w\ " is
\be
\psi _H(x)=\theta (x)\theta (1-2x)-\theta (2x-1)\theta (1-x).  \label{psih}
\ee
with boundary values defined as $\psi _H(0)=1,\; \psi_H(1/2)=-1,\;
\psi _H(1)=0$.
This is the Haar \w\ \cite{haar} known since 1910 and used in the functional
\an\ . Namely this example has been considered in the previous Section fot
the histogram decomposition.
Both the scaling function $\scl _H(x)$ and the "mother \w\ " $\psi _H(x)$
are shown in Fig. 2. It is the first one of a family of compactly supported
orthonormal \w s $_M\psi : \; \psi _H=_1\psi $.
It possesses the locality property since its support $2M-1=1$ is compact.

The dilated and translated versions of the scaling function $\scl $ and
the "mother \w " $\psi $
\be
\scl _{j,k}=2^{j/2}\scl (2^jx-k),    \label{sclj}
\ee
\be
\psi _{j,k}=2^{j/2}\psi (2^jx-k)     \label{psij}
\ee
form the orthonormal basis as can be (easily for Haar \w s) checked\footnote{We
return back to the general case and therefore omit the index $H$
because the same formula will be used for other \w s.}. The choice of $2^j$
with the integer valued $j$ as a scaling factor leads to the unique and
selfconsistent procedure of computing the \w\ \co s. In principle, there
exists an algorithm of derivation of the compact support \w s with an arbitrary
rational number in place of 2. However, only for this factor, it has been shown
that there exists the explicit algorithm with the regularity of a \w\ increasing
linearly with its support. For example, for the factor 3 the regularity index
only grows logarithmically. The factor 2 is probably distinguished here
as well as in music where octaves play a crucial role. If the dilation factor
is 2, then the Fourier transform of the "mother" \w\ is essentially localized
between $\pi $ and $2\pi $. However, for some practical applications, a sharper
frequency localization is necessary, and it may be useful to have \w\ bases
with a narrower bandwidth. The fractional dilation \w\ bases provide one of the
solutions of this problem but there also exist other possibilities.

The Haar \w\ oscillates so that
\be
\int _{-\infty }^{\infty }dx\psi (x)=0.    \label{osci}
\ee
This condition is common for all the \w s. It is called the oscillation or
cancellation condition. From it, the origin of the name \w\
becomes clear. One can describe a "wavelet"  as a function that oscillates
within some interval like a wave but is then localized by damping outside this
interval. This is a necessary condition for \w s to form an unconditional
(stable) basis. We conclude that for special choices of \co s $h_k$ one gets
the specific forms of "mother" \w s and orthonormal bases.

One may decompose any function $f$ of $L^2(R)$ at any resolution level $j_n$
in a series
\be
f=\sum _{k}s_{j_n,k}\scl _{j_n,k}+\sum _{j\geq j_n,k}d_{j,k}\psi _{j,k}.  \label{fdec}
\ee
At the finest resolution level $j_n=j_{max}$ only $s$-\co s are left, and one
gets the scaling-function representation
\be
f(x)=\sum _ks_{j_{max},k}\scl _{j_{max},k}.    \label{fjma}
\ee
In the case of the Haar \w s it corresponds to the initial experimental histogram
with finest resolution. Since we will be interested in its \an\ at varying
resolutions, this form is used as an initial input only. The final
representation of the same data (\ref{fdec}) shows all the fluctuations in the
signal.
The \w\ \co s $s_{j,k}$ and $d_{j,k}$ can be calculated as
\be
s_{j,k}=\int dxf(x)\scl _{j,k}(x),   \label{ssjk}
\ee
\be
d_{j,k}=\int dxf(x)\psi _{j,k}(x).   \label{ddjk}
\ee
However, in practice their values are determined from the fast \w\ transform
described below.

In reference to the particular case of the Haar \w\ , considered above,
these \co s are often referred
as sums ($s$) and differences ($d$), thus related to mean values and
fluctuations.

For physicists familiar with experimental histograms, it generalizes the
particular example discussed in the previous Section. The first sum in
(\ref{fdec}) with the scaling functions $\scl _{j,k}$ shows the
average\footnote{The averaging is done with the weight functions
$\scl _{j,k}(x)$.} values of $f$ within the
dyadic intervals $[k2^{-j},(k+1)2^{-j})$, and the second term contains all the
fluctuations of the function $f$ in this interval. They come from ever smaller
intervals which correspond to larger values of the scale parameter $j$. One
would say that it "brings into focus" the finer details of a signal. This
touching in of details is regularly spaced, taking account of dimension - the
details of dimension $2^{-j}$ are placed at the points $k2^{-j}$. At the
lowest (most sparse) level $j_0$ the former sum consists of a single term with
the overall weighted average $\langle f\rangle =s_{j_0,k_0}$, where $k_0$ is
the center of the histogram. The second sum in (\ref{fdec}) shows fluctuations
at all the scales. At the
next, more refined level $j_1>j_0$, there are two terms in the first sum which
show the average values of $f$ within half-intervals with their centers
positioned at $k_1,\;\; k_2$. The number of terms in the second sum becomes
less by one term which was previously responsible for the fluctuations at the
half-interval scale. The total number of terms in the expansion stays unchanged.
Here we just mention that according to (\ref{fdec}) the number of
terms in each sum depends on a definite resolution level. Changing the level
index by 1, we move some terms to another sum, and each of these
representations are "true" representations of the histogram at different
resolution levels.

Formally a similar procedure may be done the other way
round by going to the sparser resolutions $j<j_0$. Even if we "fill out" the
whole support of $f$, we can still keep going with our averaging trick.
Then the average value of $f$
diminishes, and one can neglect the first sum in (\ref{fdec}) in the
$L^2$ sense because its $L^2$-norm (\ref{vvfv}) tends to zero. In the histogram
example, it decreases as $\vert \langle f\rangle \vert \propto N^{-1}$ and
$\vert \langle f\rangle \vert ^2\propto N^{-2}$ whereas the integration region
is proportional to $N$, i.e., $\vert \vert \langle f\rangle \vert \vert ^2
\propto N^{-1}\rightarrow 0$ for $N\rightarrow \infty $. That is why
only the second term in (\ref{fdec}) is often considered, and the result is
often called as the \w\ expansion. It also works if $f$
is in the space $L^p(R)$ for $1<p<\infty $ but it can not be done if it
belongs to $L^1(R)$ or $L^{\infty }(R)$. For example, if $f=1$ identically,
all \w\ \co s $d_{j,k}$ are zero, and only the first sum matters. For the
histogram
interpretation, the neglect of this sum would imply that one is not interested
in its average value but only in its shape determined by fluctuations at
different scales. Any function can be approximated to a precision $2^{j/2}$
(i.e., to an arbitrary high precision at $j\rightarrow -\infty $) by a finite
linear combination of Haar \w s.

The Haar \w s are also suitable for the studies of functions belonging to $L^p$
spaces, i.e., possessing higher moments.

Though the Haar \w s provide a good tutorial example of an orthonormal basis,
they suffer from several
deficiences. One of them is the bad analytic behavior with the abrupt change
at the interval bounds, i.e., its bad regularity properties. By this we mean
that moments of the Haar \w\ are different from zero - only the integral
(\ref{osci}) of the function itself is zero. The Haar \w\ does not have good
time-frequency localization. Its Fourier transform decays like
$\vert \omega \vert ^{-1}$ for $\omega \rightarrow \infty $.

It would be desirable to build
up \w s with better regularity. The advantage of them compared to the
Haar system shows up in the smaller number of the \w\ \co s which are large
enough to account for and in their applicability to a wider set of
functional spaces besides $L^2$. The former feature is related to the fact
that the \w\ \co s are significantly different from zero only near
singularities of $f$ (strong fluctuations!). Therefore \w\ series of standard
functions with isolated singularities are "sparse" series in contrast to the
Fourier series which are usually dense ones for rather regular functions.
The latter feature
allows us to get an access to local and global regularities of functions
under investigation. The way to this program was opened by the \mr\ \an\ .

\section{Multiresolution analysis and Daubechies \w s}

The relation (\ref{fdec}) shows that a general function $f$ can be approximated
by a sequence of very simple functions $\scl _{j,k},\;\; \psi _{j,k}$.
The above example has demonstrated that the Haar functions are local and cover
the whole space $L^2(R)$ by using the translation $k$. They are orthogonal
for different resolution scales $j$. The transition from $j$ to $j+1$ is
equivalent to the replacement of $x$ by $2x$, i.e., to the rescaling which
allows for the \an\ to be done at various resolutions.

However the Haar \w s are oversimplified and not regular enough. The goal is
to find a general class of those functions which would satisfy the
requirements of locality, regularity and oscillatory behavior. Note that in
some particular cases the orthonormality property sometimes can be
relaxed. They should be simple enough in the
sense that they are of being sufficiently explicit and regular to be completely
determined by their samples on the lattice defined by the factors $2^j$.

The general approach which respects these properties is known as the \mr\
approximation. A rigorous mathematical definition is given in Appendix 1.
Here we just describe its main ingredients.

The \mr\ \an\ consists of a sequence of successive approximation spaces
$V_j$ which are scaled and invariant under integer translation versions of
the central functional space $V_0$. To explain the meaning of these spaces
in a simple example, we show in
Fig. 3 what the projections of some function on the Haar spaces $V_0$, $V_1$
might look like. One easily recognizes the histogram representation of this
function. The comparison of histograms at the two levels shows that the first
sum in Eq. (\ref{fdec}) provides the "blurred image" or "smoothed means" of
$f(x)$ in each interval, while the second sum of this equation adds finer and
finer details of smaller sizes. Thus the general distributions are decomposed
into the series of correctly localized fluctuations having a characteristic
form defined by the \w\ chosen.

The functions $\scl _{j,k}$ form an orthonormal basis in $V_j$. The orthogonal
complement of $V_j$ in $V_{j+1}$ is called $W_j$. The subspaces $W_j$ form a
mutually orthogonal set. The sequence of $\psi _{j,k}$ constitutes an
orthonormal basis for $W_j$ at any definite $j$. The whole collection of
$\psi _{j,k}$ and $\scl _{j,k}$ for all $j$ is an orthonormal basis for $L^2(R)$.
 This ensures us that we have constructed a \mr\ \an\ approach, and the
functions $\psi _{j,k}$ and $\scl _{j,k}$ constitute the small and large
scale filters, correspondingly. The whole procedure of the \mr\ \an\ can be
demonstrated in graphs of Fig. 4.

In accordance with the above declared goal, one can define the notion of \w s
(see Appendix 1) so that the functions $2^{j/2}\psi (2^jx-k)$ are the \w s
(generated by the "mother" $\psi $), possessing the regularity,
the localization and the oscillation properties.

At first sight, from our example with Haar wavelets, it looked as
if one is allowed to choose coefficients $h_k$ at his own wish.
This impression is, however, completely wrong. The general
properties of scaling functions and wavelets define these
coefficients in a unique way in the framework of the
multiresolution analysis approach.

Let us show how the program of the \mr\ \an\ works in practice when applied to
the problem of finding out the \co s of any filter $h_k$ and $g_k$. They can be
directly obtained from the definition and properties of the discrete \w s. These
\co s are defined by the relations (\ref{sclx}) and (\ref{psix})
\begin{equation}
 \scl(x)=\sqrt 2 \sum_{k}h_k \scl(2x-k); \qquad
\psi(x)=\sqrt 2 \sum_{k}g_k \scl(2x-k),  \label{scps}
\end{equation}
where $\sum_k\vert h_k\vert ^2<\infty $.
The orthogonality of the scaling functions defined by the relation
\begin{equation}
\int dx \scl(x)\scl(x-m) = 0
\end{equation}
leads to the following equation for the \co s:
\begin{equation}
 \sum_{k}h_k h_{k+2m}=  \delta_{0m}. \label{CoeffDefFirst}
\end{equation}
The orthogonality of \w s to the scaling functions
\begin{equation}
\int dx \psi(x)\scl(x-m) = 0
\end{equation}
gives the equation
\begin{equation}
\sum_{k}h_k g_{k+2m}= 0,
\end{equation}
having a solution of the form
\begin{equation}
  g_k=(-1)^k h_{2M-1-k}.\label{gFromh}
\end{equation}
Another condition of the orthogonality of \w s to all polynomials up to
the power $(M-1)$, defining its regularity and oscillatory behavior
\begin{equation}
\int dx x^n \psi(x) = 0, \qquad n=0, ..., (M-1),
\end{equation}
provides the relation
\begin{equation}
\sum_{k}k^n g_k = 0,
\end{equation}
giving rise to
\begin{equation}
\sum_{k} (-1)^k k^n h_k = 0, \label{CoeffDefSecond}
\end{equation}
when the formula (\ref{gFromh}) is taken into account.

The normalization condition
\begin{equation}
\int dx \scl(x) = 1
\end{equation}
can be rewritten as another equation for $h_k$:
\begin{equation}
\sum_{k} h_k = \sqrt 2.   \label{CoeffDefLast}
\end{equation}

Let us write down the equations (\ref{CoeffDefFirst}),(\ref{CoeffDefSecond}),
(\ref{CoeffDefLast}) for $M=2$ explicitly:
\begin{eqnarray*}
h_0 h_2 + h_1 h_3 = 0,\\
h_0 - h_1 + h_2 - h_3 = 0,\\
-h_1 + 2h_2 - 3h_3 = 0,\\
h_0+h_1+h_2+h_3 = \sqrt 2.
\end{eqnarray*}
The solution of this system is
\begin{equation}
h_3 = \frac{1}{4\sqrt 2}(1 \pm \sqrt{3}),\quad
h_2=\frac{1}{2\sqrt 2}+h_3, \quad
h_1=\frac {1}{\sqrt 2}-h_3, \quad
h_0 = \frac{1}{2\sqrt 2}-h_3,
\end{equation}
that, in the case of the minus sign for $h_3$, corresponds to the well known
filter
\begin{equation}
h_0 =\frac{1}{4\sqrt 2}(1+\sqrt{3}), \quad
h_1 =\frac{1}{4\sqrt 2}(3+\sqrt{3}), \quad
h_2 =\frac{1}{4\sqrt 2}(3-\sqrt{3}), \quad
h_3 =\frac{1}{4\sqrt 2}(1-\sqrt{3}).
\end{equation}

These \co s define the simplest $D^4$ (or $_2\psi $) \w\ from the famous family
of orthonormal Daubechies \w s with finite support. It is shown in the upper
part of Fig. 5 by the dotted line with the corresponding scaling function shown
by the solid line. Some other higher rank \w s are also shown there.
It is clear from this Figure (especially, for $D^4$) that
\w s are smoother in some points than in others. The choice of the plus sign
in the expression for $h_3$ would not change the general shapes of the scaling
function and \w\ $D^4$. It results in their mirror symmetrical forms obtained by
a simple reversal of the signs on the
horizontal and vertical axes, correspondingly. However, for higher rank \w s
different choices of signs would correspond to different forms of the \w\ .
After the signs are chosen, it is clear that compactly supported
\w s are unique, for a given \mr\ \an\, up to a shift in the argument
(translation) which is inherently there. The dilation factor must be rational
within the framework of the \mr\ \an\ . Let us note that $_2\scl $ is H\"{o}lder
continuous with the global exponent $\alpha =0.55$ (see Eq. (\ref{hold}) below)
and has different local H\"{o}lder exponents on some fractal sets. Typically,
\w s are more regular in some points than in others.

For the filters of higher order in $M$,
i.e., for higher rank Daubechies \w s, the \co s can be obtained in an
analogous manner. It is however necessary to solve the equation of the
$M$-th power in this case. Therefore, the numerical values of the \co s can be
found only approximately, but with any predefined accuracy.
The \w\ support is equal to $2M-1$. It is wider than for the Haar \w s.
However the regularity properties are better. The higher order \w s are
smoother compared to $D^4$ as seen in Fig. 5. The Daubechies \w\ with
$r$ vanishing moments has $\mu r$ continuous derivatives where
$\mu \approx 0.2$ as was estimated numerically. This means that $80\%$ of
zero moments are "wasted". As the regularity $r$ increases, so does the
support in general. For sufficiently regular functions, Daubechies \w\ \co s
are much smaller ($2^{-Mj}$ times) than the Haar \w\ \co s, i.e., the signal
can be compressed much better with Daubechies \w s. Since they are more
regular, the synthesis is more efficient also.

One can ask the question whether the regularity or the number of vanishing moments
is more important. The answer depends on the application, and is not always
clear. It seems that the number of vanishing moments is more important for
stronger compression which increases for larger number of vanishing moments,
while regularity can become crucial in inverse synthesis to smooth the
errors due to the compression (omission of small \co s).

In principle, by solving the functional equation (\ref{sclx}) one can find
the form of the scaling function and, from (\ref{psix}), the form of
the corresponding "mother \w\ ". There is no closed-form analytic formula for
the compactly supported $\scl (x),\; \psi (x)$ (except for the Haar case).
Nevertheless one can compute their plots, if they are continuous, with
arbitrarily high precision using a fast cascade algorithm with the \w\
decomposition of $\scl (x)$ which is a special case of a refinement scheme
(for more details, see \cite{daub}). Instead of a refinement cascade one can
compute $\scl (2^{-j}k)$ directly from Eq. (\ref{sclx}) starting from
appropriate $\scl (n)$. However, in practical calculations the above \co s
$h_k$ are used only without referring to the shapes of \w s.

Except for the
Haar basis, all real orthonormal \w\ bases with compact support are asymmetric,
i.e., they have neither symmetry nor antisymmetry axis (see Fig. 5). The
deviation of a \w\ from symmetry is judged by how much the phase of the
expresion $m_0(\omega )=\sum _k h_ke^{-ik\omega }$
deviates from a linear function. The "least asymmetric" \w s are constructed
by minimizing this phase. Better symmetry for a \w\ necessarily implies better
symmetry for the \co s $h_k$ but the converse statement is not always true.

\section{Fast wavelet transform and coiflets}

After calculation of the \co s $h_k$ and $g_k$, i.e., the choice of a definite
\w\ , one is able to perform the \w\ \an\ of a signal $f(x)$ because the \w\
orthonormal basis ($\psi _{j,k};\; \scl _{j,k}$) has been defined. Any function
$f \in L^2(R)$ is completely characterized by the \co s of its decomposition
in this basis and may be decomposed according to the formula (\ref{fdec}).
Let us make the sum limits in this formula more precise. The function $f(x)$
may be considered at any $n$-th resolution level $j_n$. Then the separation
of its average values and fluctuations at this level looks like
\be
f(x)=\sum _{k=-\infty }^{\infty }s_{j_n,k}\scl _{j_n,k}(x)+\sum _{j=j_n}^{\infty }
\sum _{k=-\infty }^{\infty }d_{j,k}\psi _{j,k}(x).  \label{flim}
\ee
At the infinite interval, the first sum may be omitted as explained above,
and one gets the pure \w\ expansion.
As we stressed already, the \co s $s_{j,k}$ and $d_{j,k}$ carry information
about the content of the signal at various scales.
They can be calculated directly using the formulas (\ref{ssjk}),
(\ref{ddjk}). However this algorithm is inconvenient for numerical computations
 because
it requires many ($N^2$) operations where $N$ denotes a number of the sampled
values of the function. We will describe a faster algorithm. It is
clear from Fig. 6, and the fast algorithm' formulas are presented below.

In real situations with digitized signals, we have to deal with finite sets
of points. Thus, there always exists the finest level of resolution where
each interval contains only a single number. Correspondingly, the sums over
$k$ will get the finite limits. It is convenient to reverse the level indexation
assuming that the label of this fine scale is $j=0$.
It is then easy to compute the \w\ \co s for more sparse resolutions $j\geq 1$.

The \mr\ \an\ naturally leads to an hierarchical and fast scheme for the
computation of the \w\ \co s of a given function. The functional equations
(\ref{sclx}), (\ref{psix}) and the formulas for the \w\ \co s
(\ref{ssjk}), (\ref{ddjk}) give rise, in case of Haar \w s, to the relations
 (\ref{dssr}), or for the backward transform (synthesis)
to (\ref{s2s1}).

In general, one can get the iterative formulas of the fast \w\ transform
\be
s_{j+1,k}=\sum _mh_ms_{j,2k+m},      \label{sshs}
\ee
\be
d_{j+1,k}=\sum _mg_ms_{j,2k+m}      \label{dsgs}
\ee
where
\be
s_{0,k}=\int dxf(x)\scl (x-k).         \label{sifs}
\ee
These equations yield fast algorithms (the so-called pyramid algorithms) for
computing the \w\ \co s, asking now just for $O(N)$ operations to be done.
Starting from $s_{0,k}$, one computes all other \co s provided the \co s
$h_m,\; g_m$ are known. The explicit shape of the \w\ is not used
in this case any more. The simple form of these equations is the only
justification for introducing the factor $\sqrt 2$ in the functional equation
(\ref{sclx}). In principle, the \co s $h_m,\; g_m$ could be renormalized.
However, in practice the Eqs. (\ref{sshs}) and (\ref{dsgs}) are used much more
often than others, and this normalization is kept intact. After choosing a
particular \w\ for \an\ ,i.e., choosing $h_m,\; g_m$, one uses only the Eqs.
(\ref{sshs}) and (\ref{dsgs}) for computing the \w\ \co s, and additional
factors in these equations would somewhat complicate the numerical processing.

The remaining problem lies in the initial data. If an explicit expression for
$f(x)$ is available, the \co s $s_{0,k}$ may be evaluated directly according
to (\ref{sifs}). But this is not so in the situation when only discrete values
are available. To get good accuracy, one has to choose very small bins (dense
lattice) which is often not accessible with finite steps of sampling.
In such a case, the usually adopted simplest solution consists in
directly using the values $f(k)$ of the available sample in place of the \co s
$s_{0,k}$ and start the fast \w\ transform using formulas
(\ref{sshs}), (\ref{dsgs}). This is a safe operation since the pyramid algorithm yields
perfect reconstruction, and the \co\ $s_{0,k}$ essentially represents a local
average of the signal provided by the scaling function.

In general, one can choose
\be
s_{0,k}=\sum _mc_mf(k-m).     \label{sscf}
\ee
The above supposition $s_{0,k}=f(k)$ corresponds to $c_m=\delta _{0m}$. This
supposition may be almost rigorous for some specific choices of scaling functions
named coiflets (by the name of R. Coifman whose ideas inspired I. Daubechies to
build up these \w s). It is possible to
construct \mr\ \an\ with the scaling function having vanishing moments, i.e.,
such that
\be
\int dxx^m\scl (x)=0; \;\;\;\;\;\;\;\;\;\; 0<m<M.      \label{ixms}
\ee
To construct such \w s (coiflets), one has to add to the equations for
determining the \co s $h_k$ a new condition
\be
\sum _kh_kk^m=0, \;\;\;\;\;\;\;\;\;\;\; 0<m<M,    \label{skhk}
\ee
which follows from the requirement (\ref{ixms}).

Coiflets are more symmetrical than Daubechies \w s as is seen from Fig. 7 if
compared to Fig. 5. The latter do not have the property (\ref{ixms}). The price
for this extra
generalization is that coiflets are longer than Daubechies \w s. If in the
latter case the length of the support is $2M-1$, for coiflets it is equal to
$3M-1$. The error in the estimation of $s_{j,k}$ decreases with the number of
vanishing moments as $O(2^{-jM})$.

There are other proposals to improve the first step of the iterative procedure
promoted by using $T$-polynomials by W. Sweldens \cite{swel} or the
so-called "lazy" or interpolating \w s by S. Goedecker and O. Ivanov \cite{giva}.
 The latter is most convenient for simultaneous \an\ at different resolution
levels, in particular for non-equidistant lattices.

Inverse fast \w\ transform allows for the reconstruction of the function starting
from its \w\ \co s.

\section{Choice of \w s}

Above, we demonstrated three examples of discrete orthonormal compactly
supported \w s. The regularity property, the number of vanishing moments and
the number of \w\ \co s exceeding some threshold value were considered
as possible criteria for the choice of a particular \w\ not to say about
computing facilities. Sometimes the so-called information cost functional used
by statisticians is introduced, and one tries to minimize it and thus select the
optimal basis. In particular, the entropy criterium for the probability
distribution of the \w\ \co s is also considered \cite{meye, chto}.
The entropy of $f$ relative to the \w\ basis measures the number of significant
terms in the decomposition (\ref{fdec}). It is defined by
$\exp (-\sum _{j,k}\vert d_{j,k}\vert ^2\log \vert d_{j,k}\vert ^2)$. If we
have a collection of orthonormal bases, we will choose for the \an\ of $f$
the particular basis that yields the minimum entropy.

The number of possible \w s at our disposal is much larger than the
above examples show. We will not discuss all of them just mentioning some and
referring the reader to the cited books.

\begin{itemize}
\item First, let us mention splines which lead to \w s with non-compact support
but with the exponential decay at
infinity and with some (limited) number of continuous derivatives. The special
orthogonalization tricks should be used there. The splines are intrinsically
associated with interpolation schemes for finding more precise initial
values of $s_{0,k}$ relating them to some linear combinations of the sampled
values of $f(x)$.

\item To insure both the full symmetry and exact reconstruction, one has to use the
so-called biorthogonal \w s.
It means that two dual \w\ bases $\psi _{j,k}$ and $\psi ^{j,k}$,
associated with two different \mr\ ladders, are used. They can have very
different regularity properties. A function $f$ may be represented in two
forms, absolutely equivalent until the compression is done:
\be
f=\sum _{j,k}\langle f,\psi _{j,k}\rangle \psi ^{j,k}  \label{fptp}
\ee
\be
=\sum _{j,k}\langle f,\psi ^{j,k}\rangle \psi _{j,k},  \label{ftpp}
\ee
where $\psi $ and its dual \w\ satisfy the biorthogonality requirement
$\langle \psi _{j,k}\vert \psi ^{j',k'}\rangle =\delta _{j,k;j',k'}$.
In distinction to Daubechies \w s, where regularity is tightly connected with
the number of vanishing moments, biorthogonal \w s have much more freedom.
If one of them has the regularity of the order $r$, then its dual partner \w\
has automatically at least $r$ vanishing moments.
If $\psi ^{j,k}$ is much more regular than $\psi _{j,k} $, then
$\psi _{j,k}$ has many more vanishing moments than $\psi ^{j,k}$.
It allows us to choose, e.g., very regular $\psi ^{j,k}$ and get many vanishing
moments of $\psi _{j,k}$. The large number of
vanishing moments of $\psi _{j,k}$ leads to better compressibility for reasonably
smooth $f$. If the compression has been done, the formula (\ref{fptp}) is
much more useful than (\ref{ftpp}). The number of significant terms is much
smaller, and, moreover, better regularity of $\psi ^{j,k}$ helps reconstruct
$f$ more precisely. Biorthogonal bases are close to an orthonormal basis.
Both \w s can be made symmetric. Symmetric biorthogonal \w s close to
orthonormal basis are close to coiflets. The construction of biorthogonal \w\
bases is simpler than of orthonormal bases.

\item The existence of two-scale relations is the main feature of the construction
of \w\ packets. The general idea of the \w\ packets is to iterate further the
splitting of the frequency band, still keeping the same pair of filters. The
scaling function introduced above acquires the name $w_0$, and the
packet is built up through the following iterations
\be
w_{2n}(x)=\sum _kh_kw_n(2x-k),    \label{w2nh}
\ee
\be
w_{2n+1}(x)=\sum _kg_kw_n(2x-k).    \label{w2ng}
\ee
The usual mother \w\ is represented by $w_1$. This family of \w s forms an
orthonormal  basis in $L^2(R)$ which is called the fixed scale \w\ packet basis.
Fig. 8 demonstrates the whole construction.

\item One can abandon the orthonormality property and construct non-orthogonal \w s
called frames. An important special class of frames is given by the Riesz bases
 of $L^2(R)$. A Riesz basis is a frame, but the converse is not true in general.
The frames satisfy the following requirement:
\be
A\vert \vert f\vert \vert ^2\leq \sum _{j \in J}\vert \langle f,\scl _j\rangle
\vert ^2\leq B\vert \vert f \vert \vert ^2.      \label{fram}
\ee
The constants $A$ and $B$ are called the frame bounds. For $A=B$ one calls
them as tight frames. The case $A=B=1$ corresponds to orthonormal \w s.

\item When acting by (singular) operators, one sometimes gets infinities when
usual \w s are considered. Some suitable function $b(x)$ may be used to specify
the extra conditions which will be necessary and sufficient for the result
of the (singular integral) operator action to be continuous in $L^2$. In this
case one chooses the so-called "\w s which are adapted to $b$". Any function $f$
is again decomposed as
\be
f(x)=\sum _{\lambda }a(\lambda )\psi ^{(b)}_{\lambda }(x)  \label{alpl}
\ee
but the \w\ \co s are now calculated according to
\be
a_{\lambda }=\int dxb(x)f(x)\psi ^{(b)}_{\lambda }(x).   \label{plal}
\ee
They satisfy the normalization condition
\be
\int dxb(x)\psi ^{(b)}_{\lambda }(x)\psi ^{(b)}_{\lambda '}(x)=\delta _{\lambda,
\lambda '}.    \label{dllp}
\ee
The cancellation condition now reads
\be
\int dxb(x)\psi ^{(b)}_{\lambda }(x)=0.   \label{canc}
\ee
As we see, the cancellation is also adapted to the function $b$ (in general,
to the "complex measure" $dxb(x)$).

\item Up to now we considered the \w s with the dilation factor equal 2. It is most
convenient for numerical calculations. However, it can be proved \cite{aush, daub}
that within the framework of a multiresolution \an\ , the dilation factor must
be rational, and no other special requirements are imposed. Therefore one can
construct schemes with other integer or fractional dilation factors. Sometimes
they may provide a sharper frequency localization. For \w s with the dilation
factor 2, their Fourier transform is essentially localized within one octave
between $\pi $ and $2\pi $, whereas the fractional dilation \w\ bases may
have a bandwidth narrower than one octave.

\item Moreover, one can use the continuous \w s as described at some length in
Ref. \cite{asta}. The \w\ is written in a form
\be
\psi _{a,b}(x)=\vert a\vert ^{-1/2}\psi (\frac {x-b}{a}).  \label{wcon}
\ee
The direct and inverse formulas of the \w\ transform look like
\be
W_{a,b}=\vert a\vert ^{-1/2}\int dxf(x)\psi (\frac {x-b}{a}),     \label{wabi}
\ee
\be
f(x)=C_{\psi }^{-1}\int \frac {dadb}{a^2}W_{a,b}\psi _{a,b}(x).  \label{fcps}
\ee
Here
\be
C_{\psi}=\int \frac {d\omega }{\vert \omega \vert}\vert \psi (\omega )\vert ^2=
\int \frac {d\omega }{\vert \omega \vert}\vert \int dxe^{-ix\omega }
\psi (x)\vert ^2.      \label{cpii}
\ee
Herefrom one easily recognizes that the oscillation of \w s required by
Eq. (\ref{osci}) is a general property. The vanishing Fourier transform
of a \w\ at $\omega \rightarrow 0$, which is just directly the condition
(\ref{osci}), provides a finite value of $C_{\psi }$
in (\ref{cpii}). One of the special  and often used examples of continuous \w s
is given by the second derivative of the Gaussian function which is called
the Mexican hat (MHAT) \w\ for its shape. Actually, it can be considered as a
special frame as shown by Daubechies. The reconstruction procedure (synthesis)
is complicated and can become unstable in this case.
However it is widely applied for \an\ of signals. The formula (\ref{wabi}) is
a kind of a convolution operation. That is why the general theory of the
so-called Calderon-Zygmund operators \cite{mcoi} (see Appendix 2) is, in
particular, applicable to problems of the \w\ decomposition.
\end{itemize}

\section{Multidimensional \w s}

The \mr\ \an\ can be performed in more than 1 dimensions. There are two ways
\cite{meye} to generalize it to the two-dimensional case, for example, but we
will consider the most often used construction given by tensor products.
The tensor product method
is a direct way to construct an $r$-regular \mr\ approximation which produces
multidimensional \w s of compact support. This enables us to analyze every
space of functions or distributions in $n$ dimensions whose regularity is
bounded by $r$.

The trivial way of constructing a two-dimensional orthonormal basis
starting from a one-dimensional orthonormal \w\ basis
$\psi _{j,k}(x)=2^{j/2}\psi (2^jx-k)$ is simply to take the tensor product
functions generated by two one-dimensional bases:
\be
\Psi _{j_1,k_1;j_2,k_2}(x_1,x_2)=\psi _{j_1,k_1}(x_1)\psi _{j_2,k_2}(x_2).
\label{2dim}
\ee
In this basis the two variables $x_1$ and $x_2$ are dilated independently.

More interesting for many applications is another construction, in which
dilations of the resulting orthonormal \w\ basis control both variables
simultaneously, and the \2\ \w s are given by the following expression:
\be
2^j\Psi (2^jx-k,2^jy-l), \;\;\;\;\;\;\;\;\; j,k,l \in Z,  \label{jpxy}
\ee
where $\Psi $ is no longer a single function: on the contrary, it
consists of three elementary \w s. To get an orthonormal basis of $W_0$
one has to use in this case three families
$\scl (x-k)\psi (y-l),\; \psi (x-k)\scl (y-l),\; \psi (x-k)\psi (y-l)$.
Then the \2\ \w s are
$2^j\scl (2^jx-k)\psi (2^jy-l),\; 2^j\psi (2^jx-k)\scl (2^jy-l),\;
2^j\psi (2^jx-k)\psi (2^jy-l)$.  In the \2\ plane, the \an\ is done along
the horizontal, vertical and diagonal strips with the same resolution in
accordance with these three \w s.

Fig. 9 shows how this construction looks like.
The schematic representation of this procedure in the left-hand side of the
Figure demonstrates how the corresponding \w\ \co s are distributed for
different resolution levels ($j=1$ and 2). The superscripts $d, v$ and $h$
correspond to diagonal, vertical and horizontal \co s. In the right-hand side
of the Figure, a set of geometrical objects is decomposed into two layers.
One clearly sees how vertical, horizontal and diagonal edges are emphasized
in the corresponding regions. One should notice also that the horizontal
strip is resolved into two strips at a definite resolution level.

In the general $n$-dimensional case, there exist $2^n-1$ functions which form
an orthonormal basis and admit the \mr\ \an\ of any function from $L^2(R^n)$.
The normalization factor is equal to $2^{nj/2}$ in front of the function, as
can be guessed already from the above \2\ case with this factor equal to $2^j$
in contrast to $2^{j/2}$ in one dimension.

There exists also a method to form the \w\ bases which are not reducible
to tensor products of \1\ \w s (see \cite{meye}). In dimension 1, every
orthonormal basis arises from a \mr\ approximation. In dimensions greater
than 1, it is possible to form an orthonormal basis such that there is no
$r$-regular \mr\ approximation ($r\geq 1$) from which these \w s can be obtained
\cite{mcoi}.

\section{The Fourier and wavelet transforms}

In many applications (especially, for non-stationary signals), one is
interested in the frequency content of a signal locally in time, i.e., tries
to learn which frequencies are important at some particular moment.
As has been stressed already, the \w\ transform is superior to the Fourier
transform, first of all, due to the locality property of \w s. The Fourier
transform uses sine, cosine or imaginary exponential functions as the main
basis. It is spread
over the entire real axis whereas \w\ basis is localized. It helps analyze
the local properties of a signal using \w s while the
Fourier transform does not
provide any information about the location where the scale (frequency) of a
signal changes. Decomposition into \w s allows singularities to be
located by observing the places where the \w\ \co s are (abnormally) large.
Obviously, nothing of the kind happens for the Fourier transform. Once the
\w s have been constructed they perform incredibly well in situations where
Fourier series and integrals involve subtle mathematics or heavy numerical
calculations. But \w\ \an\ cannot entirely replace Fourier \an\ , indeed,
the latter is often used in constructing the orthonormal bases of \w s needed
for \an\ with \w\ series. Many theorems of \w\ \an\ are proven with the help
of the Fourier decomposition. The two kinds of \an\ are thus complementary
rather than always competing.

The Fourier spectrum $f_{\omega }$ of a one-dimensional signal $f(t)$ having
finite energy (i.e., square-integrable) is given by
\be
f_{\omega }=\int _{-\infty }^{\infty }f(t)e^{-i\omega t}dt.   \label{fo}
\ee
The inverse transform restores the signal
\be
f(t)=\frac {1}{2\pi }\int _{-\infty }^{\infty }f_{\omega }e^{i\omega t}d\omega .
\label{ft}
\ee
It is an unitary transformation
\be
\int \vert f(t)\vert ^2dt=\frac {1}{2\pi }\int \vert f_{\omega }\vert ^2d\omega .
\label{pars}
\ee
This is the so-called Parseval identity which states the conservation of energy
between the time and the frequency domains.
The formula (\ref{fo}) asks for information about the signal $f(t)$ from both
past and future times. It is suited for a stationary signal when the frequency
$\omega $ does not depend on time. Thus its time-frequency band is well located
in frequency and practically unlimited in time, i.e., Eq. (\ref{fo}) gives a
representation of the frequency content of $f$ but not of its time-localization
properties. Moreover, the signal $f(t)$ should decrease fast enough at
infinities the integral (\ref{fo}) to be meaningful.

The attempt to overcome these difficulties and improve time-localization while
still using the same basis functions is
done by the so-called windowed Fourier transform. The signal $f(t)$ is considered
within some time interval (window) only. In practice, one has to restrict the
search for the optimal window to the easily generated windows. A simple way of
doing so amounts to multiplying $f(t)$ by a simplest compactly supported window
function, e.g., by $g_w=\theta (t-t_i)\theta (t_f-t)$, where $\theta $ is the
commonly used Heaviside
step-function different from zero at positive values of its argument only,
$t_i, t_f$ are the initial and final cut-offs of the signal (more complicated
square-integrable window functions $g$, well concentrated in time, e.g., a
Gaussian or canonical coherent states\footnote{In quantum mechanics, they are
introduced for quantizing the classical harmonic oscillator. In signal \an\ ,
they are known by the name of Gabor functions.}, can be used as well). Thus
one gets
\be
f_{\omega ,w}=\int _{-\infty }^{\infty }f(t)g(t)e^{-i\omega t}dt. \label{fow}
\ee
In its discrete form it can be rewritten as
\be
f_{\omega ,w}=\int f(t)g(t-nt_0)e^{-im\omega _0t}dt,  \label{fodi}
\ee
where $\omega _0, t_0>0$ are fixed, and $m, n$ are some numbers which define
the scale and location properties.
The time localization of the transform is limited now but the actual window is
fixed by different functions in time and frequency which do not depend on the
resolution scale and have fixed widths. Moreover, the orthonormal basis of the
windowed Fourier transform can be constructed only for the so-called Nyquist
density (corresponding to the condition $\omega _0t_0=2\pi $; see Section 14),
whereas there is no such restriction imposed on \w s. At this critical value,
frames are possible, but the corresponding functions are either poorly
localized, or have poor regularity. This result is known as the Balian--Low
phenomenon. In practical applications
of the windowed Fourier transform, to achieve better localization one should
choose $\omega _0t_0<2\pi $ thus destroying orthonormality.

The difference between the \w\ and windowed Fourier transforms
lies in the shapes of the analyzing functions $\psi $ and $g$. All
$g$, regardless of the value of $\omega $, have the same width. In
contrast, the \w s $\psi $ automatically provide the time
(location) resolution window adapted to the problem studied, i.e.,
to its essential frequencies (scales). Namely, let $t_0, \delta $
and $\omega _0, \delta _{\omega }$ be the centers and the
effective widths of the \w\ basic function $\psi (t)$ and its
Fourier transform. Then for the \w\ family $\psi _{j,k}(t)$
(\ref{psij}) and, correspondingly, for \w\ coefficients, the
center and the width of the window along the $t$-axis are given by
$2^j(t_0+k)$ and $2^j\delta $. Along the $\omega $-axis they are
equal to $2^{-j}\omega _0$ and $2^{-j}\delta _\omega $. Thus the
ratios of widths to the center position along each axis do not
depend on the scale. It means that the \w\ window resolves both
the location and the frequency in fixed proportions to their
central values. For the high-frequency component of the signal it
leads to quite large frequency extension of the window whereas the
time location interval is squeezed so that the Heisenberg
uncertainty relation is not violated. That is why \w\ windows can
be called as Heisenberg windows. Correspondingly, the
low-frequency signals do not require small time intervals and
admit wide window extension along the time axis. Thus \w s
localize well the low-frequency "details" on the frequency axis
and the high-frequency ones on the time axis. This ability of \w s
to find a perfect compromise between the time localization and the
frequency localization by choosing automatically the widths of the
windows along the time and frequency axes well adjusted to their
centers location is crucial for their success in signal \an\ . The
\w\ transform cuts up the signal (functions, operators etc) into
different frequency components, and then studies each component
with a resolution matched to its scale providing a good tool for
time-frquency (position-scale) localization. That is why \w s can
zoom in on singularities or transients (an extreme version of very
short-lived high-frequency features!) in signals, whereas the
windowed Fourier functions cannot. In terms of the traditional
signal \an\ , the filters associated with the windowed Fourier
transform are constant bandwidth filters whereas the \w s may be
seen as constant relative bandwidth filters whose widths in both
variables linearly depend on their positions.

In Fig. 10 we show the difference between these two approaches.
It demonstrates the constant shape of the windowed Fourier transform region
and the varying shape (with a constant area) of the \w\ transform region.
The density of localization centers is homogeneous for the windowed Fourier
transform whereas it changes for the \w\ transform so that at low frequencies
the centers are far apart in time and become much denser for high frequencies.

From the mathematical point of view, it is important that orthonormal \w s give
good unconditional\footnote{For unconditional bases, the order in which the
basis vectors are taken does not matter. All of the known constructions of
unconditional bases of \w s rely on the concept of the multiresolution \an\ .
} bases for other spaces than that of
the square integrable functions, out-performing the Fourier basis
functions in this respect. It is applied in the subsequent Sections to a
characterization of such functions using only the absolute values of \w\ \co s.
In other words, by looking only at the absolute values of \w\ \co s we can
determine to which space this function belongs. This set of spaces is much wider
than in the case of the Fourier transform by which only Sobolev spaces\footnote{
The function $f$ belongs to the Sobolev space $W^s(R)$ if its Fourier transform
provides the finite integrals
$\int d\omega (1+\vert \omega \vert ^2)^s\vert f(\omega )\vert ^2$.}
can be completely characterized.

As we mentioned already, the \w\ \an\ concentrates near the singularities
of the function analyzed. The corresponding \w\ \co s are negligible in the
regions where the function is smooth. That is why the \w\ series with a plenty
of non-zero \co s represent really pathological functions, whereas "normal"
functions have "sparse" or "lacunary" \w\ series and easy to compress.
On the other hand, Fourier series of the usual functions have a lot of
non-zero \co s, whereas "lacunary" Fourier series represent pathological
functions.
Let us note at the end that, nevertheless, the Fourier transform is
systematically used in the proof of many theorems in the theory of \w s.
It is not at all surprising because they constitute the stationary signals
by themselves.

\section{Wavelets and operators}

The study of many operators acting on a space of functions or distributions
becomes simple when suitable \w s are used because these operators can
be approximately diagonalized with respect to this basis. Orthonormal \w\
bases provide the unique example of a basis with non-trivial diagonal,
or almost-diagonal, operators. The operator action on the \w\
series representing some function does not have uncontrollable sequences, i.e.,
\w\ decompositions are robust. One can describe precisely what happens to
the initial series under the operator action and how it is transformed.
In a certain sense, \w s are stable under the operations of integration and
differentiation.
That is why \w s, used as a basis set, allow us to solve differential equations
characterized by widely different length scales found in many areas of
physics and chemistry. Moreover, \w s reappear as eigenfunctions of certain
operators.

To deal with operators in the \w\ basis it is convenient, as usual, to
use their matrix representation. For a given operator $T$ it is represented
by the set of its matrix elements in the \w\ basis:
\be
T_{j,k;j,'k'}=<\psi_{j,k}^{*}|T|\psi_{j',k'}>.   \label{opma}
\ee
For linear homogeneous operators, their matrix representation can be explicitly
calculated \cite{beyl}.

It is extremely important that it is sufficient to first calculate the matrix
elements at some ($j$-th) resolution level. All other matrix elements can
be obtained from it using the standard recurrence relations. Let us derive the
explicit matrix elements $r_{j,l;j,l'}$ of the homogeneous operator $T$ of the
order $\alpha $:
\be
    r_{j,l;j,l'}=<\varphi_{j,l}|T|\varphi_{j,l'}>.  \label{varv}
\ee
Using the recurrence relations between the scaling functions at the given and
finer resolution levels, one gets the following equation relating the matrix
elements at neighboring levels:
\be
   r_{j,l;j,l'}=<(\sum_k h_k \varphi_{j+1,2l-k})|T|(\sum_{k'} h_{k'}
       \varphi_{j+1,2l'-k'})>=\sum_k\sum_{k'} h_k h_{k'} r_{j+1,2l-k;j+1,2l'-k'}.
       \label{rdif}
\ee
For the operator $T$ having the homogeneity index $\alpha $ one obtains
\be
r_{j,l;j,l'}=2^{\alpha}\sum_k\sum_{k'} h_k h_{k'} r_{j,2l-k;j,2l'-k'}. \label{rrec}
\ee
The solution of this equation defines the required \co s up to the
normalization constant which can be easily obtained from the results of the
action by the operator $T$ on a polynomial of a definite rank. For non-integer
values of $\alpha $, this is an infinite set of equations.

The explicit equation for the $n$-th order differentiation operator is
\begin{eqnarray}
r_k^{(n)}=\langle \scl (x)\vert \frac {d^{n}}{dx^{n}}\vert \scl (x-k)\rangle= \nonumber \\
\sum_{i,m}h_ih_m\langle \scl (2x+i)\vert \frac {d^{n}}{dx^{n}}\vert \scl (2x+m-k)\rangle= \nonumber \\
2^{n}\sum_{i,m}h_ih_mr_{2k-i-m}^{(n)}.   \label{rkdi}
\end{eqnarray}
It leads to a finite system of linear equations for $r_k$ (the index $n$ is
omitted):
\be
2r_k=r_{2k}+\sum _ma_{2m+1}(r_{2k-2m+1}+(-1)^nr_{2k+2m-1}),   \label{rkli}
\ee
where both $r_k$ and $a_m=\sum _ih_ih_{i+m}$ ($a_0=1$) are rational numbers in
the case of Daubechies \w s. The \w\ \co s can be found from these equations up to
a normalization constant. The normalization condition reads \cite{beyl}:
\be
\sum _k k^nr_k=n!.   \label{noco}
\ee
For the support region of the length $L$, the \co s $r_k$ differ from zero
for $-L+2\leq k\leq L-2$, and the solution exists for $L\geq n+1$. These
\co s possess the following symmetry properties:
\be
r_k=r_{-k}
\ee
for even $k$, and
\be
r_k=-r_{-k}
\ee
for odd values of $k$.

Here, as an example, we show in the Table the matrix elements of the first and
second order differential operators in the Daubechies \w\ basis with four
vanishing moments ($D^8$).

\medskip

Table.

\begin{tabular}{|r|c|c|c|} \hline
  k  & $h_k$ & $\langle \scl(x)|\nabla|\scl(x-k) \rangle$ & $\langle \scl(x)|\nabla^2|\scl(x-k) \rangle $ \\ \hline
  -6 & 0          & 0.00000084 & 0.00001592  \\
  -5 & 0          &-0.00017220 &-0.00163037  \\
  -4 & 0          &-0.00222404 &-0.01057272  \\
  -3 & 0          & 0.03358020 & 0.15097289  \\
  -2 &-0.07576571 &-0.19199897 &-0.69786910  \\
  -1 &-0.02963552 & 0.79300950 & 2.64207020  \\
   0 & 0.49761866 & 0          &-4.16597364  \\
   1 & 0.80373875 &-0.79300950 & 2.64207020  \\
   2 & 0.29785779 & 0.19199897 &-0.69786910  \\
   3 &-0.09921954 &-0.03358020 & 0.15097289  \\
   4 &-0.01260396 & 0.00222404 &-0.01057272  \\
   5 & 0.03222310 & 0.00017220 &-0.00163037  \\
   6 & 0          &-0.00000084 & 0.00001592  \\ \hline
\end{tabular}

\medskip

For a continuous linear operator $T$ represented by a singular integral
\be
Tf(x)=\int K(x,y)f(y)dy     \label{tfxi}
\ee
with some definite conditions imposed on the kernel $K$ (see Appendix 2), there
exists the important theorem (called $T(1)$ theorem) which states a necessary and
sufficient condition for the extension of $T$ as a continuous linear operator on
$L^2(R^n)$ (more details and the elegant wavelet proof of the theorem are
given in Ref. \cite{mcoi}).

Let us note that a standard difficulty for any spectral method is a
representation of the operator of multiplication by a function. As an example
for physicists, we would mention the operator of the potential energy in
the Schr\"{o}dinger equation. However, it is well known that
this operation is trivial and diagonal in the real space. Therefore for such
operations one should deal with a real space, and after all operations done
there return back. Such algorithm would need $O(N)$ operations only.

\section{Nonstandard matrix multiplication}

There are two possible ways to apply operators onto functions within \w\
theory. They are called the standard and nonstandard matrix forms.

For smooth enough functions most \w\ \co s are rather small. For a wide
class of operators, most of their matrix elements are also quite small.
Let us consider the structure of the elements of the matrix
representation of some operator $T$ that are large enough.
The matrix elements satisfy the following relations
\be
T_{j,k;j',k'} \rightarrow 0 ,\; where \;\; |k-k'|\rightarrow \infty ,   \label{tjk1}
\ee
\be
T_{j,k;j',k'} \rightarrow 0 ,\; where \;\; |j-j'|\rightarrow \infty.\label{tjk2}
\ee
The topology of the distribution of these matrix elements within the matrix
can be rather complicated. The goal of the nonstandard form is to replace the
latter equation (\ref{tjk2}) by another, more rigorous one:
\be
 T_{j,k;j',k'} = 0 ,\; if \;\; j\neq j'.   \label{tjk0}
\ee
It avoids taking the matrix elements between the different resolution levels. To deal
with it, one should consider, instead of the \w\ space, the overfull space
with the basis containing both \w s and scaling functions at various resolution
levels.

Let us consider the action of the operator $T$ on the function $f$ which
transforms it into the function $g$:
\be
g=Tf.   \label{gtfx}
\ee
Both $g$ and $f$ have the \w\ representations with the \w\ \co s
$(_fs_{j,k};\;_fd_{j,k})$ and $(_gs_{j,k};\;_gd_{j,k})$. At the finest resolution
level $j_n$ only $s$-\co s differ from zero, and the transformation looks like
\be
_gs_{j_n,k}=\sum _{k'}T_{SS}(j_n,k;j_n,k')_fs_{j_n,k'}.   \label{gsjk}
\ee
At the next level one gets in both the standard and nonstandard approaches
\be
_gs_{j_n-1,k}=\sum _{k'}T_{SS}(j_n-1,k;j_n-1,k')_fs_{j_n-1,k'}+
\sum _{k'}T_{SD}(j_n-1,k;j_n-1,k')_fd_{j_n-1,k'},   \label{gsj1}
\ee
\be
_gd_{j_n-1,k}=\sum _{k'}T_{DS}(j_n-1,k;j_n-1,k')_fs_{j_n-1,k'}+
\sum _{k'}T_{DD}(j_n-1,k;j_n-1,k')_fd_{j_n-1,k'},   \label{gdj1}
\ee
where $T_{SS}(j_n,k;j_n,k')=\int dx {\bar \scl }_{j_n,k}(x)T\scl _{j_n,k'}(x)$
and the replacement of subscripts $S\rightarrow D$ corresponds to the substitute
$\scl \rightarrow \psi $ in the integrals. There is coupling between all
resolution levels because all $s$-\co s at this $(j_n-1)$-th level should be
decomposed by the fast \w\ transform into $s$ and $d$-\co s at higher levels.
Therefore even for the almost diagonal initial step the standard matrix
acquires rather complicated form as demonstrated in Fig. 11 for 4-level
operations (similar to those discussed above in the case of the Haar \w s).
It is thus inefficient for numerical purposes.

As we see in Fig. 11, at the final stage of the standard approach
we have to deal with the \w\ representation corresponding to the formula
(\ref{fdec}) with only one $s$-\co\ left in the vectors which represents the
overall weighted average of the functions (the $SS$-transition from $f$ to $g$
is given by the upper left box in the Figure). At the same time, in the process
of approaching it from the scaling-function representation (\ref{fjma}),
(\ref{fscl}) we had to deal with average values at intermediate levels
decomposing them at each step into $s$ and $d$ parts of further levels.
These intermediate $s$-\co s have been omitted since they were replaced
by $s$ and $d$-\co s at the next levels. That is why the matrix of the standard
approach looks so complicated.

To simplify the form of the matrix, it was proposed \cite{bcro} to use the
redundant set of the \w\ \co s. Now, let us keep these average values in the
form of corresponding $s$-\co s of the intermediate levels both in initial and
final vectors representing functions $f$ and $g$. Surely, we will deal with
redundant vectors which are
larger than necessary for the final answer. However, it should not bother us
because we know the algorithm to reduce the redundant vector to a non-redundant
form. At the same time, this trick simplifies both the form of the transformation
matrix and the computation. This non-standard form is shown in Fig. 12.
Different levels have been completely decoupled because there are no blocks in
this matrix which would couple them. The block with $SS$-elements has been
separated, and in place of it the zero matrix is inserted. The whole matrix
becomes artificially enlarged. Correspondingly, the vectors, characterizing
the functions $f$ and $g$, are also enlarged. Here all
intermediate $s$-\co s are kept for the function $f$ (compare the vectors in
the right-hand sides of Figs. 11 and 12).
Each $S_{j+1}$ is generated from $S_j$ and $D_j$. That is where the coupling of
different levels is still present. In the transformation matrix
all $SS$-elements are zero except for the lowest one $S_0S_0$. All other
$SD,\; DS,\;DD$ matrices are almost diagonal due to the finite support of
scaling functions and \w s. The redundant form of the $g$-function vector of
Fig. 12 may be reduced to its usual \w\ representation of Fig. 11 by splitting
up any $S_j$ into $S_{j-1}$ and $D_{j-1}$ by a standard method. Then these
$S_{j-1}$ and $D_{j-1}$ are added to the corresponding places in the vector.
This iterative procedure allows, by going from $S_{j-1}$ down to $S_0$, to get the
usual \w\ series of the function $g$. We get rid of all $s$-\co s apart from
$S_0$. The computation becomes fast.

\section{Regularity and differentiability}

The \an\ of any signal includes finding the regions of its regular and
singular behavior. One of the main features of \w\ \an\ is its capacity of
doing a very precise \an\ of the regularity properties of functions.
When representing a signal by a \w\ transform, one
would like to know if and at which conditions the corresponding series is
convergent and, therefore, where the signal is described by the differentiable
function or where the singularities appear. For certain singular functions,
essential information is carried by a limited number of their \w\ \co s. Such
an information may be used in the design of numerical algorithms.
We start with the traditional H\"{o}lder conditions of regularity
and, in the next Section, proceed to their generalization following
Ref. \cite{jmey}.

The H\"{o}lder definition of the pointwise regularity at a point $x_0$ of a
real-valued function $f$ defined on $R^n$ declares that this function belongs
to the space $C^{\alpha }(x_0)$ ($\alpha >-n$) if there exists a polynomial
$P$ of order at most $\alpha $ such that
\be
\vert f(x)-P(x-x_0)\vert \leq C\vert x-x_0\vert ^{\alpha },  \label{hold}
\ee
where $C$ is a constant.

The supremum of all values of $\alpha $ such that (\ref{hold}) holds is called
the H\"{o}lder exponent of $f$ at $x_0$ and denoted $\alpha (x_0)$\footnote{In
principle, the expression $\vert x-x_0\vert ^{\alpha }$ in the right-hand side
of Eq. (\ref{hold}) can be replaced by a more general function satisfying
definite conditions and called a modulus of continuity but we shall use only
the above definition. In general, a modulus of continuity defines the largest
deviation from that best polynomial approximation of a function $f$ which is
characterized by the set of smallest deviations compared to other polynomial
approximations.}.

Correspondingly, a point $x_0$ is called a strong $\alpha $-singularity
$(-n<\alpha \leq 1)$ of $f$ if, at small intervals, the following inequality
is valid
\be
\vert f(x)-f(y)\vert \geq C\vert x-y\vert ^{\alpha }   \label{sing}
\ee
for a relatively large set of $x$'s and $y$'s close to $x_0$.

Such definitions work quite well for a rather smooth function. However, in
the case of a function drastically fluctuating from point to point they are
difficult to handle at each point because, e.g., the derivative of $f$ may
have no regularity at $x_0$, the condition (\ref{hold}) may become unstable
under some operators such as Hilbert transform etc. The generalized
definition of pointwise regularity will be given in the next Section by
introducing two-microlocal spaces. Here, we concentrate on global properties
of the function $f$.

The uniform regularity of a function $f$ at positive non-integer values of
$\alpha $ consists in the requirement for the Eq. (\ref{hold}) to hold for
all real $n$-dimensional $x_0$ with the uniform value of the constant $C$.
At first sight, this condition looks rather trivial because for smooth functions
 the uniform regularity coincides with the pointwise regularity,
which is everywhere the same. To understand that this is
non-trivial, one could consider, e.g., the function $f(x)=x\sin 1/x$ with
pointwise exponents $\alpha (0)=1, \;\; \alpha (x_0)=\infty $ for any fixed
$x_0$ different from 0,
and uniform H\"{o}lder exponent $\alpha =1/2$ which appears for the set of
points $x_0\propto \vert x-x_0\vert ^{1/2}\gg \vert x-x_0\vert $.

The requirement of a definite uniform regularity of a $n$-dimensional function
$f$ on the whole real axis at positive non-integer values of $\alpha $ can be
stated in terms of \w\ \co s as an inequality
\be
\vert d_{j,k}\vert \leq C2^{-(\frac {n}{2}+\alpha)j}.  \label{djku}
\ee
Thus from the scale behavior of \w\ \co s one gets some characteristics of
uniform regularity of the function. In particular, the linear dependence
of the logarithms of \w\ \co s on the scale $j$ would indicate on the scaling
properties of a signal, i.e., on the fractal behavior whose parameters are
determined from higher moments of \w\ \co s (see below). For pointwise
regularity determining the local characteristics, the similar condition is
discussed in the next Section.

Now, let us formulate the conditions under which \w\ series converge at some
points, or are differentiable. It has been proven that

\begin{itemize}

\item if $f$ is square integrable, the \w\ series of $f$ converges almost
everywhere;

\item the \w\ series of $f$ is convergent on a set of points equipotent to $R$ if
\be
\vert d_{j,k}\vert =2^{-j/2}\eta _j,   \label{djk2}
\ee
where $\eta _j$ tends to 0 when $j$ tends to $+\infty $;

\item the function $f$ is almost everywhere differentiable and the derivative of
the \w\ series of $f$ converges almost everywhere to $f'$, if the following
condition is satisfied
\be
\sum \vert d_{j,k}\vert ^22^{2j}\leq \infty ;  \label{djk3}
\ee

\item the function $f$ is differentiable on a set of points equipotent to $R$
and, at these points the derivative of its \w\ series converges to $f'$, if
\be
\vert d_{j,k}\vert \leq 2^{-3j/2}\eta _j,   \label{djk4}
\ee
where $\eta _j$ tends to 0 when $j$ tends to $+\infty $.

\end{itemize}

Limitations imposed on \w\ \co s are generalized in the next Section to include
the pointwise
regularity condition when \m\ spaces are considered. These better estimates
do not hold just uniformly, but hold locally (possibly, apart from a set of
points of small Hausdorff dimension which can be determined by \w\ \an\ ).

Note finally that one can also derive global regularity of $f$ from the
decay in $\omega $ of the absolute value of its windowed Fourier transform,
if the window function $g$ is chosen to be sufficiently smooth. In most cases,
however, the value of the uniform H\"{o}lder exponent computed from  the
Fourier \co s will not be optimal. Nothing can be said from Fourier \an\
about the local regularity, in contrast to \m\ \w\ \an\ discussed below.

\section{Two-microlocal analysis}

The goal of the two-microlocal \an\ is to reveal the pointwise behavior
of any function from the properties of its \w\ \co s. The local regularity
of the function is thus established.

The scalar product of an analyzed \1\ function and a \w\ maps this
function into \2\ \w\ \co s which reveal both scale and location properties
of a signal. In the above definitions of the pointwise regularity of the function
at $x_0$ determined by the H\"{o}lder exponent only local but not scaling
properties of the signal are taken into account. The problem of determining
the exact degree of the H\"{o}lder regularity of a function is easily solved if this
regularity is everywhere the same, because in such a case it usually coincides
with the uniform regularity. The determination of the pointwise H\"{o}lder
regularity, however, becomes much harder if the function changes wildly from
point to point, i.e., its scale (frequency) characteristics depend strongly on
the location (time). In that case we have to deal with a very non-stationary
signal. To describe the singularity of $f(x)$ at $x_0$ only by
the H\"{o}lder exponent $\alpha (x_0)\leq 1$ one looks for the order of
magnitude of the difference $\vert f(x)-f(x_0)\vert $ when $x$ tends to
$x_0$, without taking into account, e.g., the possible high-frequency
oscillations of $f(x)-f(x_0)$, i.e., its scale behavior.

The properties of the \w\ \co s as functions of both scale
and location\footnote{Let us note that the word "location" (and, correspondingly,
"scale") does not necessarily imply a single dimension but could describe the
position (and size or frequency) in any $n$-dimensional space.} provide a
unique way of describing the
pointwise regularities. It is more general than the traditional H\"{o}lder
approach because it allows us to investigate, characterize and easily
distinguish some specific local behaviors such as approximate selfsimilarities
and very strong oscillatory features like those of the indefinitely oscillating
functions which are closely related to the so-called "chirps" of the form
$x^{\alpha }\sin (1/x^{\beta })$ (reminding of bird, bat or dolphin sonar
signals with very sharp oscillations which accelerate at some point
$x_0$). Chirps are well known to everybody dealing with modern radar and sonar
technology. In physics, they are known to the authors, e.g., in theoretical
considerations of the dark matter radiation and of the gravitational waves.
Sometimes, similar dependence can reveal itself even in the more traditional
correlation \an\ \cite{dr93, dhwa, dgary}.
The large value of a second derivative of the phase function of a
frequency modulated signal is their typical feature. Such special behavior with
frequency modulation laws hidden in a given signal can be revealed
\cite{jmey} with the help of the \m\ \an\ . Actually, there is no universally
accepted definition of a chirp. Sometimes, any sine of a non-linear polynomial
function of time is also called a chirp.

The \m\ space $C^{s,s'}(x_0)$ of the real-valued $n$-dimensional functions $f$
(distributions) is defined by the following simple decay condition on their
\w\ \co s $d_{j,k}$
\be
\vert d_{j,k}(x_0)\vert \leq C2^{-(\frac {n}{2}+s)j}(1+\vert 2^{j}x_0-k\vert )^{-s'},
\label{djkm}
\ee
where $s$ and $s'$ are two real numbers. This is a very important extension of
the H\"{o}lder conditions. The \m\ condition is a local counterpart of the
uniform condition (\ref{djku}). It is very closely related to the
pointwise regularity condition (\ref{hold}) because it expresses, e.g., the
singular behavior of the function itself at the point $x_0$ in terms of the
$k$-dependence of its \w\ \co s at the same point. Such an estimate is stable
under derivation and fractional integration.

For $s'=0$, the space thus defined is the global H\"{o}lder space $C^s(R^n)$.

If $s'>0$, these conditions are weaker at $x_0$
than far from $x_0$ so that they describe the behavior of a function which is
irregular at $x_0$ in a smooth enviroment, i.e., the regularity of $f$ gets
worse as $x$ tends to $x_0$. If $s'$ is positive, the inequality (\ref{djkm})
implies that $\vert d_{j,k}\vert \leq C2^{-(\frac {n}{2}+s)j}$, so that $f$
belongs to $C^s(R^n)$. The H\"{o}lder norms do not give any
information about this singularity because one has to consider them on domains
that exclude $x_0$, and the H\"{o}lder regularity of $f$
deteriorates when $x$ gets close to $x_0$. That is why we need a uniform
regularity assumption when $s'$ is positive.

If $s'<0$, the converse situation
takes place when $x_0$ is a point of regularity in a nonsmooth enviroment.

Thus the inequality
Eq. (\ref{djkm}) clarifies the relationship between the pointwise behavior of a
function and the size properties of its \w\ \co s. If $s>0$ and $s'>s$, then
$C^{s,s'}(x_0)$ is included in $C^s(x_0)$. One can prove
that the elements of the space $C^{s,s'}(x_0)$, for $s'>-s$, are functions
whereas the elements of $C^{s,-s}(x_0)$ are (in general) not functions but
distributions (and, moreover, quite "wild distributions" for which the local
H\"{o}lder condition (\ref{hold}) cannot hold and should be generalized by
multiplying the right-hand side by the factor $\log 1/\vert x-x_0\vert $).
However, if $f$ belongs to the space $C^s(x_0)$ then it belongs to
$C^{s,-s}(x_0)$ too. In particular, if at $x$ close to $x_0$ the following
inequality $\vert f(x)\vert \leq C\vert x-x_0\vert ^s$ is valid for $s\leq 0$,
imposing some limit on the nature of the singularity at this point,
then the estimates for \w\ \co s are given by
\be
\vert d_{j,k}\vert \leq C2^{-(\frac {n}{2}+s)j}\vert k-2^jx_0\vert ^s, \label{djks}
\ee
if the support of $\psi _{j,k}$ is at least at a distance $2^{-j}$ from $x_0$, and
\be
\vert d_{j,k}\vert \leq C2^{-(\frac {n}{2}+s)j}, \label{djkf}
\ee
if the support of $\psi _{j,k}$ is at a distance less than $2^{-j}$ from $x_0$,
which demonstrate the above statement.

The \m\ spaces have some stability
properties. In particular, the pseudodifferential operators of order 0 are
continuous on these spaces (in contrast with the usual pointwise H\"{o}lder
regularity condition which is not preserved under the action of these operators,
for example, of the Hilbert transform). The position of the points of regularity
of a function $f$ which belongs to the space $C^{s,s'}(x_0)$ is essentially
preserved under the action of singular integral operators such as the Hilbert
transform. This property leads to a pointwise regularity result for solutions
of partial differential equations which is formulated \cite{jmey} by the following

{\it Theorem}. \\
Let $\Lambda $ be a partial differential operator of order $m$, with smooth
\co s and elliptic at $x_0$. If $\Lambda f=g$ and $g$ belongs to $C^{s,s'}(x_0)$,
then $f$ belongs to $C^{s+m,s'}(x_0)$.

Thus, if $\Lambda f=g$, and if $g$ is a function that belongs to $C^s(x_0)$,
then there exists a polynomial $P$ of degree less than $s+m$ such that,
for $\vert x-x_0\vert\leq 1$,
\begin{center}
$\vert f(x)-P(x)\vert \leq C\vert x-x_0\vert ^{s+m}$.
\end{center}
All the above conditions look as if the \w s were eigenvectors of the
differential operators $d^{\alpha },\; \vert \alpha \vert \leq r$, with
corresponding eigenvalues $2^{j\vert \alpha \vert }$.

The more general functional spaces which
admit the power dependence of \w\ \co s on the scale $j$ defined by an extra
parameter $p$ are considered in Ref. \cite{jmey}. In particular, those are
the Sobolev spaces where the integrability is required up to some power of both
the function itself and its derivatives up to the definite order. In this case,
the additional factor $j^{2/p}$ appears in the right-hand side of the inequality
(\ref{djkm}), or, in other words, in addition to the linear term in the
exponent one should consider another logarithmically dependent term. Thus
there is no rigorous selfsimilarity (fractality) of the function any more.
However, the \w\ \an\ allows us to determine the fractal dimensions of the
sets of points where the function $f$ is singular.

For continuous \w s, the condition equivalent to Eq. (\ref{djkm}) takes the form
\be
\vert W(a,b)\vert \leq Ca^{s}(1+\frac {\vert b-x_0\vert}{a})^{-s'}. \label{wabm}
\ee
The definite upper limits on the behavior of $\vert W(a,b)\vert $ can also be
obtained if $f$ is integrable in the neighborhood of the origin (see \cite{jmey}).
They are somewhat complicated, and we do not show them here.

It is instructive to examine the meaning of the conditions of the type (\ref{wabm}).
Let us consider the cone in the $(b, a)$ half-plane defined by the condition
$\vert b-x_0\vert <a$. Within this cone we have $\vert W(a,b)\vert=O(a^s)$ as
$a\rightarrow 0$. Outside the cone, the behavior is governed by the distance of
$b$ to the point $x_0$. These two behaviors are generally different and have to
be studied independently. However, it is shown in \cite{mhwa} that
non-oscillating singularities may be characterized by the behavior of the
absolute values of their \w\ \co s within the cone. It is also shown in
\cite{mhwa} that rapidly oscillating singularities, which we consider below,
cannot be characterized by the behavior of their \w\ transform in the cone.

The \m\ methods proved especially fruitful in the \an\ of oscillating
trigonometric and logarithmic chirps.

A simplest definition of trigonometric chirps at the origin $x=0$ could
simply be read
\be
f(x)=x^{\alpha }g(x^{-\beta }),   \label{trig}
\ee
where $g$ is a $2\pi $-periodic $C^r$ function with vanishing integral. This
type of behavior leads to the following expansion of continuous \w\ \co s at
small positive $b\leq \delta $ on curves $a=\lambda b^{1+\beta }$
\be
W(\lambda b^{1+\beta },b)=b^{\alpha }m_{\lambda }(b^{-\beta }),  \label{wtri}
\ee
where $m_{\lambda }$ is a $2\pi $-periodic function with a definite norm.
It shows that rapidly oscillating singularities cannot be characterized by
the behavior of their \w\ transform in the cone as in the previous examples.
In this case, the \w\ \co s should be carefully analyzed not in the cone but on definite
curves because they are concentrated along these ridges. For small enough
scales, such a ridge lies outside the influence cone of the singularity.
A typical example is given by the function $f(x)=\sin 1/x$ whose instantaneous
frequency tends to infinity as $x\rightarrow 0$. The \w\ \co\ modulus is
maximum on a curve of equation $b=Ca^2$ for some constant $C$ depending
only on the \w\, and this curve is not a cone.
Therefore it is not sufficient to study the behavior of \w\ \co s inside
the cone to characterize the rapidly oscillating singularities. Let us note,
that in case of noisy signals, the contribution of the deterministic part of
the signal may be expected to be much larger than that of a noise just near
the ridges of the \w\ transform because the \w\ transform of the noise is
spread in the whole $(j,k)$-plane. It can be thus used for denoising the signal.

Logarithmic chirps have an approximate scaling invariance near the point $x_0$.
A function $f$ has a logarithmic chirp of order ($\alpha ,\lambda $) and of
regularity $\gamma \geq 0$ at the origin $x=0$ if for $x>0$ there exists
a $\log \lambda $-periodic function $G(\log x)$ in $C^{\gamma}(R)$ such that
\be
f(x)-P(x)=x^{\alpha }G(\log x).    \label{fpl}
\ee
The continuous \w\ \co s of it satisfy the condition
\be
W(a,b)=a^{\alpha }H(\log a,b/a),   \label{wlog}
\ee
where $H$ is $\log \lambda $-periodic in the first argument and its behavior
in $b/a$ is restricted by some decay conditions. In other words, if the
following scaling property is observed
\be
C(\lambda a,\lambda b)=\lambda ^{\alpha }C(a,b)    \label{scal}
\ee
for $\lambda <1$ and small enough $a$ and $b$, then $f$ has a logarithmic chirp
of order ($\alpha ,\lambda $). More general and rigorous statements and proofs
can be found in Ref. \cite{jmey}. Thus we conclude that the type and behavior
of a chirp can be determined from the behavior of its \w\ \co s.

As an example of the application of the above statements, let us mention the
continuous periodic function $\sigma (x)$ represented by the Riemann series
\be
\sigma (x)=\sum _1^{\infty }\frac {1}{n^2}\sin \pi n^2x,   \label{riem}
\ee
which belongs to the space $C^{1/2}$. It is best analyzed by a continuous \w\
transform using the specific complex \w\ $\psi (x)=(x+i)^{-2}$ proposed by Lusin
in 1930s in the functional analysis. Then the \w\ transform of the Riemann
series is given by the well known Jacobi Theta function. The behavior of \w\ \co s
determines the chirp type and its parameters. With the help of the \m\ \an\ it
was proven that this function has trigonometric chirps at some rational values of
$x=x_0$ which are ratios of two odd numbers (also, its first derivative exists
only at these points) and logarithmic chirps at quadratic irrational numbers
(near these points its \w\ transform possesses the scaling invariance
properties).

For practical purposes, however, it may happen that very large values of $j$
are needed to determine H\"{o}lder exponents reliably from the above conditions.
For the global H\"{o}lder exponent, no assumptions about regularity of \w s
is required whereas determination of the local H\"{o}lder exponents requires
a more detailed approach.

A nice illustration of application of \w\ \an\ for studies of local regularity
properties of the function $f(x)$ is given in \cite{daub} for the following
function
\be
f(x)=2e^{-\vert x\vert}\theta (-x-1)+e^{-\vert x\vert}\theta (x+1)\theta (1-x)
+e^{-x}[(x-1)^2+1]\theta (x-1),    \label{lreg}
\ee
which is infinitely differentiable everywhere except at $x=-1, 0, 1$ where,
respectively, $f, f', f''$ are discontinuous. One computes for each of the
three points the maxima of \w\ \co s $A_j=max _k\vert W_{j,k}\vert $ at
various resolution levels $3<j<10$ and plots
$\log A_j/\log 2$ versus $j$. The linear dependences on $j$ are found at these
points. The slopes at $x=-1, 0, 1$ are, correspondingly, -0.505 ; -1.495 ;
-2.462, leading with pretty good accuracy of 1.5$\%$ to rather precise
estimates of the H\"{o}lder exponents 0, 1, 2.

Even better accuracy in determining the local regularity of a function
can be achieved with the help of redundant \w\ families (frames)  where the
translational non-invariance is much less pronounced, the \w\ regularity
plays no role and only the number of its vanishing moments is important.

If orthonormal \w s are used, then their regularity can become essential.
Typically, they are continuous, have a non-integer uniform H\"{o}lder exponent
and different local exponents at different points. In fact, there exists a
whole hierarchy of (fractal) sets in which these \w s have different H\"{o}lder
exponents. There is a direct relation between the regularity of the function
$\psi $ and the number of its vanishing moments. The more moments are vanishing,
the more smooth is the function $\psi $. For example, for widely used Daubechies
\w s with a compact support the degree of regularity increases linearly with the
maximum number of vanishing moments $N$ for higher-tap \w s as $\mu N$ with
$\mu =0.2$ and, correspondingly, with the support width, as was already
mentioned in Section 4. These properties
justify the name "mathematical microscope", which is sometimes bestowed on
the \w\ transform.
Let us note that the choice of the $h_k$ that leads to maximal regularity is,
however, different from the choice with maximal number of vanishing moments
for $\psi $. How well the projections of a multiresolution approximation
$V_j$ converge to a given function $f$ depends on the regularity of the
functions in $V_0$.

There exists a special class of \w s called vaguelets. They possess
the regularity properties characterized by the following restrictions:
\be
\vert \psi _{j,k}(x)\vert \leq C2^{nj/2}(1+\vert 2^jx-k\vert )^{-n-\alpha },
\label{vag1}
\ee
\be
\vert \psi _{j,k}(x')-\psi _{j,k}(x)\vert \leq C2^{j(n/2+\beta )}
\vert x'-x\vert ^{\beta }    \label{vag2}
\ee
with $0<\beta <\alpha <1$ and a constant $C$.
Surely, the cancellation condition (\ref{osci}) must be also satisfied.
The norm of any function $f$ is then limited by its \w\ \co s as
\be
\vert \vert f\vert \vert \leq C'(\sum _{j,k}\vert d _{j,k}\vert ^2)^{1/2}
\label{vagn}
\ee
with a constant $C'$. Any continuous linear operator $T$ on $L^2$ which
satisfies the condition $T(1)=0$ transforms an orthonormal \w\ basis into
vaguelets.

In practice, the regularity of a \w\ can become especially important during the
synthesis when after omission of small \w\ \co s it is better to deal with a
rather smooth $\psi $ to diminish possible mistakes at the restoration stage.
On the contrary, for \an\ it seems to be more important to have \w s with
many vanishing moments to "throw out" the smooth polynomial trends and reveal
potential singularities. Also, the large number of vanishing moments leads to
better compression but can enlarge mistakes at inverse procedure of
reconstruction because of worsened regularity of \w s. The use of the
biorthogonal \w s helps a lot at this stage because among two dual \w s
one has many vanishing moments whereas another possesses good regularity
properties. By choosing the appropriate inversion formula one can minimize
possible mistakes.

\section{Wavelets and fractals}

Some signals (objects) possess the self-similar (fractal) properties (see,
e.g., \cite{mand, fede, pvul, dref}). It
means that by changing the scale one observes at a new scale the features
similar to those previously noticed at other scales. This property leads
to power-like dependences. The formal definition of a (mono)fractal Hausdorff
dimension $D_F$ of a geometrical object is given by a condition
\be
0<lim _{\epsilon \rightarrow 0}N(\epsilon )\epsilon ^{D_F}<\infty ,  \label{frac}
\ee
which states that $D_F$ is the only value for which the product of the
minimal number of the (covering this object) hypercubes $N(\epsilon )$ with a
linear size $l=\epsilon $ and of the factor $\epsilon ^{D_F}$ stays constant
for $\epsilon $ tending to zero. In the common school geometry of homogeneous
objects, it coincides with the topological dimension. The probability
$p_i(\epsilon )$ to belong to a hypercube $N_i(\epsilon )$ is proportional to
$\epsilon ^{D_F}$, and the sum of moments is given by
\be
\sum _ip_i^q(\epsilon )\propto \epsilon ^{qD_F}.   \label{momq}
\ee
The fractal dimension is directly related to the H\"{o}lder exponents.

Moreover, for more general objects called multifractals, the "fractal exponents"
$\alpha (x_0)$ (see Eq. (\ref{hold})) vary from point to point. The Hausdorff
dimension of the set of points $x_0$, where $\alpha (x_0)=\alpha _0$, is a
function of $d(\alpha _0)$ whose graph determines multifractal properties of
a signal, let it be Brownian motion, a fully developed turbulence or a
purely mathematical construction of the Riemann series. Thus the weights of
various fractal dimensions inside a multifractal differ for different
multifractals, and the value $D_F$ is now replaced by $D_{q+1}$ which depends
on $q$. It is called the Renyi (or generalized) dimension \cite{reny}. Usually,
it is a decreasing function of $q$. The fractal (Hausdorff), information and
correlation dimensions are, correspondingly, obtained from the Renyi dimension
at $q=-1; 0; 1$. The difference between the topological and Renyi dimensions is
called the anomalous dimension (or codimension). Fractals and multifractals are
common among the purely mathematical constructions (the Cantor set, the
Serpinsky carpet etc) and in the nature (clouds, lungs etc).

The pointwise H\"{o}lder exponents are now determined using \w\ \an\ . As we
have seen, all \w s of a given family $\psi  _{j,k}(x)$ are similar to the
basic \w\ $\psi (x)$ and derived from it by dilations and translations. Since
\w\ \an\ just consists in studying the signal at various scales by calculating
the scalar product of the analyzing \w\ and the signal explored, it is well
suited to reveal the fractal peculiarities. In terms of \w\ coefficients it
implies that their higher moments behave in a power-like manner with the
scale changing. The \w\ \co s are less sensitive to noise because they measure,
at different scales, the average fluctuations of the signal.

Namely, let us consider the sum $Z_q$ of the $q$-th moments
of the coefficients of the \w\ transform at various scales $j$
\be
Z_q(j)=\sum _k\vert d_{j,k}\vert ^q ,    \label{zq}
\ee
where the sum is over the maxima of $\vert d_{j,k}\vert $. Then it was shown
\cite{mbar, adat} that for a fractal signal this sum should behave as
\be
Z_q(j)\propto 2^{j[\tau (q) +\frac {q}{2}]} , \label{zqj}
\ee
i.e.,
\be
\log Z_q(j)\propto j[\tau (q)+\frac {q}{2}].   \label{lzq}
\ee
Thus the necessary condition for a signal to possess fractal properties is the
linear dependence of $\log Z_q(j)$ on the level number $j$. If this requirement
is fulfilled the dependence of $\tau $ on $q$ shows whether the signal is
monofractal or multifractal. Monofractal signals are characterized by a single
dimension and, therefore, by a linear dependence of $\tau $ on $q$, whereas
multifractal ones are described by a set of such dimensions, i.e., by non-linear
functions $\tau (q)$. Monofractal signals are homogeneous, in the sense that
they have the same scaling properties throughout the entire signal. Multifractal
signals, on the other hand, can be decomposed into many subsets characterized by
different local dimensions, quantified by a weight function. The \w\ transform
removes lowest polynomial trends that could cause the traditional box-counting
techniques to fail in quantifying the local scaling of the signal. The function
$\tau (q)$ can be considered as a scale-independent measure of the fractal
signal. It can be further related to the Renyi dimensions, Hurst and H{\"o}lder
(at $q=1$ as is clear from the examples in the previous Sections)
exponents (for more detail, see Refs. \cite{dwdk, taka}). The range of validity
of the multifractal formalism for functions can be elucidated \cite{jaff} with
the help of the \m\ methods generalized to the higher moments of \w\ \co s.
Thus, \w\ \an\ goes very far beyond the limits of the traditional \an\ which
uses the language of correlation functions (see, e.g., \cite{dwdk}) in
approaching much deeper correlation levels.

Let us note that $Z_2(j)$ is just the dispersion (variance) of \w\ coefficients
whose average is equal to zero. For positive values of $q$, $Z_q(j)$ reflects
the scaling of the large fluctuations and strong singularities, whereas for
negative $q$ it reflects the scaling of the small fluctuations and weak
singularities, thus revealing different aspects of underlying dynamics.

\section{Discretization and stability}

In signal \an\ , real-life applications produce only sequences of numbers due to
the discretization of continuous time signals. This procedure is called the
sampling of analog signals. Below we consider its implications. At first
sight, it seems that in this case the notions of singularities and H\"{o}lder
exponents are meaningless. Nevertheless, one can say that the behavior of \w\
\co s across scales provides a good way of describing the regularity of
functions whose samples coincide with the observations at a given resolution.

First, we treat the doubly infinite sequence $f^{(d)}=\{f_n\}=\{f(n\Delta t)\}$
for $-\infty <n<\infty $ obtained by sampling a continuous (analog) signal at
the regularly spaced values $t_n=n\Delta t$. If within each $n$-th interval
$\Delta t$ the function $f$ can be replaced by the constant value $f(n\Delta t)$
then for small enough $\Delta t$ one gets
\be
f_{\omega }=\int _{-\infty}^{\infty }f(t)e^{-i\omega t}dt\approx \Delta t\sum
_{n=-\infty }^{\infty }f(n\Delta t)e^{-in\omega \Delta t}.  \label{fdis}
\ee
The inversion formula reads
\be
f_n=\frac {1}{2\pi }\int _{-\infty }^{\infty }f_{\omega }e^{in\omega \Delta t}
d\omega .   \label{finv}
\ee
The function $f_{\omega }$ is periodic with the period $2\pi /\Delta t$. It means
that one can consider it only within the frequency interval $[-\pi /\Delta t,
\pi /\Delta t)$. Moreover, for real signals even the interval $[0, \pi /\Delta t)$
is sufficient. For time-limited signals the summation in Eq. (\ref{fdis}) is done
from $n_{min}=0$ to $n_{max}=N-1$ for $N$ sampled values of a signal. For fast
Fourier transform algorithms, it asks for $O(N\log N)$ computations.

To get the relation between Fourier transforms of the continuous time signal
$f$ and of the sequence of samples $f^{(d)}$ we rewrite the formula (\ref{finv})
\begin{eqnarray}
f_n=\frac {1}{2\pi }\sum _{k=-\infty}^{\infty }\int _{(2k-1)\pi /\Delta t}
^{(2k+1)\pi /\Delta t}f_{\omega }e^{in\omega \Delta t}d\omega =  \nonumber
\end{eqnarray}
\begin{eqnarray}
\frac {1}{2\pi }
\int _{-\pi /\Delta t}^{\pi /\Delta t}\sum _{k=-\infty }^{\infty }
f_{\omega+2k\pi /\Delta t}e^{in\omega \Delta t}d\omega
=\frac {1}{2\pi }
\int _{-\pi /\Delta t}^{\pi /\Delta t}f_{\omega }^{(d)}e^{in\omega \Delta t}
d\omega ,    \label{fnfo}
\end{eqnarray}
wherefrom one gets
\be
f_{\omega }^{(d)}=\sum _{k=-\infty }^{\infty }f_{\omega +2k\pi /\Delta t}.
\label{fofd}
\ee
Thus the Fourier transform of the discrete sample contains, in general, the
contributions from the Fourier transform of the continuous signal not only at
the same frequency $\omega $ but also at countably infinite set of frequencies
$\omega +2k\pi /\Delta t$. These frequencies are called aliases of the frequency
$\omega $. For "undersampled" sets, the aliasing phenomenon appears, i.e., the
admixture of high frequency components to lower frequencies. The frequency
$\omega _{(N)}=\pi /\Delta t$ is called the Nyquist frequency. To improve the
convergence of the series, the "oversampling" is used, i.e., $f$ is sampled
at a rate exceeding the Nyquist rate.

The band-limited function $f$ can be recovered from its samples $f_{n}$'s
whenever the sampling frequency $\omega _s\propto (\Delta t)^{-1}$ is not
smaller than the band-limit frequency $\omega _f$, i.e., whenever
$\omega _s\geq \omega _f$. This statement is known under the name of the
sampling (or Shannon-Kotelnikov) theorem.

In this case, there is no aliasing since only one aliased frequency is in the
band limits $[-\omega _f, \omega _f]$. If $\omega _s<\omega _f$, then the
function $f$ cannot be recovered without additional assumptions.

The rigorous proof of the theorem exists, but it is quite clear already from
intuitive arguments that one cannot restore the high-frequency content of a
signal by sampling it with lower frequency. Therefore, to get knowledge of
high-frequency components one should sample a signal with frequency exceeding
all frequencies important for a given physical process. Only then the
restoration of a signal is stable. One can reduce the sampling frequency
$\omega _s$ all the way down to $\omega _f$ without losing any information.
It allows to subsample the signal by keeping only smaller sets of data, i.e.,
to get a shorter sampled signal.

In case of \w\ \an\ , to have a numerically stable reconstruction algorithm for
$f$ from $d_{j,k}$ one should be sure that $\psi _{j,k}$ constitute a frame.
For better convergence, one needs frames close to tight frames, i.e., those
satisfying the condition $(\frac {B}{A}-1)\ll 1$. The orthonormal \w\
bases have good time-frequency localization. In principle, if $\psi $
itself is well localized both in time and in frequency, then the frame
generated by $\psi $ will share that property as well.

Discrete \w s are quite well suited for the description of functions by their mean
values at the equally spaced points. However in practical real-life
applications, apart from this projection of a function, one has to deal with
a finite interval of its definition and with a finite number of resolution
levels. As was mentioned in Section 5, one usually rescales the "units" of
levels by assuming that the
label of the finest available scale is $j=0$, and the coarser scales have
positive labels $1\leq j\leq J$. The fine-level \co s are defined by the
sampled values of $f(x)$ as $s_{0,k}=f(k)$, i.e., instead of the function $f(x)$
one considers its projection $P_0f$. The higher-level \co s are found
from the iterative relations (\ref{sshs}), (\ref{dsgs}),
i.e., with the help of fast \w\
transform without direct calculation of integrals $\int f(x)\psi _{j,k}(x)dx$
and $\int f(x)\scl _{j,k}(x)dx$. Therefore the approximate representation
of the function $f(x)$ which corresponds to the redefined versions of
Eqs. (\ref{fsc0}) and (\ref{fdec}) can be written as
\be
f(x)\approx P_0f=\sum _{j=1}^J\sum _kd_{j,k}\psi _{j,k}+\sum _ks_{J,k}\scl _{J,k},
\label{disf}
\ee
where the sum over $k$ is limited by the interval in which $f(x)$ is defined.
Moreover, to save the computing time, one can use not a complete set of \w\ \co s
$d_{j,k}$ but only a part of them omitting small \co s not exceeding some
threshold value $\epsilon $. This standard estimation method is called
estimation by \co\ thresholding.
If the sum in (\ref{disf}) is taken only over
such \co s $\vert d_{j,k}\vert >\epsilon $ and the number of the omitted
\co s is equal to $n_o$, then the function $f_{\epsilon }$ which approximates
$f(x)$ in this situation will differ from $f(x)$ by the norm as
\be
\vert \vert f(x)-f_{\epsilon }(x)\vert \vert <\epsilon n_o^{1/2}.  \label{nore}
\ee
Thus for a function, which is quite smooth in the main domain of its definition
and changes drastically only in very small regions, the threshold value
$\epsilon $ can be extremely small. Then it admits large number of \w\ \co s
to be omitted with rather low errors in final approximations. Instead of this
so-called hard-shrinkage procedure, one can use a soft-shrinkage thresholding
\cite{djoh} which shifts positive and negative \co s to their common origin
after the omission procedure, i.e., replaces non-omitted $d_{j,k}$ in the
formula (\ref{disf}) by
\be
d_{j,k}^{(\epsilon )}=sign (d_{j,k})(\vert d_{j,k}\vert -\epsilon ). \label{djk'}
\ee
It has been proven that such an approach leads to optimal min-max estimators.

One can find the \co s of the expansion (\ref{disf}) with use of the fast \w\
transform since the \co s $s_{0,k}$ are fixed as the discrete values of $f(x)$.
In iterative schemes, the error will however accumulate and their precision
will not be sufficiently high. Much better accuracy can be achieved if the
interpolation \w s are used. In this case, the values of the function on the
homogeneous grid $f(k)$ are treated as $s$-\co s for the interpolation basis,
and initial values of $s_{0,k}$ are formed by some linear combinations of them
with \co s which depend on the shapes of \w s considered.

The more elaborate procedure of weighting different \w\ \co s before their
omission called quantization is often used instead of this simplified
procedure of the \co\ thresholding (see also Section 15.4). According to experts
estimates, different weights (importance) are ascribed to different \co s
from the very beginning. These weights depend on the final goals of compression.
Only then the thresholding is applied.
Many orthogonal \w\ bases are infinitely more robust with respect to this
quantization procedure than the Fourier trigonometric basis. Inspite of this,
the lack of symmetry for Daubechies \w s with compact support does not always
satisfy the experts because some visible defects appear. This problem can be
cured when one uses symmetric biorthogonal \w s having compact support.

\section{Some applications}

In this section we describe mostly those applications of wavelets which are
closest to our personal interests (the brief summary is given in Ref.
\cite{dine}). Even among them we have to choose those where,
in our opinion, the use of \w s was crucial for obtaining principally new
information unavailable with other methods (see Web site www.a\w.com).

\subsection{Physics}

Physics applications of \w s are so numerous that it is impossible to
describe even the most important of them here in detail (see, e.g., \cite{asta, aapi, comp,
torr}). They are used both in purely theoretical studies in functional
calculus, renormalization in gauge theories, conformal field theories,
nonlinear chaoticity, and in more practical fields like quasicrystals,
metheorology, acoustics, seismology, nonlinear dynamics of accelerators,
turbulence, structure of surfaces, cosmic ray jets, solar wind, galactic
structure, cosmological density fluctuations, dark matter, gravitational waves
 etc. This list can easily be made longer. We discuss
here two problems related to use of wavelets for solving the differential
equations with an aim to get the electronic structure of an extremely
complicated system, and for pattern recognition in multiparticle production
processes in high energy collisions.

\subsubsection{Solid state and molecules}

The exact solution of a many-body problem is impossible, and one is applying
some approximate methods for solid state problems, e.g., the density functional
theory \cite{pyan}. However, electronic spectra of complex atomic systems are
so complicated that, even within this approach, it is impossible, in practice,
to decipher them by commonly used methods. For example, it would require to
calculate about $2^{100}$ Fourier coefficients to represent the effective
potential of the uranium atom (and even more, in case of uranium dimers).
This is clearly an unrealistic task. Application of the \w\ \an\ methods makes
it possible to resolve this problem \cite{igoe, giva}. The potential of the
uranium dimer is extremely singular. It varies by more than 10 orders of magnitude.
Its reconstruction has now become a realistic problem. Quite high precision has
been achieved with the help of \w s as seen in Fig. 13.

The solution of the density functional theory equations by the \w\ methods
was obtained by several groups \cite{1234, 2345, twan, iant, 3456}. The method
described in Refs \cite{iant, 3456} has been applied to a large variety of
different materials.
It possesses good convergence properties as is shown in Figs. 14 and 15.
They demonstrate how fast is the decrease of energy (or wave function)
recipies with respect to the number of iterations (Fig. 14) or of the absolute
values of the \w\ \co s (Fig. 15) with respect to their index (when the \co s
are ordered according to their magnitude). Such a fast decrease allows us to
consider much more complicated systems than it was possible before. This
method has been successfully tested for solid hydrogen crystals at high
pressure, manganate and hydrogen clusters \cite{iant, 3456}, 3d-metals and their
clusters etc. The density matrix can be effectively represented with the
help of Daubechies \w s \cite{giv2}.

\subsubsection{Multiparticle production processes}

First attempts to use \w\ \an\ in multiparticle production go back to
P. Carruthers \cite{carr, lgca, gglc} who used \w s for diagonalisation of
covariance matrices of some simplified cascade models. The proposals of
correlation studies in high multiplicity events with the help of \w s were
promoted \cite{sboh, huan}, and also, in particular, for special correlations
typical for the disoriented chiral condensate \cite{shth, nand}. It was
recognized \cite{adko} that \w\ \an\ can be used for pattern recognition in
individual high multiplicity events observed in experiment.

High energy collisions of elementary particles result in production of many
new particles in a single event. Each newly created particle is depicted
kinematically by its momentum vector, i.e., by a dot in the three-dimensional
phase space. Different patterns formed by these dots in the phase space
would correspond to different dynamics.  To understand this dynamics is a
main goal of all studies done at accelerators and in cosmic rays. Especially
intriguing is a problem of the quark-gluon plasma, the state of matter with
deconfined quarks and gluons which could exist during an extremely short
time intervals. One hopes to create it in collisions of high energy nuclei.
Nowadays, the data about Pb-Pb collisions are available where, in a single event,
more than 1000 charged particles are produced. We are waiting for RHIC
accelerator in Brookhaven and LHC in CERN to provide events with up to 20000
new particles created. However we do not know yet which patterns will be
drawn by the nature in individual events.
Therefore the problem of phase space pattern recognition in an
event-by-event \an\ becomes meaningful.

It is believed that  the detailed
characterization of each collision event could reveal the rare new phenomena,
and it will be statistically reliable due to a large number of particles
produced in a single event.

When individual events are imaged visually, the human eye has a tendency
to observe different kinds of intricate patterns with dense clusters
and rarefied voids. Combined with search for maxima (spikes) in pseudorapidity
\footnote{The pseudorapidity is defined as $\eta =-\log \tan (\theta /2)$,
where $\theta $ is a polar angle of particle emission.}
(polar angle) distributions, it has lead to indications on the so-called
ring-like events in cosmic ray studies \cite{addk, alex, masl, arat, dtre}
and at accelerators \cite{maru, adam}. However, the observed
effects are often dominated by statistical fluctuations. The method of
factorial moments was proposed \cite{bpes} to remove the statistical background
but it is hard to apply in event-by-event approach. The \w\ \an\ avoids smooth
polynomial trends and underlines the fluctuation patterns. By choosing the
strongest fluctuations, one hopes to get those which exceed the statistical
component. The \w\ transform of the pseudorapidity spectra of cosmic ray high
multiplicity events of JACEE collaboration was done in Ref. \cite{sboh}.

In Ref. \cite{adko} \w s were first applied to analyze patterns formed in the
phase space of the accelerator data on individual high
multiplicity events of Pb-Pb interaction at energy 158 GeV per nucleon.
With emulsion technique used in experiment the angles of emission of particles
are measured only, and the two-dimensional (polar+azimuthal amgles) phase
space is considered therefore. The experimental statistics is rather low but
acceptance is high and homogeneous that is important for the proper pattern
recognition. To simplify the \an\ ,
the two-dimensional target diagram representing the
polar and azimuthal angles of created charged particles was split into 24
azimuthal angle sectors of $\pi /12$ and in each of them particles were
projected onto the polar angle $\theta $ axis. Thus
one-dimensional functions of the polar angle (pseudorapidity) distribution
of these particles in 24 sectors were obtained. Then the \w\ coefficients
were calculated in all of these sectors and connected together afterwards
(the continuous MHAT \w\
was used). The resulting pattern showed that many particles are concentrated
close to some value of the polar angle, i.e., reveal the ring-like structure
in the target diagram. The interest to such patterns is related to the fact
that they can result from the so-called gluon Cherenkov radiation
\cite{dre1, dre2} or, more generally, from the gluon bremsstrahlung at a finite
length within quark-gluon medium (plasma, in particular). More elaborate
two-dimensional \an\ with Daubechies ($D^8$) \w s was done recently \cite{dikk} and confirmed these conclusions with jet
regions tending to lie on some ring-like formations. It is seen, e.g. in Fig. 16
where dark regions correspond to large \w\ coefficients of the large-scale
particle fluctuations in two of the events analyzed. The resolution levels
$6\leq j\leq 10$ were left only after the event \an\ was done to store
the long-range correlations in the events and get rid of short-range ones and
background noise. Then the inverse restoration was done to get the event
images with these dynamic correlations left only, and this is what is seen
in Fig. 16. It directly demonstrates that large-scale correlations chosen
have a ring-like (ridge) pattern. With larger statistics, one will be able to
say if they correspond to theoretical expectations. However preliminary results
favor positive conclusions \cite{dikk}. It is due to the two-dimensional \w\ \an\
that for the first time the fluctuation structure of an event is shown
in a way similar to the target diagram representation of events on the
two-dimensional plot.

Previously, some attempts \cite{addk, aaaa, cddh} to consider such events with
different methods of treating the traditional projection and correlation
measures revealed just that such substructures lead to spikes in
the polar angle (pseudorapidity) distribution and are somewhat jetty. Various
Monte Carlo simulations of the process were compared to the data and failed
to describe this jettiness in its full strength. More careful \an\
\cite{dlln, agab} of large statistics data on hadron-hadron interactions
(unfortunately, however, for rather low multiplicity) with
dense groups of particles separated showed some "anomaly" in the angular
distribution of these groups awaited from the theoretical side. Further \an\
using the results of \w\ transform is needed when many high multiplicity events
become available. The more detailed review of this topic is given in
Ref. \cite{duf0}.

\subsection{Aviation (engines)}

Multiresolution wavelet analysis has proved to be an extremely useful
mathematical method for analyzing complicated physical signals at various
scales and definite locations. It is tempting to begin with analysis
of signals depending on a single variable\footnote{The trick of reducing
the two-dimensional function to a set of one-dimensional ones was described
in the previous subsection devoted to particle production.}. The time variation
of the pressure in an aircraft compressor is one of such signals. The aim of
the \an\ of this signal is motivated by the desire to find the precursors of a
very dangerous effect (stall+surge) in engines leading to their destruction.

 An axial multistage compressor is susceptible to the formation of a rotating
stall which may be precipitated by a distorted inlet flow induced by abrupt
manoeuvres of an aircraft (helicopter) or flight turbulence. The aerodynamic
instability behind the rotating
 blades is at the origin of this effect. The instability regions called
stall cells separate from the blades and rotate with a speed about 60$\%$ of
the rotor speed. Thus they are crossed by the blades approximately
one time every 1.6 rotor revolutions. It limits abruptly the flow and thus the
pressure delivery to the downstream plenum (the combustion chamber) where the
fuel is burnt up and gas
jet is formed. The high prestall pressure build-up existing in the combustion
chamber tends to push the flow backwards through the compressor. If the flow
is reversed one calls it "deep surge". In many cases it is attested by
flames emerging from the engine inlet. It can have serious consequences for
the engine life and operation, not to say about an aircraft and its passengers
if it happens at in-flight conditions.
That is why the search for precursors of these dangerous phenomena is very
important. Attempts to predict the development of a rotating stall and a
subsequent surge with a velocity measuring probe such as a hot wire
anemometer \cite{geor} provide a precursor or warning of only about 10 msec
what is not enough for performing any operations which would preclude surge
development.

Multiresolution wavelet analysis of pressure variations in a gas turbine
compressor reveals much earlier precursors of stall and surge processes
\cite{dfin}. Signals from 8 pressure sensors positioned at various places
with the compressor were recorded and digitized at 3 different modes of its
operation (76$\%$, 81$\%$, 100$\%$ of nominal rotation speed) in stationary
conditions with interval 1 msec during 5--6 sec before the stall and
the surge developed. An instability was induced
by a slow injection of extra air into the compressor inlet. After a few
minutes, this led to a fully blown instability in the compressor.
The interval of last 5--6 sec was wavelet-analyzed. Since the signal fluctuates
in time, so too does the sequence of \w\ coefficients at any given scale.
A natural measure for this variability is the standard deviation (dispersion)
of \w\ coefficients as a function of scale. A scale of $j=5$ showed
the remarkable drop about 40$\%$ of the dispersion (variance) of the wavelet
coefficients for more than 1 sec prior the malfunction develops. The dispersion
is calculated according to the standard expression
\begin{equation}
\sigma (j,M)=\sqrt {\frac {1}{M-1}\sum _{k=0}^{M-1}[d_{j,k}-\langle d_{j,k}
\rangle]^2} ,   \label{sigm}
\end{equation}
where $M$ is the number of wavelet coefficients at the scale $j$ within some
time interval which was chosen equal to 1 sec.

In Fig. 17 the time variation of the pressure in the compressor at one of the
indicated above modes of engine operation is shown by
a very irregular line. Large fluctuations of the pressure at the right-hand side
denote the surge onset. It is hard to guess about any warning of this drastic
instability from the shape of this curve. The wavelet coefficients of this
function were calculated with Daubechies 8-tap ($D^8$) wavelets.
The dashed curve in the same Figure
shows the behavior of the dispersion of wavelet coefficients as a function
of time at the scale level $j=5$ which happens to be most sensitive to the
surge onset\footnote{It shows that the correlations among
successive values of the sampled signal are likely to occur on intervals of
the order of 30 msec. Let us note that similar correlation lengths are typical
for a sampled speech signal as well.}. Its precursor is seen as the maximum
and the subsequent drop of the standard deviation
by about 40$\%$ which appears about 1 - 2 sec prior the malfunction denoted by
the large increase of both the pressure and dispersion at the right-hand side.
Initial part of the dispersion plot is empty because the prescribed interval
for the compilation of the representative distribution of \w\ coefficients is
chosen equal to 1 sec.
The randomly reordered (shuffled) sample of the same values of the pressure
within the pre-surge time interval does not show such a drop of the
dispersion as demonstrated by the dot-dashed line in the Figure. It
proves the dynamic origin of the effect. Further analyzed characteristics of
this process (fractality, high-moments behavior) are discussed in \cite{dfin}.
No fractal properties of the signal have been found. The scale $j=5$ falls
off the linear dependence of the partition function on $j$ necessary for fractal
behavior. Let us note that the multifractality of the signal may fail because
of accumulation of points where a chirp-like behavior happens. If it is
the origin of this effect, it should provide a guide to the description
of the dynamics of the analyzed processes. The time intervals before and after
appearence of the precursor were analyzed separately.
Higher moments of \w\ coefficients as functions of their rank $q$
behave in a different way at pre- and post-precursor time. It indicates on
different dynamics in these two regions.

This method of wavelet analysis can be applied to any
engines, motors, turbines, pumps, compressors etc. It provides significant
improvement in the diagnostics of the operating regimes of existing
engines, which is important for preventing their failure and, consequently,
for lowering associated economic losses. Two patents on diagnostics and
automatic regulation of engines dated 19.03.1999 have
been obtained by the authors of Ref. \cite{dfin}.

There are other problems in aviation which could be approached with \w\ \an\ ,
e.g., combustion instabilities, \an\ of ions in the jet from the combustion
chamber or more general problem of metal aging and cracks etc.

\subsection{Medicine and biology}

Applications of the \w\ \an\ to medicine and biology (see, e.g.,
\cite{auns, akay, llsp}) follow
also the same procedures as discussed above . They are either deciphering
information hidden in one-dimensional functions (analysis of heartbeat intervals,
ECG, EEG, DNA sequences etc) or pattern recognition (shapes of biological
objects, blood cell classification etc).

\subsubsection{Hearts}

Intriguing results were obtained \cite{tfte} when the multiresolution \w\ \an\
was applied to the sequence of time intervals between human heartbeats. It has
been claimed that the {\it clinically} significant measure of the presence
of heart failure is found just from analysis of heartbeat intervals alone without
knowledge of the full electrocardiogram plot while the previous approaches provided
{\it statistically} significant measures only. Series of about 70000 interbeat
intervals were collected for each of 27 patients treated. They were plotted
versus interval number. Signal fluctuations were \w\ transformed and dispersions
of \w\ coefficients at different scales
were calculated as in the case of aviation engines considered above but with
averaging over the whole time interval because it is not necessary now to
study the evolution during this time interval. Thus each patient was
characterized
by a single number (dispersion) at a given scale. It happened that the sets
of numbers for healthy and heart failure patients did not overlap at the
scale\footnote{By coincidence it is the same as for aviation engines.}
$j=5$ as seen in Fig. 18. This is considered as the clinically significant
measure in distinction to the statistically significant measure when these
sets overlap partly. The dispersions for healthy patients are larger
probably corresponding to a higher flexibility of healthy hearts. The fluctuations
are less anti-correlated for heart failure than for healthy dynamics.

This analysis was questioned by another group \cite{agis, iagh} stating that the
above results on the scale-dependent separation between healthy and pathologic
behavior depend on the database analyzed and on the analyzing \w\ . They
propose new scale-independent measures -- the exponents characterizing the
scaling of the partition function of the \w\ coefficients of the heartbeat
records. These exponents reveal the multifractal behavior for healthy human
heartbeat and loss of multifractality for a life-threatening condition,
congestive heart failure. This distinction is ascribed to nonlinear features
of the healthy heartbeat dynamics. The authors claim that the multifractal
approach to \w\ \an\ robustly discriminates healthy subjects from
heart-failure subjects. In our opinion, further studies are needed.

Earlier, the Fourier \an\ of the heart rate variability disclosed the
frequency spectrum of the process. Three frequency regions play the main role.
The high frequency peak is located about 0.2 Hz, and the low frequency peak
is about 0.1 Hz. The superlow frequency component reminds $1/f$-noise with its
amplitude strongly increasing at low frequencies. There are some attempts
\cite{ylia} to develop theoretical models of such a process with the help of
\w\ transform.

The \w\ \an\ of one-dimensional signals is useful also for deciphering the
information contained in ECG and EEG. The functions to be analyzed there are
more complicated than those in above studies. Some very promising
results have been obtained already. In particular, it was shown that anomalous
effects in ECG appear mostly at rather large scales (low frequencies) whereas
the normal structures are inclined to comparatively small scales (high frequencies).
It corresponds to the above results on heartbeat intervals. Such \an\ of ECG was
reported, e.g., in Ref. \cite{irpm}. We shall not describe it in detail here
because these studies are still in their initial stage only. The time-frequency
\an\ of EEG can be found, e.g., in Ref. \cite{bcrs}. It can locate the source
of the epileptic activity and its propagation in the brain. The general
description of EEG methodics see, e.g., in Refs. \cite{elec, blan}.

The \w\ \an\ has been used also for diagnostics of the embrion
state during the pregnancy period \cite{llsp}.
The matching pursuit method was developed to fit the properties of the signal
as well as possible.

\subsubsection{DNA sequences}

The \w\ transform with continuous \w s has been used for fractal \an\ of DNA
sequences \cite{abgm, aabm} in attempt to reveal the nature and the origin
of long range correlations in these sequences. It is still of debate whether
such correlations are different for protein-coding (exonic) and noncoding
(intronic, intergenic) nucleotide sequences. To graphically portray these
sequences in the form of one-dimensional functions, the so-called "DNA walk"
\an\ was applied with the conversion of the 4-letter text of DNA into a binary
set \cite{pbgh}. Applying \w\ transform to 121 DNA sequences
(with 47 coding and 74 noncoding regions) selected in the human genome, the
authors of Refs. \cite{abgm, aabm} have been able to show that there really
exists the difference between the two subsequences. It is demonstrated by the
presence of long range correlations in noncoding sequences while the coding
sequences look like uncorrelated random walks. To show this, the averaged
partition  (generating) functions of \w\ coefficients (see Eq.(\ref{zq})) over these
two statistical samples were calculated. Both of them scaled with the exponent
predicted for homogeneous Brownian motions, i.e., $\tau (q)=qH-1$. The main
difference between the noncoding and coding sequences is the value of $H$,
namely, $H_{nc}=0.59\pm 0.02$ and $H_c=0.51\pm 0.02$ which distinguish
uncorrelated and correlated subsamples. Moreover, it has been shown that
$\tau (q)$ spectra extracted from both sets are surprisingly in remarkable
agreement with the theoretical prediction for Gaussian processes if the
probability density of \w\ coefficients at fixed scales are plotted versus
these coefficients scaled with their r.m.s. value at the same scale. The
results of \w\ \an\ clearly show that the purine (A, G) versus pyrimidine
(C, T) content\footnote{A, C, G, T denote the usual four letter alphabet of any
DNA text.}
of DNA is likely to be relevant to the long range correlation
properties observed in DNA sequences.

\subsubsection{Blood cells}

Another problem which can be solved with the help of wavelet analysis is the
recognition of different shapes of biological objects. By itself, it has very
wide range of applicability. Here, we consider the blood cell classification
scheme according to the automatic \w\ \an\ developed by us, and a
particular illustration is given for erythrocytes classification. The automatic
search, stability in determining the cells shapes and high speed of processing
can be achieved with the computer \w\ \an\ . The main idea of the method relies
on the fact that at a definite resolution scale \w\ \an\ reveals clearly the
contours of the blood cells what allows to classify them. In Fig. 19 we
demonstrate  how it became possible to improve the image resolution by such a
method. The low quality image of blood cells has been transformed in its \w\
image with clearly seen contours of the cells.

The shapes of erythrocytes differ for various types of them. Depending on
their shape, they can play either positive or negative role for a human being.
Therefore their differentiation is very important. After microscope \an\ of
a dry blood smear and registration of blood cells in a computer, the \w\ \an\
of the set of registered blood cells has been done.

However the extreme random irregularities at some points of a cell contour
can prevent from performing the \an\ . Therefore a special smearing procedure
was invented to avoid such points without loss of crucial peculiarities
typical for a definite type of cells. After that, the set of blood cells
with somewhat smeared contours is ready for \w\ \an\ . It consists in the \w\
correlation \an\ which shows different behavior of the correlation measures
depending on the particular cell shapes. Using the expert classification of
cells in a definite sample, the \w\ characteristics of correlations typical
for a particular class were found and then inserted in the computer program.
This software was used for classification of blood cells obtained from other
patients, and results were cross-checked by experts. Their positive
conclusions are encouraging.
The whole procedure is now done fast and automatically without human
intervention. In Fig. 20 the blood cells of various kinds are shown.


\subsection{Data compression}

Data compression is needed if one wants, e.g., to store the data spending as
low memory capacity as possible or to transfer it at a low cost using smaller
packages. The example of FBI using \w\ \an\ for pattern recognition and saving
in that way a lot of money on computer storage of fingerprints is well known.
This is done by omitting small \w\ coefficients after the direct \w\ transform
was applied. Surely, to restore the information one should be confident in
stability and good quality of the inverse transformation. Therefore, both \an\ and
synthesis procedures are necessary for the data compression and its subsequent
reconstruction. Above, we have shown how successful is \w\ \an\ in solving many
problems. Due to the completeness of the \w\ system it is well suited for the
proper inverse transform (synthesis) as well (see, e.g., Ref. \cite{abmd}).

The approach to the solution of the problem strongly depends on the actual
requirements imposed on the final outcome. There are at least three of them.
If one wants to keep the quality of the restored image (film) practically
as good as the initial one, the compression should not be very strong.
It is required, e.g., if the experts should not distinguish the compressed
and uncompressed copies of a movie when they are shown on two screens in
parallel. Another requirement would be important if one wants to compress
an image as strongly as possible leaving it still recognizable. It is
required, e.g., if one needs to transfer the information in a line with
limited capacity. Finally, one can require the whole procedure of \an\ and
synthesis to be done as fast as possible. It is necessary, e.g., if the
information must be obtained immediately but at lower cost. These three
situations ask for different choice of \w s to optimize \an\ and synthesis.
In all cases \w\ \an\ has an advantage over the coding methods which use the
windowed Fourier transform but the quantitative estimate of this
advantage varies with the problem solved.

Let us remind that any image in the computer should be digitized and saved
as the bitmap or, in other words, as a matrix, each element of which
describes the color of the point in the original image. The number of elements
of the matrix (image points) depends on the resolution chosen in the
digital procedure. It is a bitmap that is used for the subsequent
reproduction of the image on a screen, a printer etc. However, it is not
desirable to store it in such a form because it would ask for the huge
computer capacity. That is why, at present, the numerous coding algorithms
(compression) of a bitmap are developed, whose effectiveness depends
on the image characteristics\footnote{It is evident that to store the
"Black square", one does not have to deal with a matrix of all black dots
but must store just the three numbers showing the width, the height and the
color. This is the simplest example of the image compression.}. All these
algorithms belong to the two categories - they are either the coding with
a loss of information or without any loss of it (in that case the original
bitmap can be completely recovered by the decoding procedure). In a more
general applications, the last algorithm is often called as the data
arxivation.

As an example, we consider the compression algorithm with a loss of information.
"The loss of information" means in this case that the restored image is not
completely identical to the original one but this difference is practically
indistinguishable by a human eye\footnote{The "compression quality" is usually
characterized by a parameter which varies from 0 to 100. Here 100 implies the
minimal compression (the best quality), and the restored image is practically
indistinguishable from the initial one, while 0 means the maximum compression
which still allows to distinguish some details of the original image in the
recovered one.}.

At present, most of the computer stored images (in particular, those used in
Internet) with the continuous tone\footnote{that is the images contained many
slightly different colors. Ordinarily to store such images in a computer 16 millions
colors per pixel are used}
are coded with the help of the algorithm
JPEG (for the detailed description, see \cite{wall}). Main stages of this
algorithm are as follows. One splits the image in the matrices with 8 by 8 dots.
For each matrix, the discrete cosine-transform is performed. The obtained
frequency matrices are subjected to the so-called "quantization" procedure
when the most crucial for the visual recognition elements are chosen according
to a special "weight" table compiled beforehand by specialists after their
collective decision was achieved. This is the only stage where the "loss" of
information happens. Then the transformed matrix with the chosen frequencies
(scales) becomes compact and it is coded by the so-called entropy method (also
called arithmetic or Huffmann method).

Applied above algorithm differs from the described one by use of the \w s
instead the windowed cosine-transform and by the transform of the whole
image instead the 8 by 8 matrix only. Fig. 21 demonstrates the original image
and its two final images restored after the similar compression according to
JPEG and \w\ algorithms. It is easily seen that the quality of the \w\ image
is noticeably higher than for JPEG at practically the same size of the coded
files. The requirement of the same quality for both algorithms leads to the
file sizes 1.5 - 2 times smaller for the \w\ algorithm that could be crucial
for the transmission of the image, especially if the transmission line
capacity is limited.

\subsection{Microscope focusing}

Surely, the problem of microscope focusing is tightly related to pattern
recognition. One should resolve well focused image from that with diffused
contours. It is a comparatively easy task for \w s because in the former case
the image gradients at the contour are quite high while in the latest one
they become rather vague. Therefore the \w\ coefficients are larger when
microscope is well focused to the object and drastically decrease with
defocusing. At a definite resolution level corresponding to the contour
scale the defocusing effect is strongest. It is demonstrated in Fig. 22.
The peak of the \w\ \co s shows the most focused image. After doing the \w\ \an\
of an image, the computer sends a command to shift the microscope so that
larger values of the \w\ \co s and thus better focusing are attained. At other
levels it is somewhat less pronounced and, moreover, it is asymmetric
depending on whether the microscope
is positioned above the focus location or below it. This asymmetry has been
used for the automatic microscope focusing with a well defined direction
of its movement toward the focus location.

\section{Conclusions}

The beauty of the mathematical construction of the \w\ transformation and its
utility in practical applications attract nowadays researchers from both pure
and applied science. Moreover, commercial outcome of this research becomes
quite important. We have outlined a minor part of activity in this field.
However we hope that the general trends in the development of this subject
became comprehended and appreciated.

Unique mathematical properties of \w s made them a very powerful tool in \an\
and subsequent synthesis of any signal. The orthogonality property allows to get
an independent information from different scales. Normalization assures that
at different steps of this transformation the value of information is not
changed and mixed up. The property of localization helps get knowledge about
those regions in which definite scales (frequencies) play most important
role. Finally, the completeness of the system of \w\ functions formed by
dilations and translations results in validity of the inverse transformation.

The multiresolution \an\ leads to fast \w\ transform and together with the
procedure of the nonstandard matrix multiplication to rather effective
computing. Analytic properties of functions, local and global H\"{o}lder
indices, multifractal dimensions etc are subject to \w\ \an\ . The natural
accomodation of differential operators in this enviroment opens a way to
effective solution of differential equations.

All these properties enable us to analyze through the \w\ transform complex
signals at different scales and locations, to solve equations describing
extremely complicated nonlinear systems involving interactions at many scales,
to study very singular functions etc. The \w\ transform is easily generalized
to any dimension, and multidimensional objects can be analyzed as well.
Therefore, \w s are indispensable for pattern recognition.

Thus the \w\ applications in various fields are numerous and give nowadays
very fruitful outcome. In this review paper we managed to describe only
some of these applications leaving aside the main bulk of them. The
potentialities of \w s are still not used at their full strength.

However one should not cherish vain hopes that this machinery works
automatically in all situations by using its internal logic and does not
require any intuition. According to Meyer \cite{meye}, "no 'universal
algorithm' is appropriate for the extreme diversity of the situations
encountered". Actually it needs a lot of experience in choosing
the proper \w s, in suitable formulation of the problem under investigation,
in considering most important scales and characteristics describing the
analyzed signal, in the proper choice of the algorithms (i.e., the methodology)
used, in studying the intervening singularities, in avoiding
possible instabilities etc. By this remark we would not like to prevent
newcomers from entering the field but, quite to the contrary, to attract those
who are not afraid of hard but exciting research and experience.

\vspace{3mm}

{\bf Appendix 1}

\vspace{2mm}
The general approach which respects all properties required from \w s is known
as the \mr\ approximation and defined mathematically in the following way:

{\it DEFINITION} \cite{meye}

A multiresolution approximation of $L^2(R^n)$ is, by definition, an
increasing sequence $V_j,\; j\in Z$, of closed linear subspaces of
$L^2(R^n)$ with the following properties:

\begin{enumerate}
 \item $\bigcap_{-\infty}^{\infty} V_j=\{0\},
       \bigcup_{-\infty}^{\infty} V_j$ is dense in $L^2(R^n)$;
 \item for all $f \in L^2(R^n)$ and all $j \in Z^n$
       $$f(x) \in V_j \leftrightarrow f(2x) \in V_{j+1};$$
 \item for all $f \in L^2(R^n)$ and all $k \in Z^n$
       $$f(x) \in V_0 \leftrightarrow f(x-k) \in V_{0};$$
 \item there exists a function, $g(x) \in V_0$, such that the sequence
       $g(x-k),\; k \in Z^n$, is an orthonormal (or Riesz) basis of the
       space $V_0$.
 It is clear, that scaling versions of the function $g(x)$ form bases
 for all $V_j$.
\end{enumerate}

Thus the \mr\ \an\ consists of a sequence of successive approximation spaces
$V_j$ which are scaled and invariant under integer translation versions of
the central space $V_0$. There exists an orthonormal or, more generally,
Riesz basis in this space. The Haar \mr\ \an\ can be written in these terms as
\be
V_j=\{f \in L^2(R);\;\;  for\; all\; k \in Z: f\vert _{[k2^{-j},(k+1)2^{-j})}=
const \}.  \label{vjha}
\ee
In Fig. 3 we showed what the projections of some $f$ on the Haar spaces $V_0$,
$V_1$ might look like. The general distributions are decomposed into
series of correctly localized fluctuations of a characteristic form defined by
the \w\ chosen.

The functions $\scl _{j,k}$ form an orthonormal basis of $V_j$. The orthogonal
complement of $V_j$ in $V_{j+1}$ is called $W_j$. The subspaces $W_j$ form a
mutually orthogonal set. The sequence of $\psi _{j,k}$ constitutes an
orthonormal basis for $W_j$ at any definite $j$. The whole collection of
$\psi _{j,k}$ and $\scl _{j,k}$ for all $j$ is an orthonormal basis for $L^2(R)$.
 This ensures us that we have constructed a \mr\ \an\ approach, and the
functions $\psi _{j,k}$ and $\scl _{j,k}$ constitute the small and large
scale filters, correspondingly. The whole procedure of the \mr\ \an\ has been
demonstrated in graphs of Fig. 4.

In accordance with the above formulated goal, one can define the notion of \w s
in the following way:

{\it DEFINITION} \cite{meye}

A function $\psi (x)$ of a real variable is called a (basic) \w\ of class $m$
if the following four properties hold:

\begin{enumerate}
\item if $m=0$, $\psi (x)$ and $\scl (x)$ belong to $L^{\infty}(R)$;
if $m\geq 1, \psi (x),\; \scl (x)$
and all their derivatives up to order $m$ belong to $L^{\infty}(R)$;
\item $\psi (x),\; \scl (x)$ and all their derivatives up to order $m$
decrease rapidly as $x\rightarrow \pm \infty$;
\item $\int _{-\infty }^{\infty }dxx^n\psi (x)=0$ for $0\leq n\leq m$,
and $\int _{-\infty }^{\infty }dx\scl (x)=1$;
\item the collection of functions $2^{j/2}\psi (2^jx-k),\; 2^{j/2}\scl (2^jx-k),\;
j,k \in Z$, is an orthonormal basis of $L^2(R)$.
\end{enumerate}

Then the equation (\ref{fdec}) is valid. If $\psi $ and $\scl $ both have
compact support, then it gives a decomposition of any distribution of order
less than $m$. Moreover, the order of the distribution $f$ (the nature of its
singularities) can be calculated exactly and directly from the size of its \w\
\co s as has been shown in Sections 11, 12.
The functions $2^{j/2}\psi (2^jx-k)$ are the \w s (generated by the "mother"
$\psi $), and the conditions 1., 2., 3. express, respectively, the regularity,
the localization and the oscillatory character. One sees that the property 3.
is satisfied for the Haar \w s for $m=0$ only, i.e., its regularity is $r=0$.
In general, for each integer $r\geq 1$, there exists a \mr\ approximation
$V_j$ of $L^2(R)$ which is $r$-regular and such that the associated real-valued
functions $\scl $ and $\psi $ have compact support. As the regularity increases,
so do the supports of $\scl $ and $\psi $.
The \w\ $2^{j/2}\psi (2^jx-k)$ is "essentially concentrated" on the dyadic
interval $I=[k2^{-j},(k+1)2^{-j})$ for $j,k \in Z$. Its Fourier transform is
supported by $2^{j}(2\pi /3)\leq \vert \omega \vert \leq 2^j(8\pi /3)$. In fact,
it has a one octave frequency range.

\vspace{3mm}

{\bf Appendix 2}

\vspace{2mm}

In the \w\ \an\ of operator expressions the integral Calderon-Zygmund
operators are often used.
There are several definitions of them (see, e.g., the monograph \cite{mcoi}).
We give here their definition used by Daubechies \cite{daub}.

{\it DEFINITION}\\

A Calderon-Zygmund operator $T$ on $R$ is an integral operator
\be
Tf(x)=\int dyK(x,y)f(y)     \label{czyg}
\ee
for which the integral kernel satisfies
\be
\vert K(x,y)\vert \leq \frac {C}{\vert x-y\vert},     \label{kxyl}
\ee
\be
\vert \frac {\partial }{\partial x}K(x,y)\vert +
\vert \frac {\partial }{\partial y}K(x,y)\vert \leq \frac {C}{\vert x-y\vert ^2},
\label{dkxy}
\ee
and which defines a bounded operator on $L^2(R)$.\\

With such a definition the special care should be taken at $x=y$.

\vspace{3mm}

{\bf Appendix 3}

\vspace{2mm}

In the literature devoted to the signal processing, the so-called
Littlewood-Paley decomposition is often used. It is closely related to
the \w\ transform. Therefore we give here the dictionary which relates the
Littlewood-Paley \co s denoted as $\Delta _j(f)(x)$
to both discrete $(d_{j,k})$ and continuous $(W(a,b))$  \w\ \co s.
\be
\Delta _j(f)(x)=2^{nj/2}d_{j,2^jx}=W(2^{-j},x),  \label{ddw1}
\ee
or
\be
d_{j,k}=2^{-nj/2}W(2^{-j},2^{-j}k)=2^{-nj/2}\Delta _j(f)(2^{-j}k).  \label{ddw2}
\ee
\vspace{3mm}

{\bf Acknowledgements}

\vspace{2mm}

We are grateful to all our colleagues with whom we worked on \w\ problems.
Our special thanks are to A. Leonidov who read this paper and made important
suggestions. O.I. is especially grateful to S. Goedecker with whom many methods
were developed and discussed. V.N. is also grateful to INTAS for the support (grant 97-103).

\newpage

{\bf Figure Captions}

\vspace{2mm}

Fig. 1. The histogram and its \w\ decomposition.\\
The initial histogram is shown in the upper part of the Figure. It corresponds
to the level $j=4$ with 16 bins (Eq. (\ref{fscl})). The intervals are labelled
on the abscissa axis at their left-hand sides. The next level $j=3$ is shown
below. The mean values over two neighboring intervals of the previous level are
shown at the left-hand side. They correspond to eight terms in the first sum
in Eq. (\ref{fsc3}). At the right-hand side, the \w\ \co s $d_{3,k}$ are shown.
Other graphs for the levels $j=2, 1, 0$ are obtained in a similar way.\\

Fig. 2. The Haar scaling function $\scl (x)\equiv \scl _{0,0}(x)$ and
"mother" \w\ $\psi (x)\equiv \psi _{0,0}(x)$.\\

Fig. 3. The analyzed function (a) and its Haar projections onto two subsequent
spaces V.\\

Fig. 4. The graphical representation of the multiresolution \an\ with
decomposition of $V_{j+1}$ space onto its subspace $V_j$ and the orthogonal
complement $W_j$ iterated to the lower levels.\\

Fig. 5. Daubechies scaling functions (solid lines) and \w s (dotted lines)
for $M=2, 4$.\\

Fig. 6. The fast \w\ transform algorithm.\\

Fig. 7. Coiflets (dotted lines) and their scaling functions (solid lines)
for $M=4$. \\

Fig. 8. The \w\ packet construction.\\

Fig. 9. The example of the \w\ \an\ of a two-dimensional plot.\\
One sees that either horizontal or vertical details of the plot are more clearly
resolved in the corresponding \co s. Also, the small or large size (correlation
length) details are better resolved depending on the level chosen.\\

Fig. 10. The lattices of time-frequency localization for the \w\
transform (left)  and windowed Fourier transform (right).\\

Fig. 11. The matrix representation of the standard approach to the \w\ \an\ .\\
         The parts containing non-zero \w\ \co s are shaded.\\

Fig. 12. The nonstandard matrix multiplication in the \w\ \an\ .\\

Fig. 13. The Coulomb potential V (the solid line) as a function of the distance
and the precision of its calculation $\Delta V$ (the dashed line) for the
Uranium dimer in arbitrary units (a.u.).\\
It is drawn for the direction from one atom to another one in the left-hand side
and for the opposite direction in the right-hand side. The complete symmetry
is seen with a larger distances shown in the right-hand side.\\

Fig. 14. Fast decrease of energy recipies with respect to the number of iterations
demonstrates good convergence of the solution of the density functional theory
equations by the \w\ methods.\\

Fig. 15. The same as in Fig. 14 for the absolute values of the \w\ \co s.\\

Fig. 16. The restored images of long-range correlations in experimental target
diagrams.\\  They show the typical ring-like structure in some events of central
Pb-Pb interactions.\\

Fig. 17. The signal of the pressure sensor (the solid line) and the dispersion
of its \w\ \co s (the dashed line).\\
The time variation of the pressure in the engine compressor (the
irregular solid line) has been \w\ analyzed. The dispersion of the \w\ \co s
(the dashed line) shows the maximum and the remarkable drop prior the drastic
increase of the
pressure providing the precursor of this malfunction. The shuffled set of the
data does not show such an effect for the dispersion of the \w\ \co s
(the upper curve) pointing to its dynamic origin.\\

Fig. 18. The sets of the values of the dispersions of the \w\ \co s for
the heartbeat intervals of the healthy (white circles) and heart failure
(black circles) patients are shown. \\ They do not overlap at $j=4$.\\

Fig. 19. a) The photo of blood cell, obtained from the microscope.\\
         b) The same photo after the \w\ \an\ done.\\
The blurred image of a blood cell (left) becomes clearly visible one
(right) after the \w\ transform.\\

Fig. 20. The classification of the erythrocytes cells.\\

Fig. 21. a) The original photo (the file size is 461760 bytes).\\
         b) The photo reconstructed after compression according to the
         JPEG-algorithm.(the file size is 3511 bytes).\\
         c) The photo reconstructed after compression according to the
         \w\ algorithm (the file size is 3519 bytes).\\
 Better quality of the \w\ transform is clearly seen when comparing the
 original image (left) and two images restored after the similar
compression by the windowed Fourier transform (middle) and the \w\ transform
(right). \\

Fig. 22. The focusing line.\\
The values of the \w\ \co s calculated for different positions of the microscope
are shown on the vertical axis. The corresponding positions are numbered on
the horizontal axis. The best focusing is obtained at the maximum of the curve
(positions 6 - 10).
Large \w\ \co s correspond to the better focused image. The microscope
moves to the focus position being driven by computer commands for increase of
the \w\ \co s of the analyzed objects.

\newpage
\vspace*{-18mm}

 \hspace*{-2cm}
\newlength{\jup}
\jup=20mm

\begin{tabular}{rcrcc}
&&&&j\\
&\epsfig{file=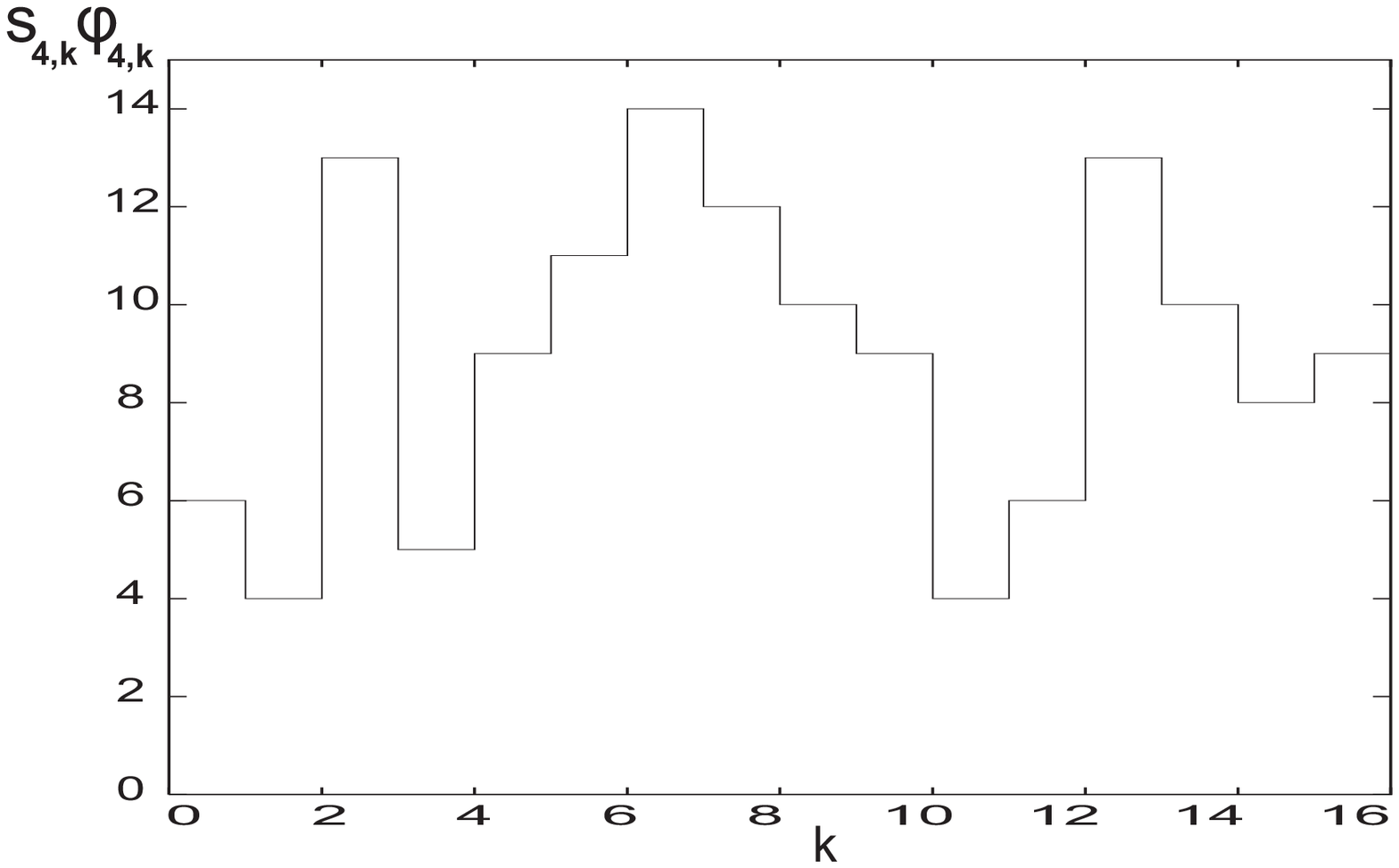,scale=0.4,clip=}& &&\raisebox{\jup}{4}\\
&\epsfig{file=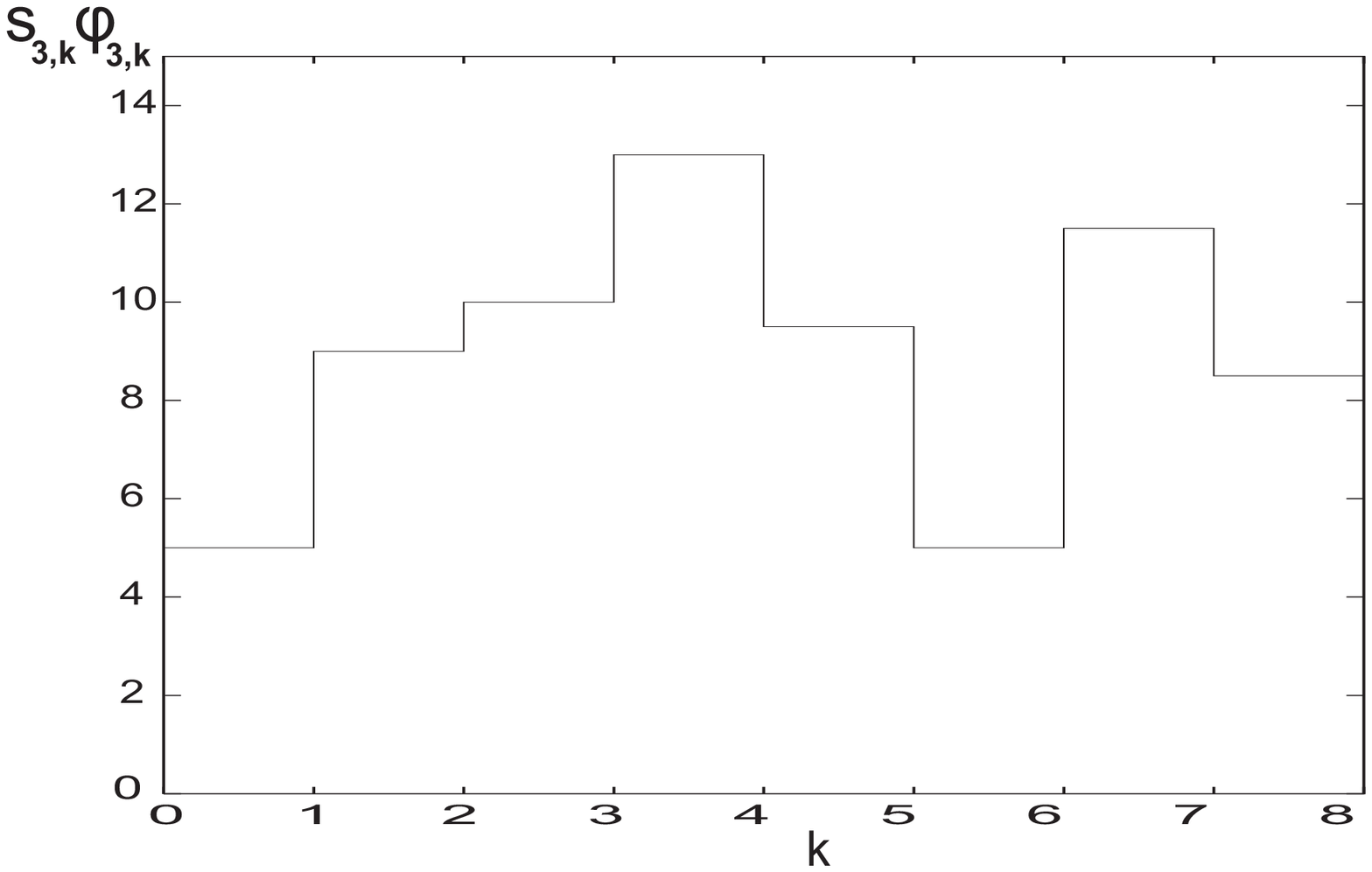,scale=0.4,clip=}&&
\epsfig{file=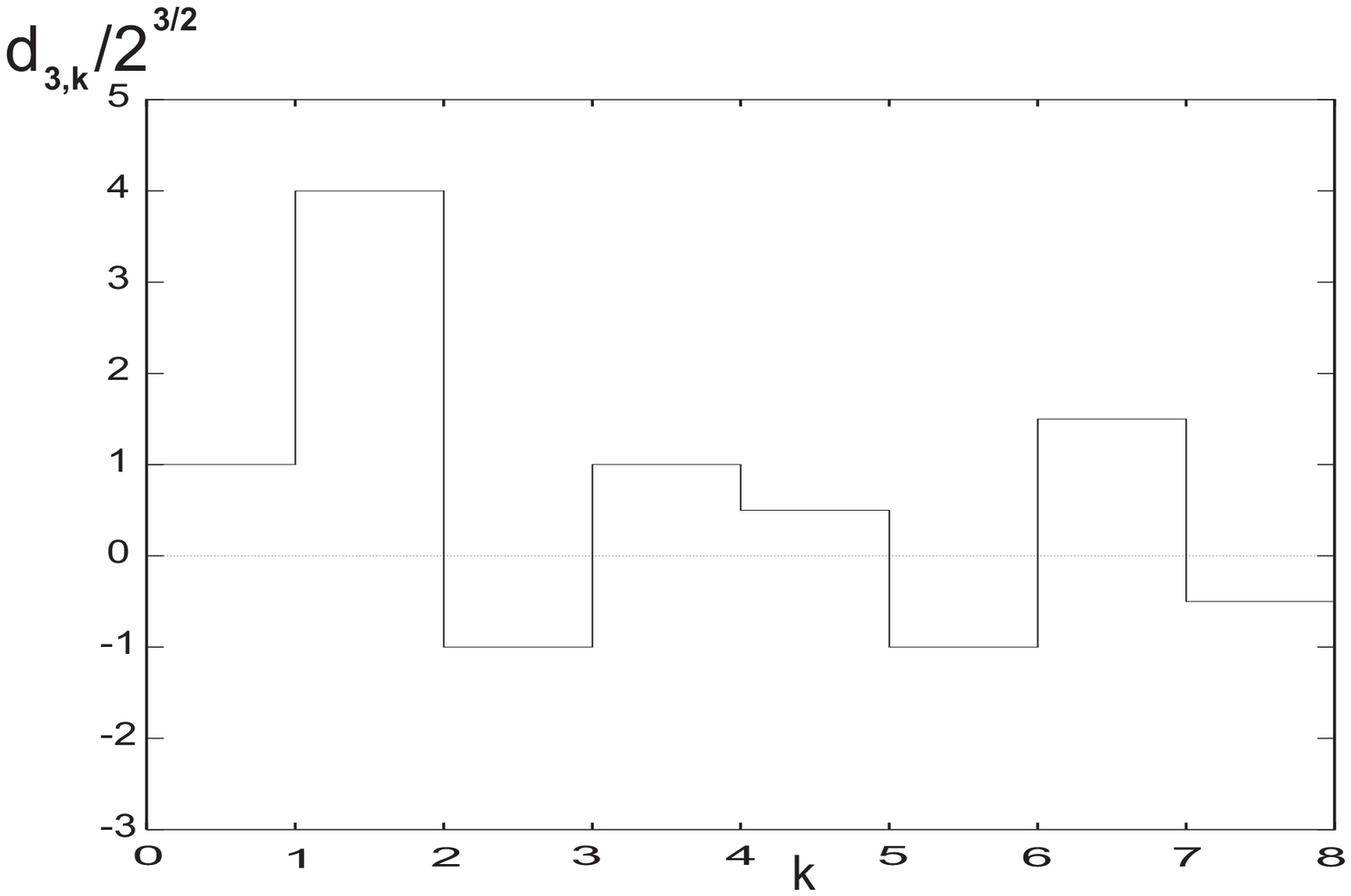,scale=0.4,clip=}&\raisebox{\jup}{3}\\
&\epsfig{file=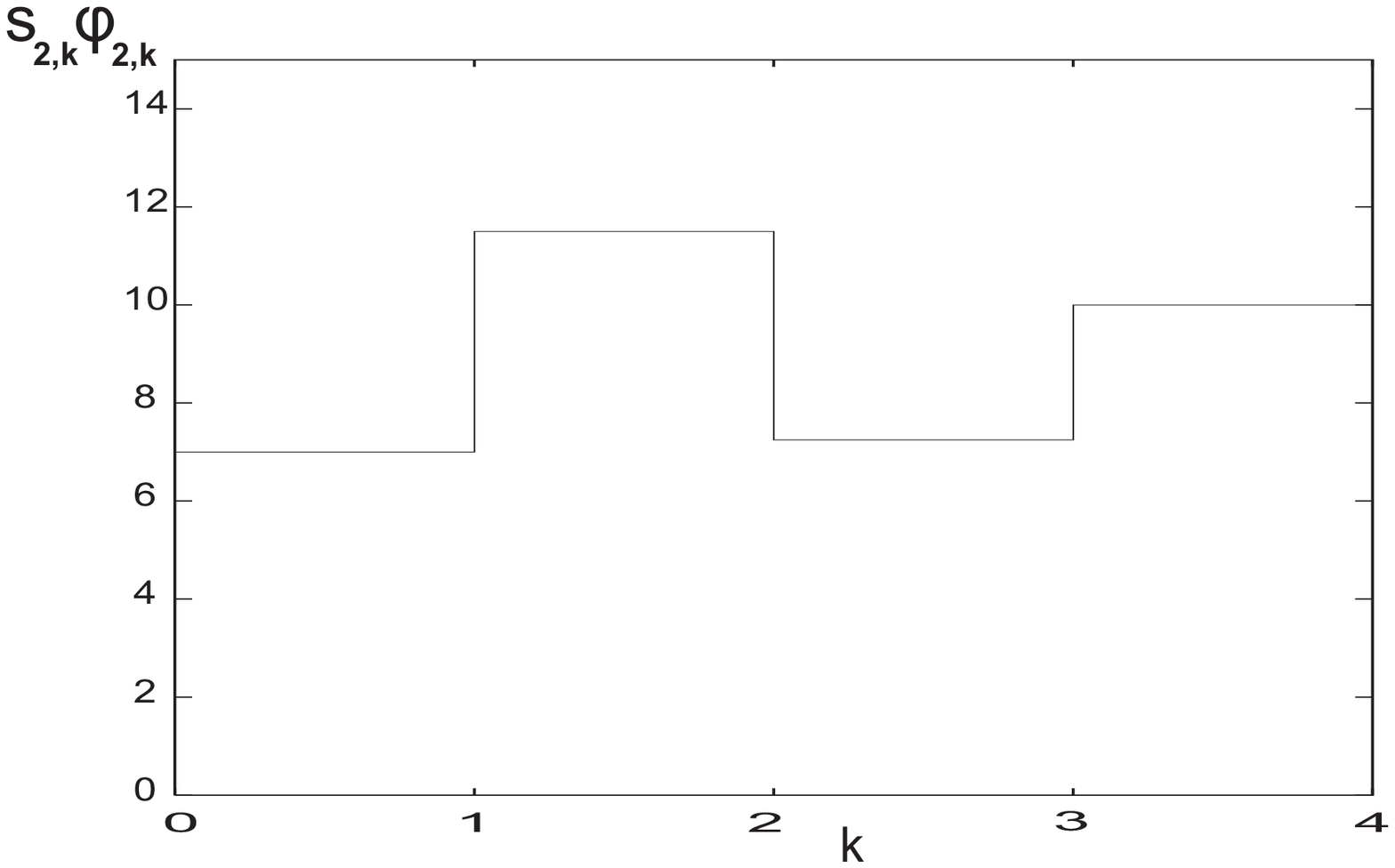,scale=0.4,clip=}&&
\epsfig{file=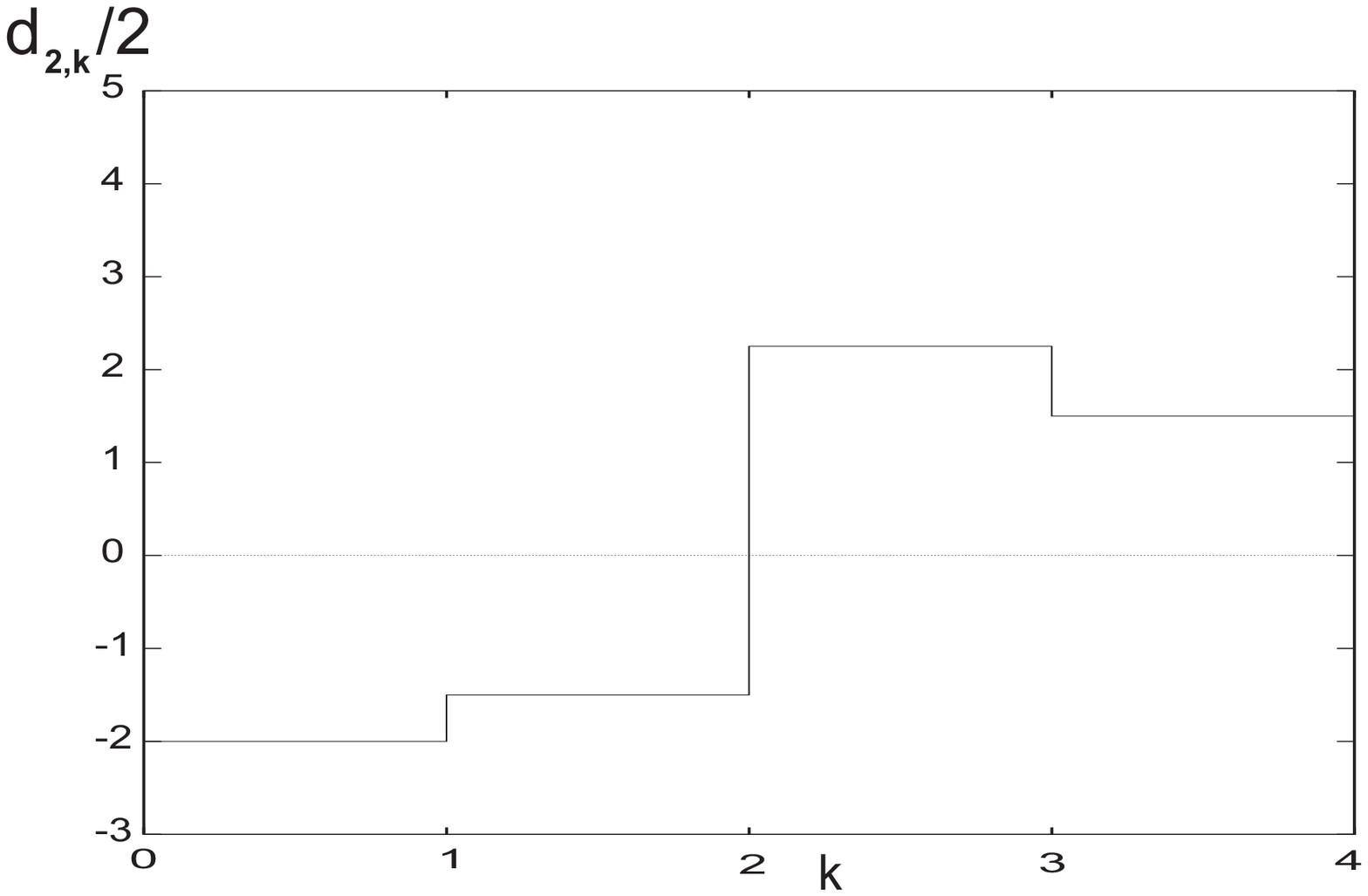,scale=0.4,clip=}&\raisebox{\jup}{2} \\
&\epsfig{file=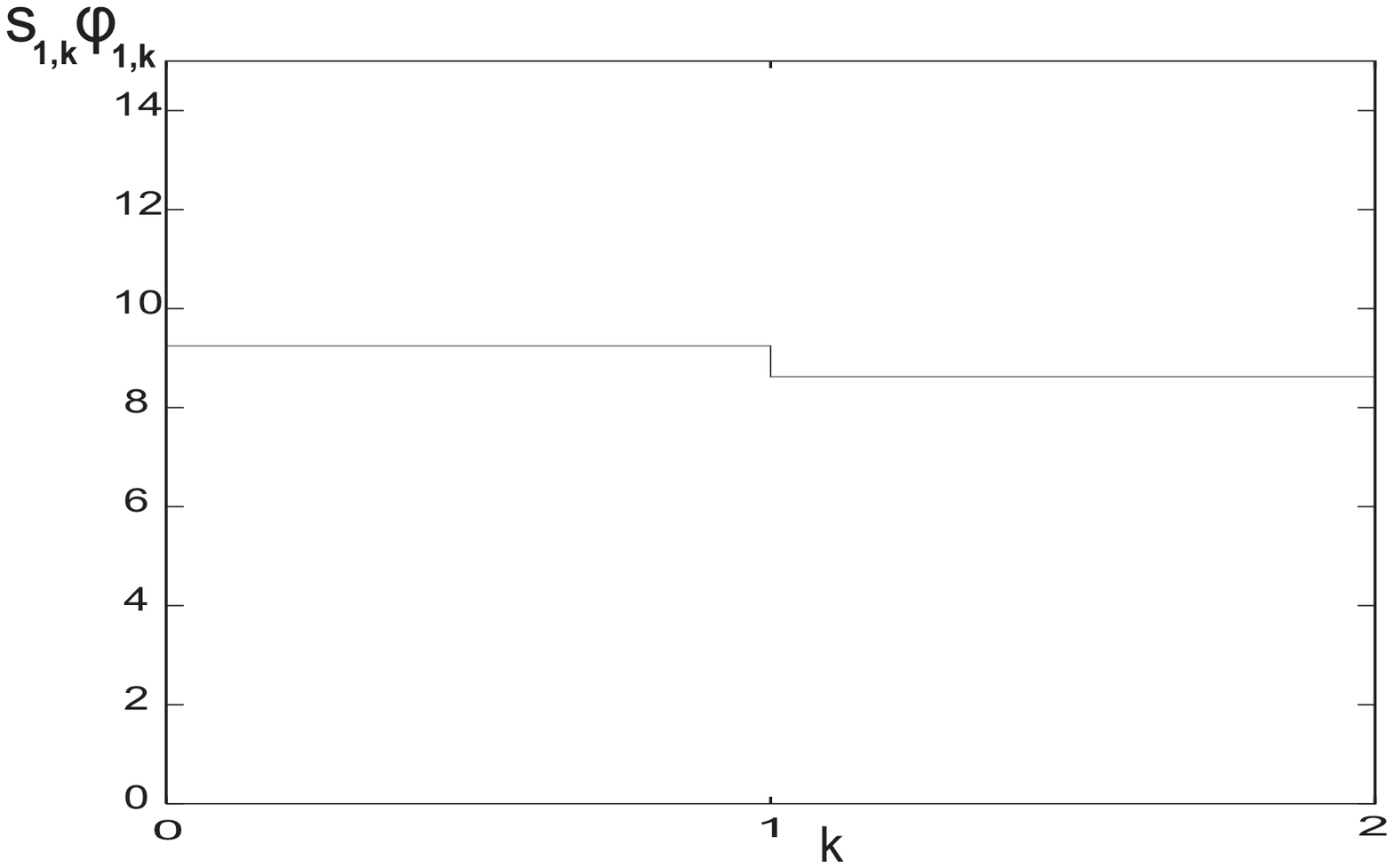,scale=0.4,clip=}&&
\epsfig{file=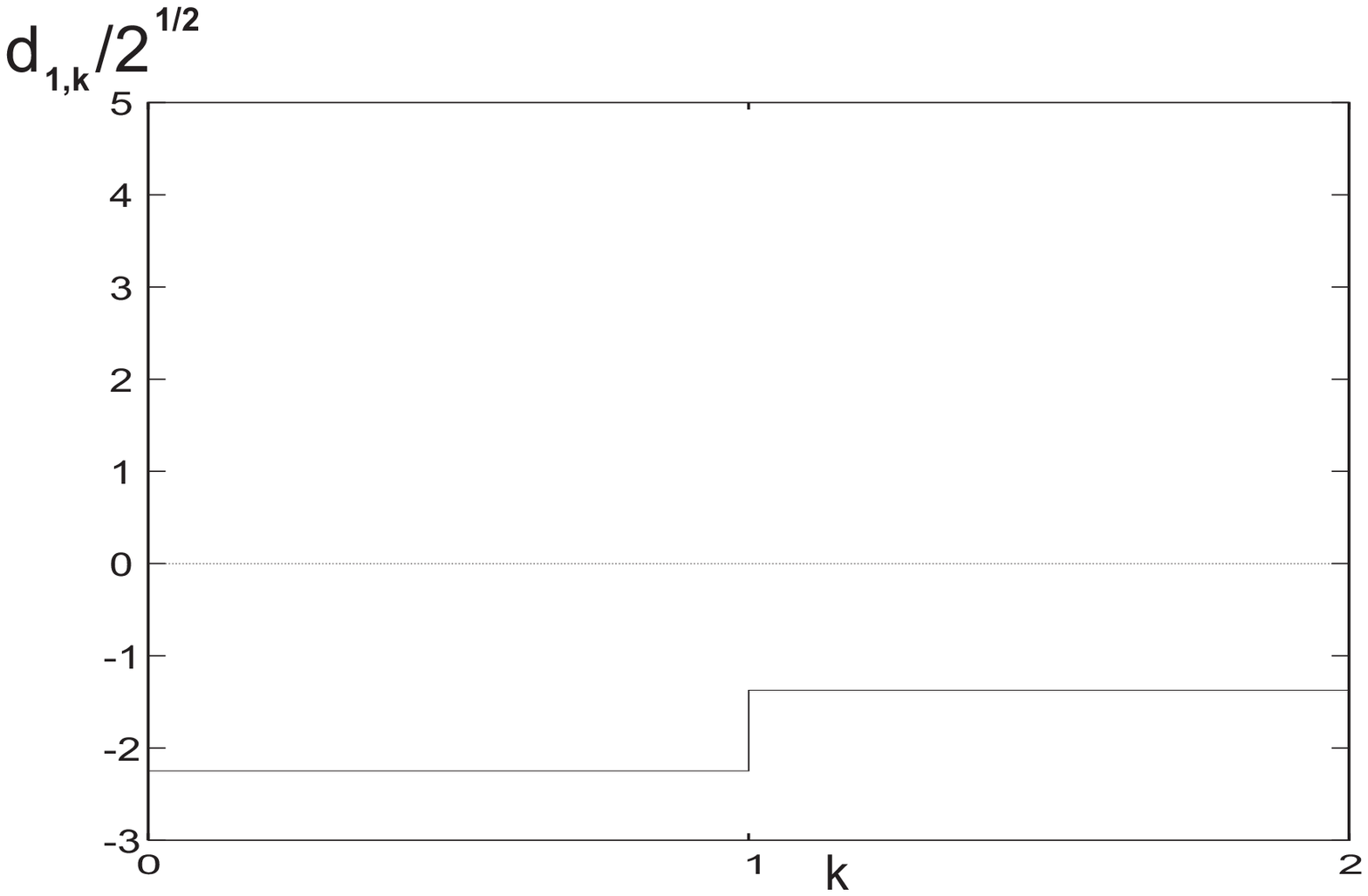,scale=0.4,clip=}&\raisebox{\jup}{1} \\
&\epsfig{file=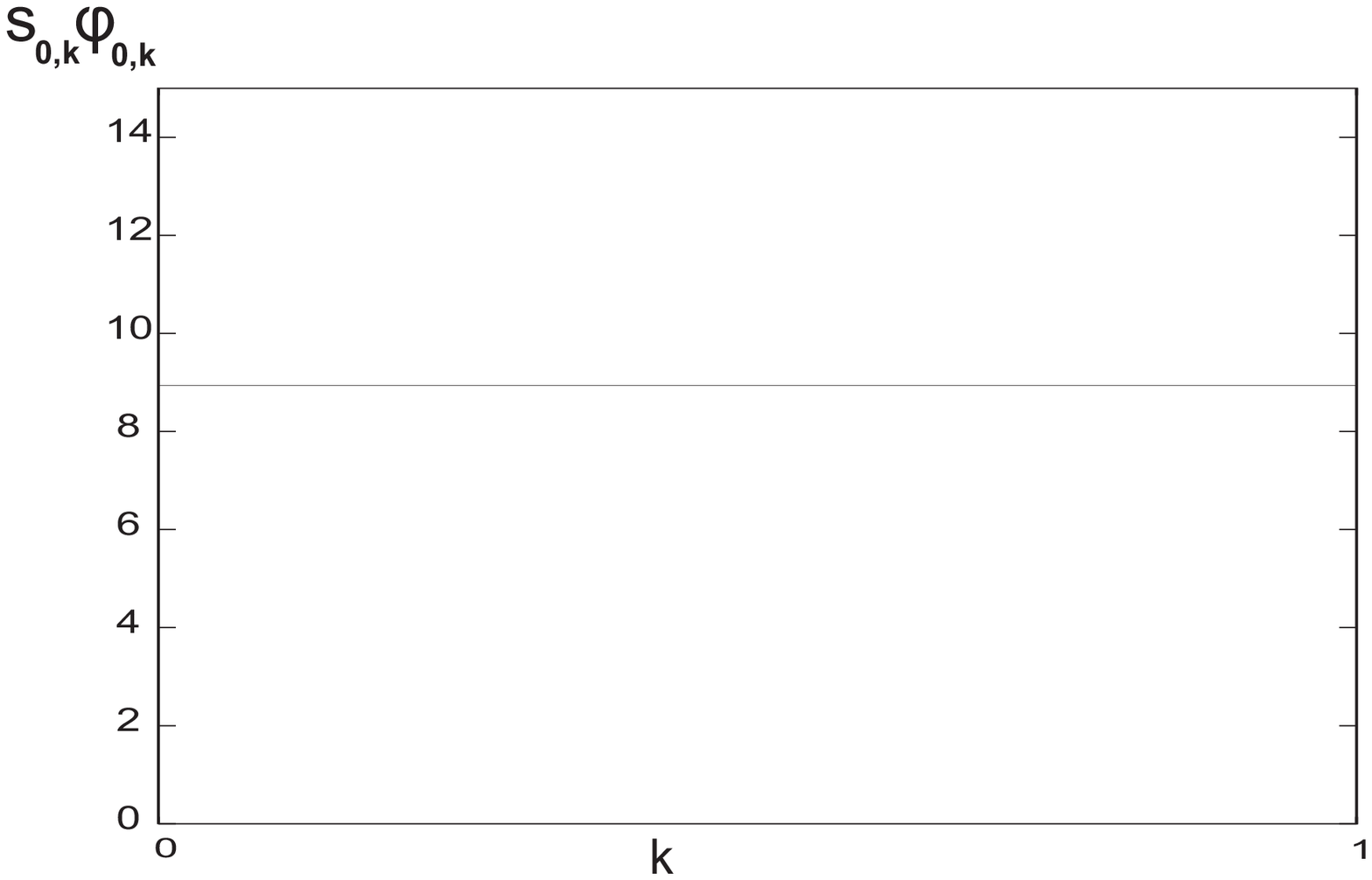,scale=0.4,clip=}&&
\epsfig{file=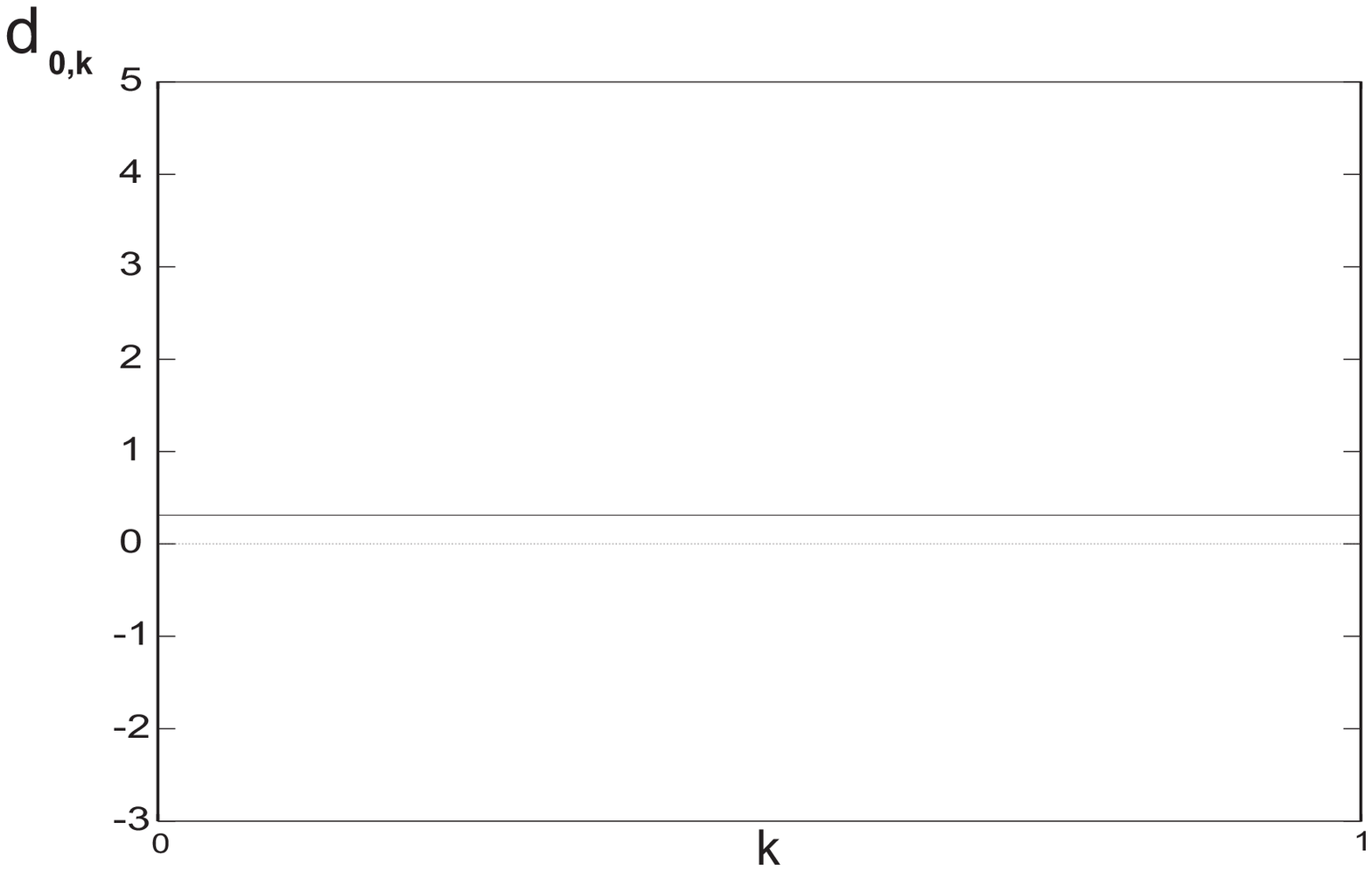,scale=0.4,clip=}&\raisebox{\jup}{0} \
\end{tabular}

Fig. 1
\newpage

\vspace*{-10mm}

 \epsfig{file=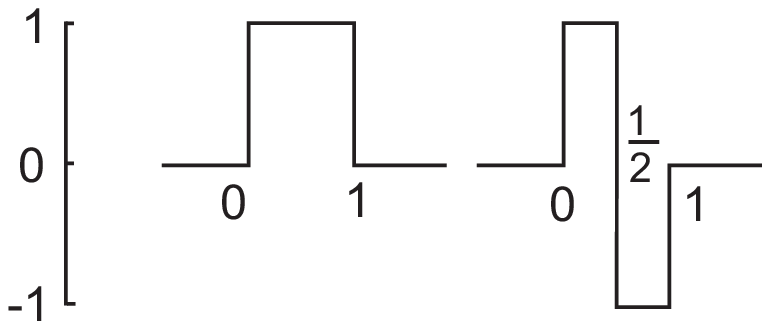,scale=1,clip=}
 Fig.2

\begin{tabular}{c}
\epsfig{file=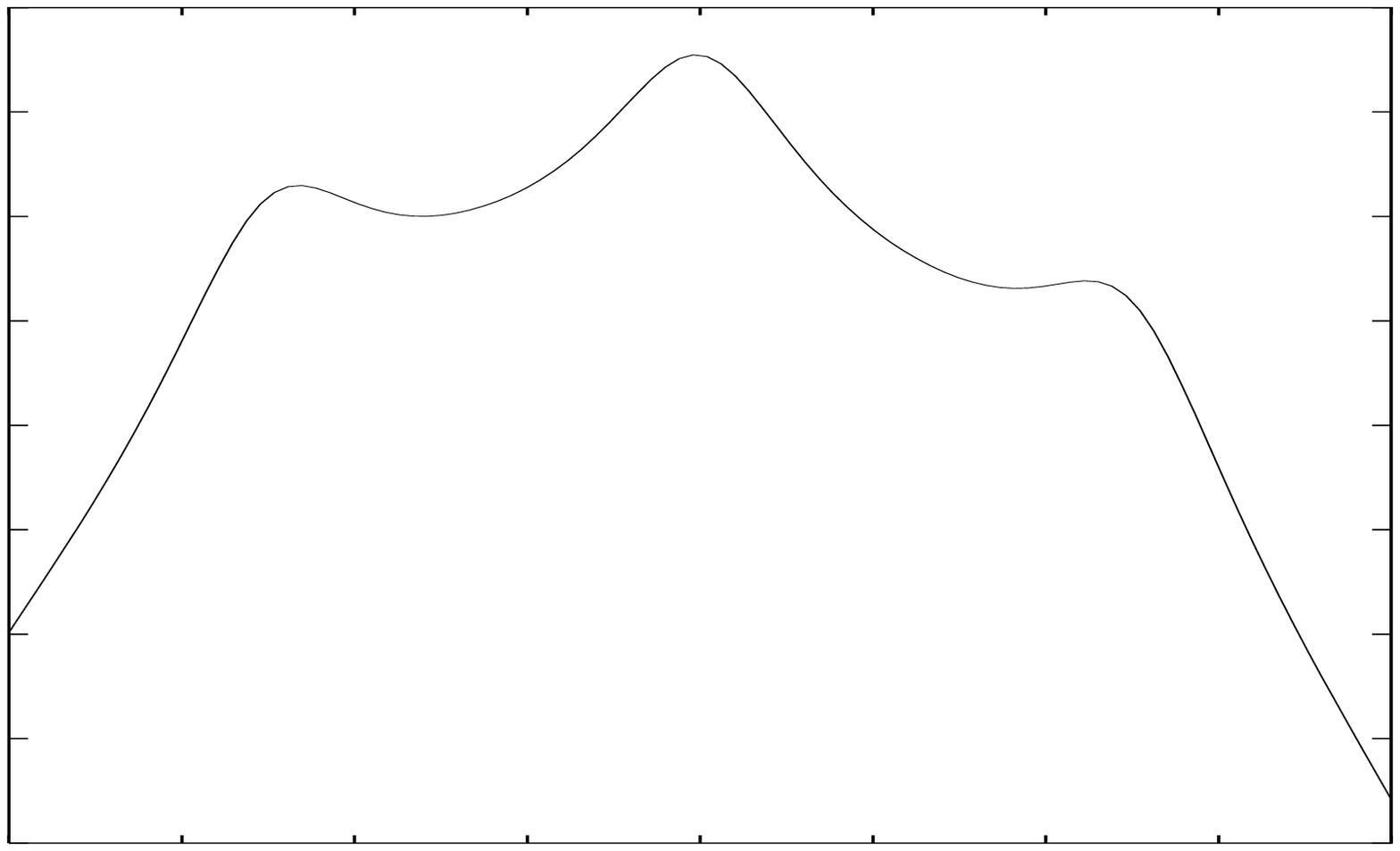,scale=0.4,clip=} \\
\epsfig{file=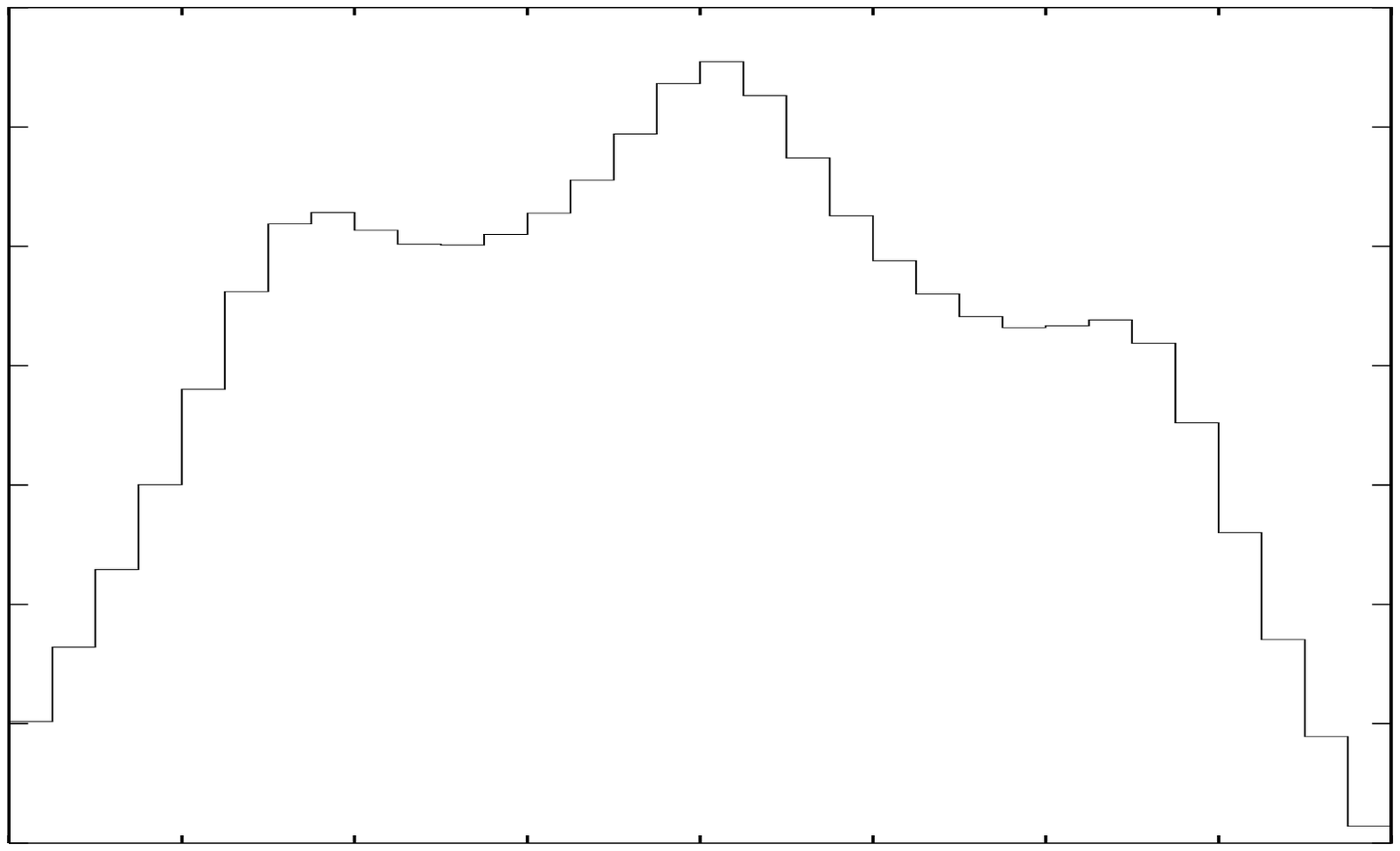,scale=0.4,clip=} \\
\epsfig{file=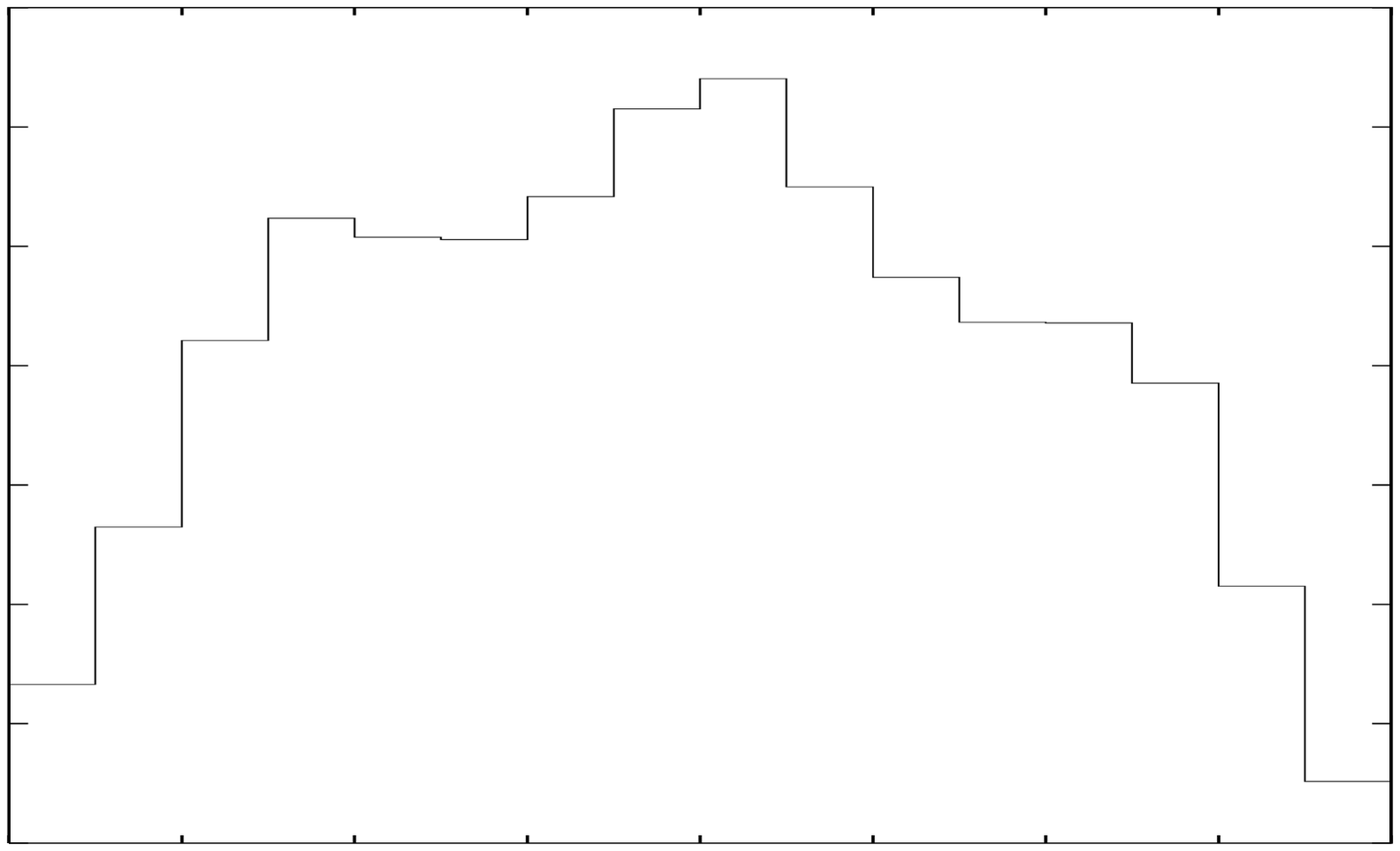,scale=0.4,clip=} \
\end{tabular}

Fig. 3

\epsfig{file=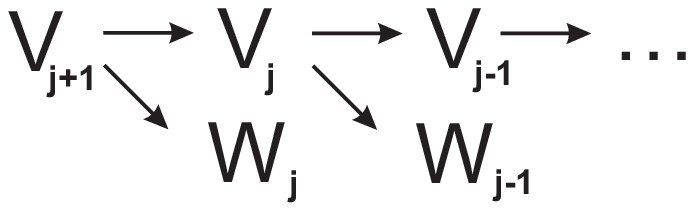,scale=1,clip=}

Fig.4

\noindent
\begin{tabular}{llll}
\epsfig{file=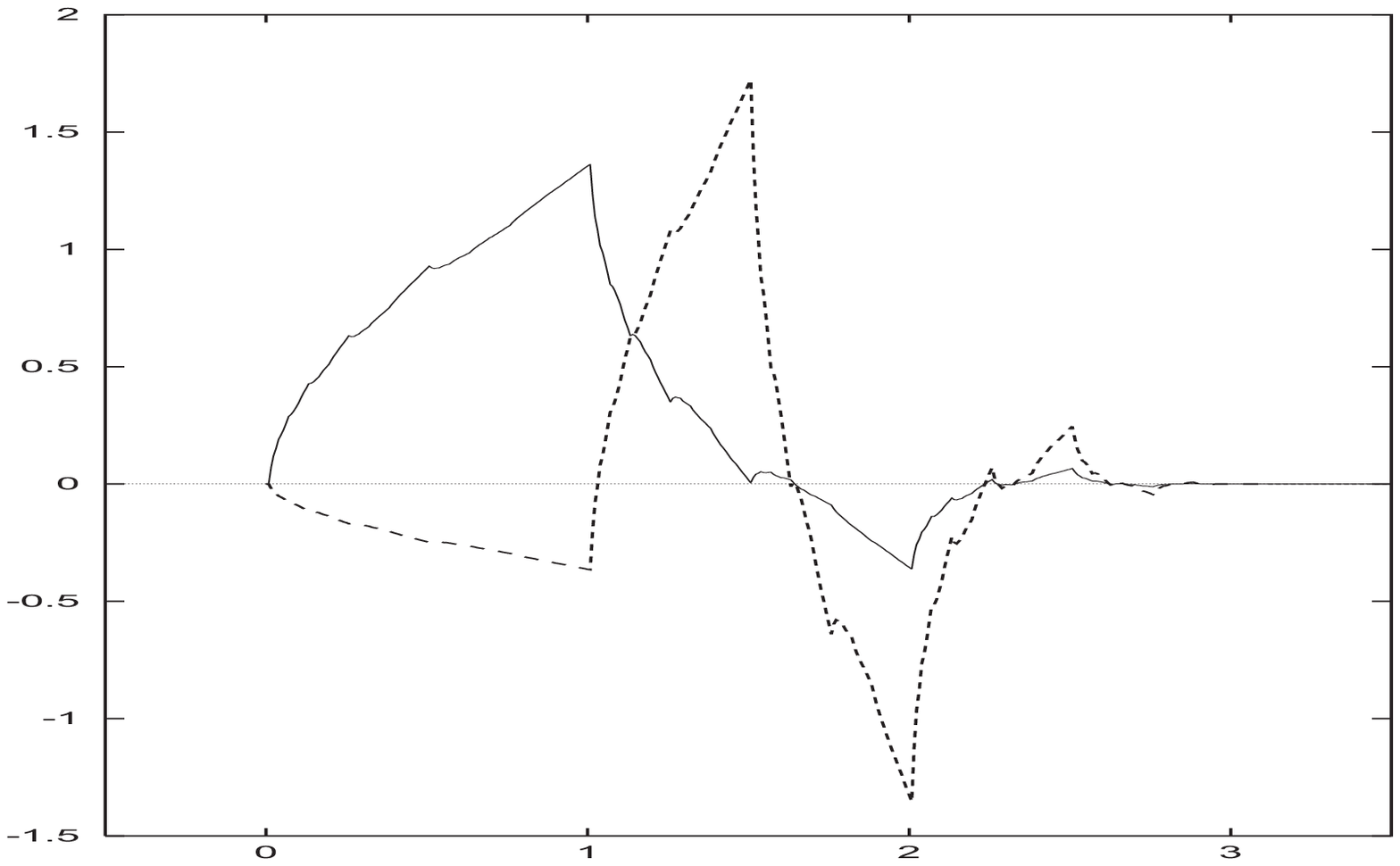 ,scale=0.4,clip=}&&\raisebox{\jup}{M=2}\\
\epsfig{file=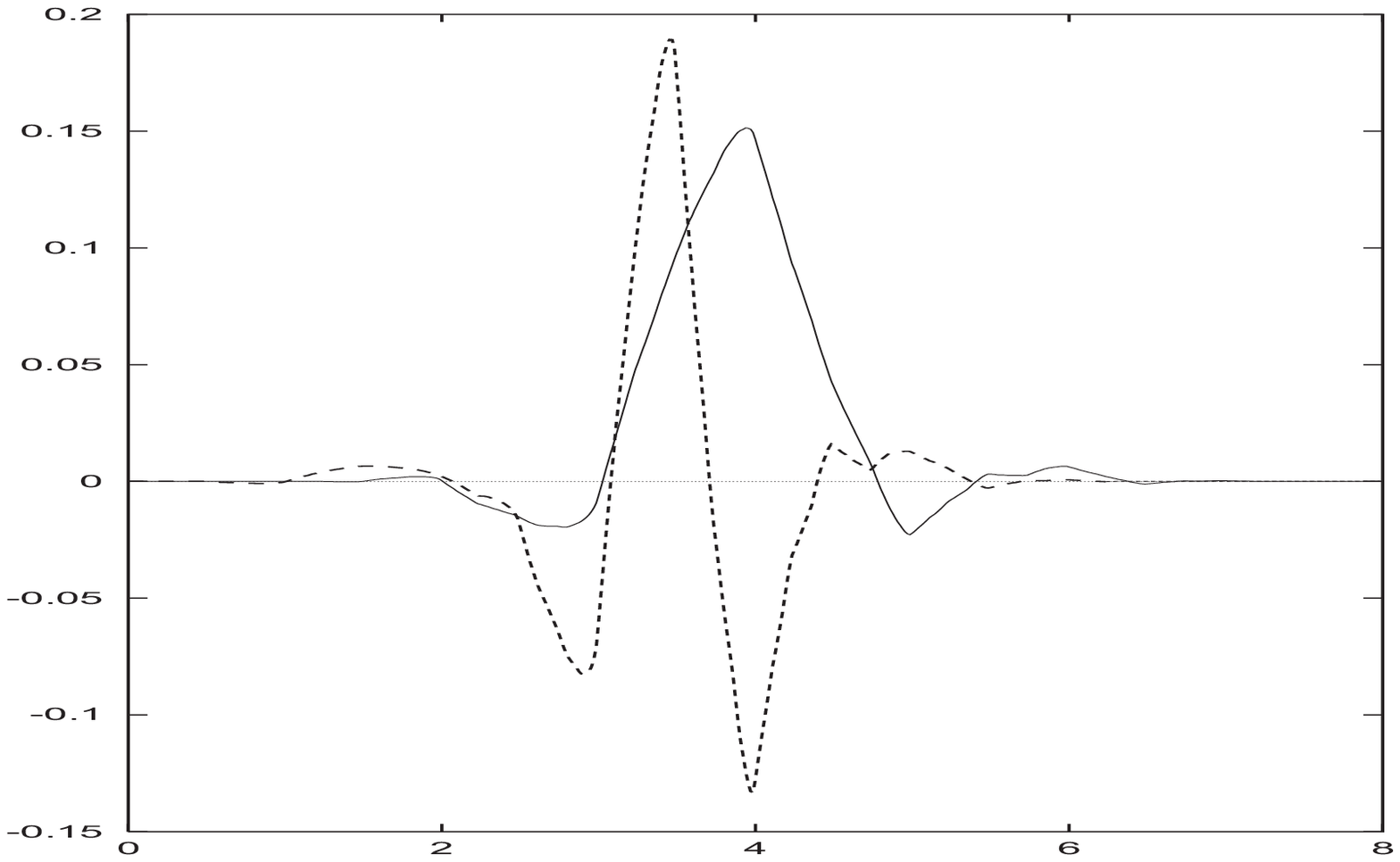 ,scale=0.4,clip=}&
\epsfig{file=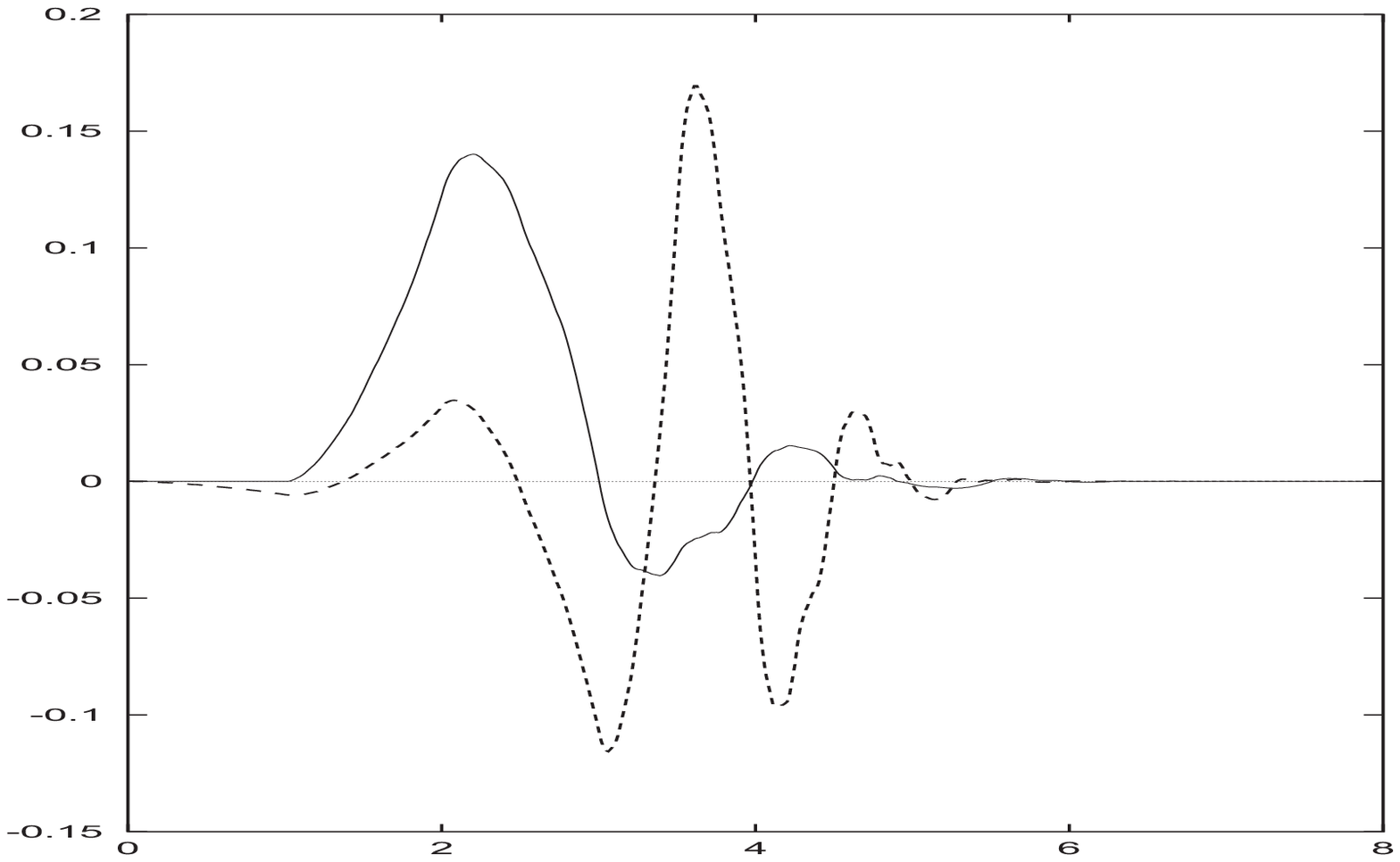 ,scale=0.4,clip=}
&\raisebox{\jup}{M=4}\
\end{tabular}

Fig.5

\epsfig{file=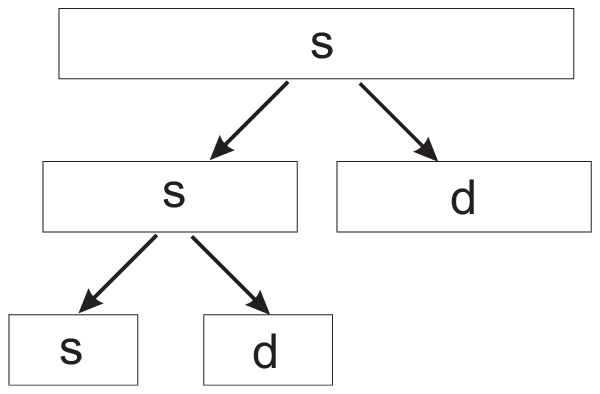,scale=1,clip=}

Fig.6

\bigskip

\noindent
\begin{tabular}{ll}
\epsfig{file=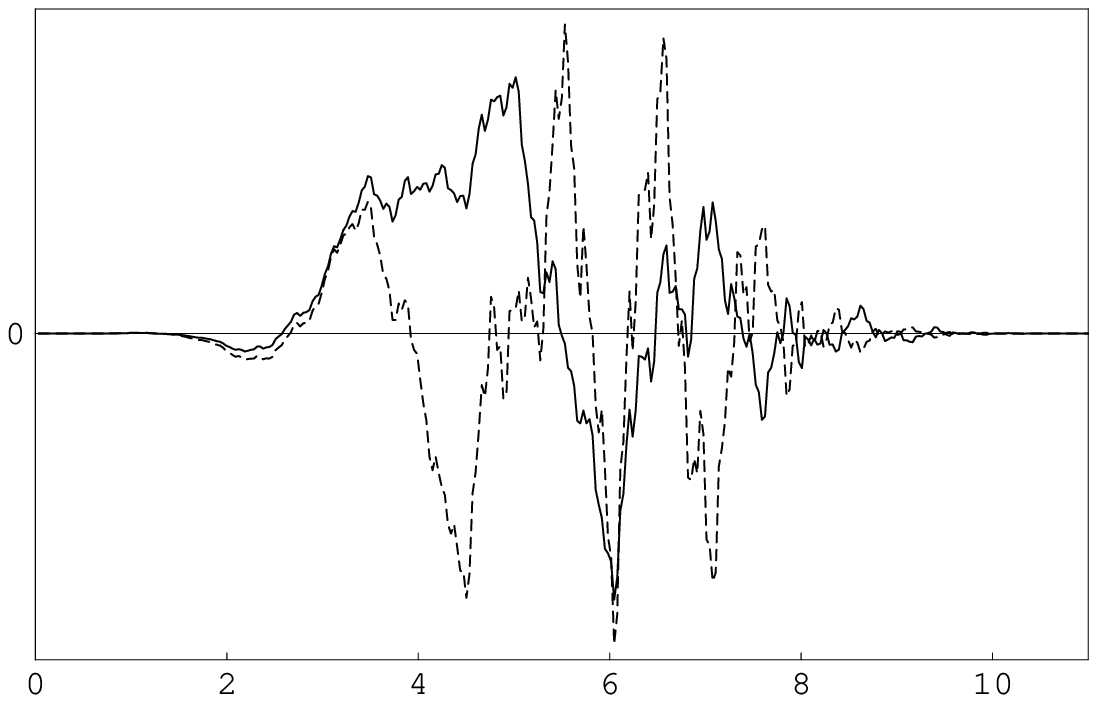 ,scale=0.7,clip=}&
\epsfig{file=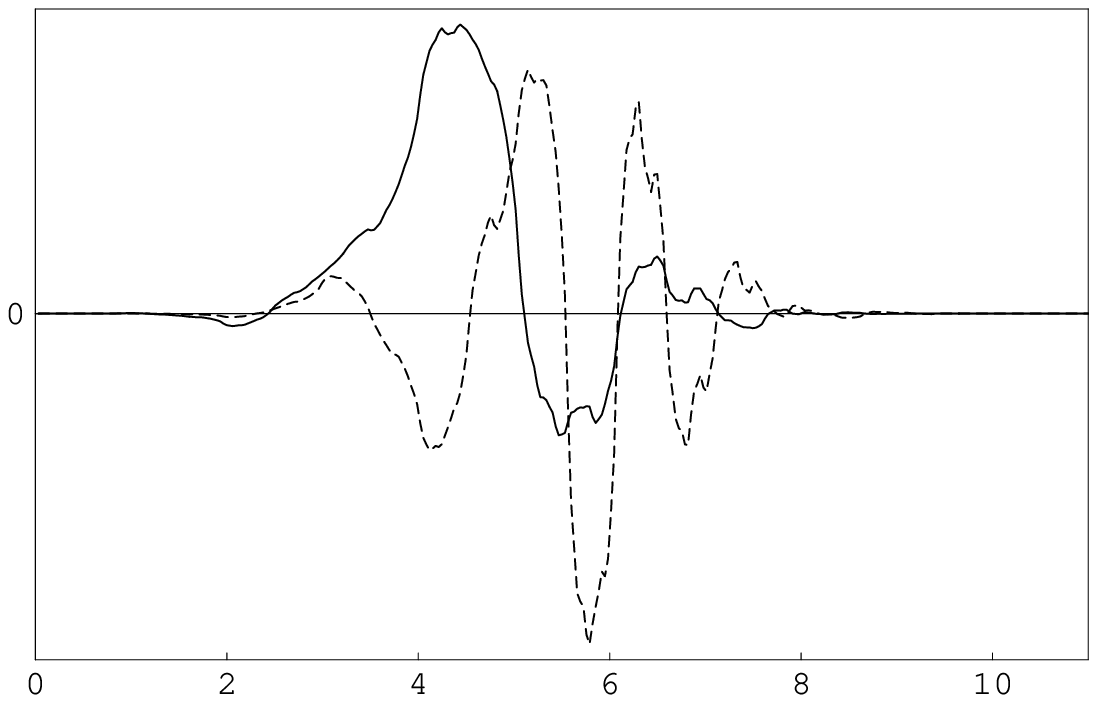 ,scale=0.7,clip=} \\
\epsfig{file=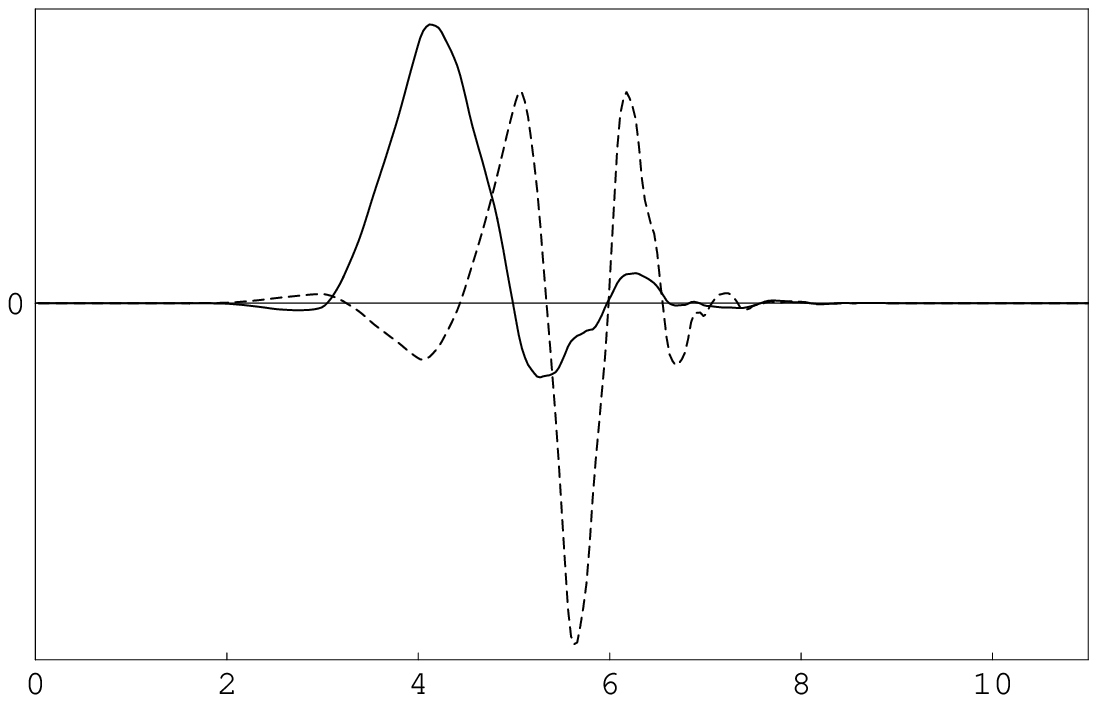 ,scale=0.7,clip=}&
\epsfig{file=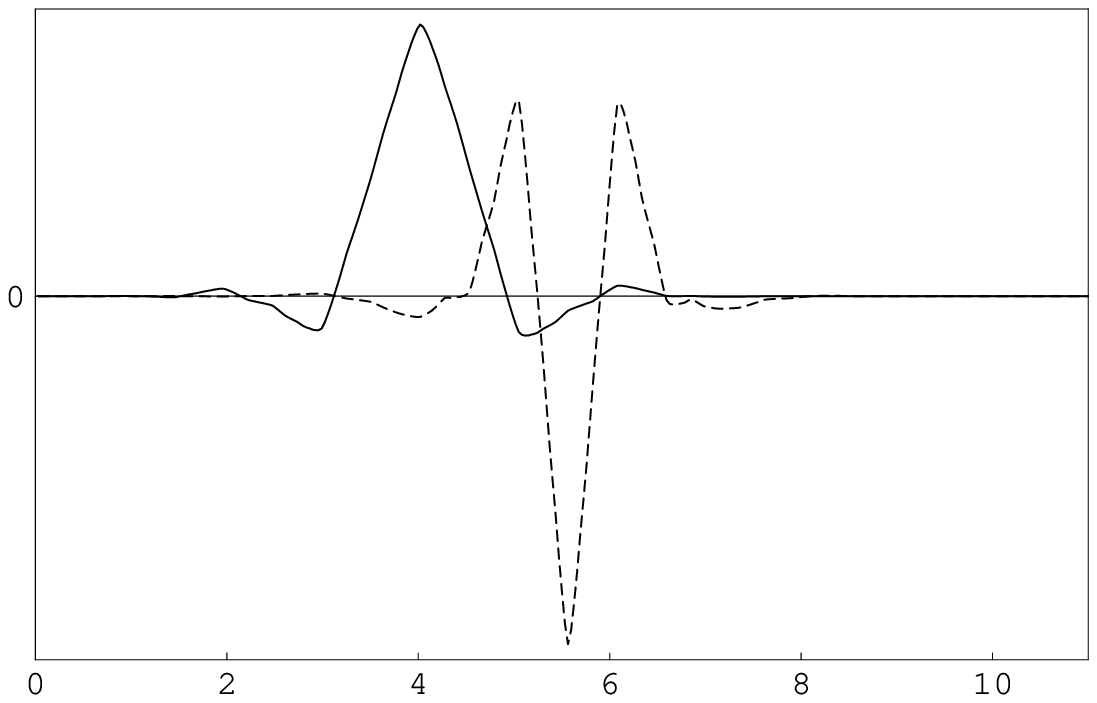 ,scale=0.7,clip=} \
\end{tabular}

Fig.7

\epsfig{file=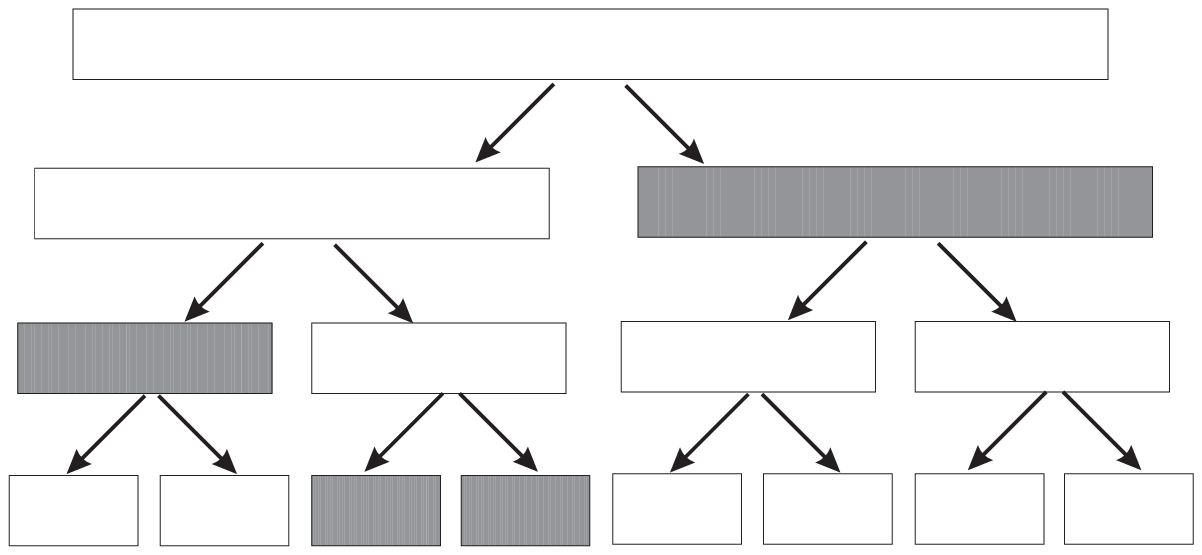,scale=1,clip=}

Fig.8

\epsfig{file=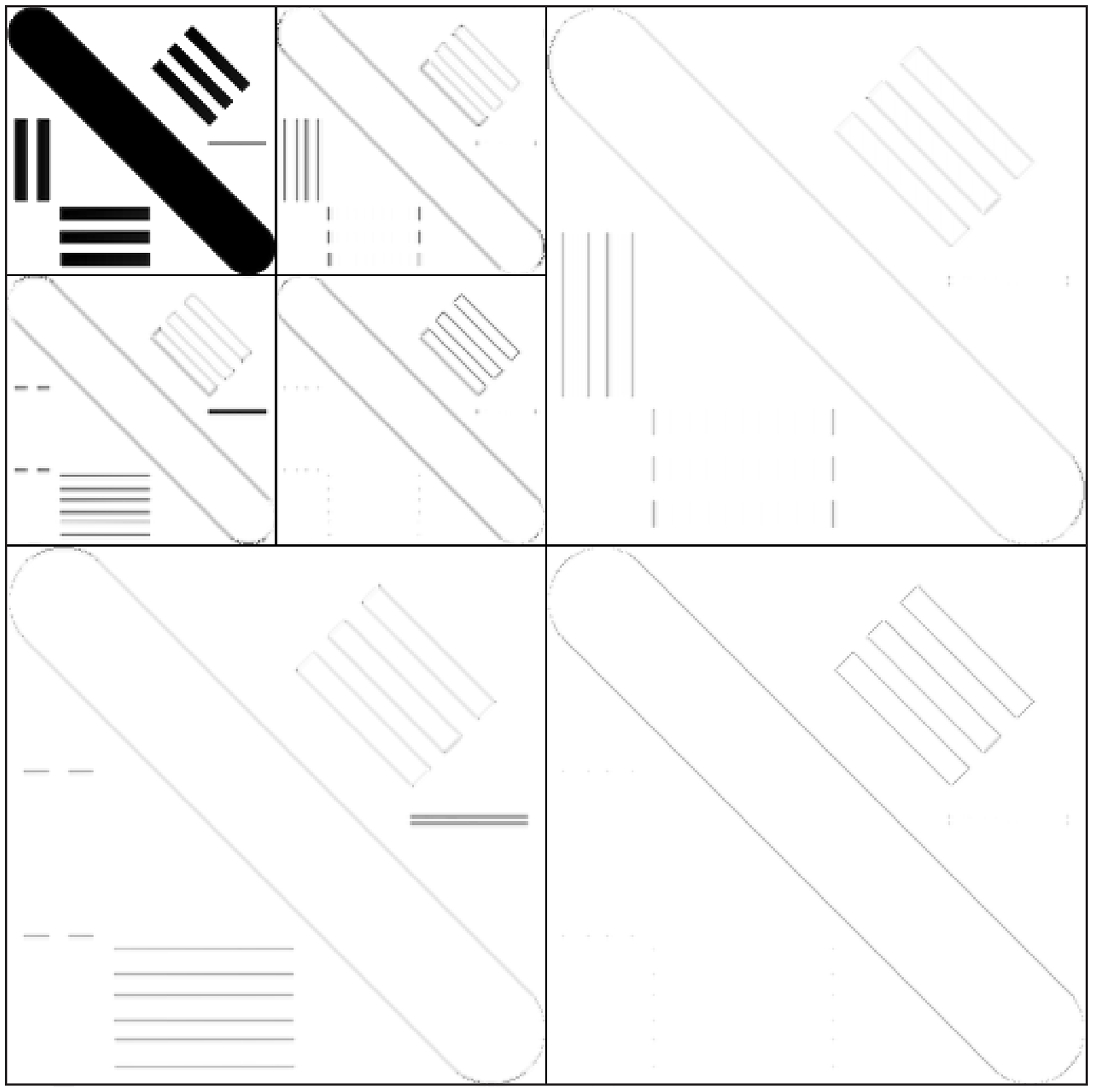,scale=0.5}

Fig.9

\noindent
\begin{tabular}{cc}
\epsfig{file=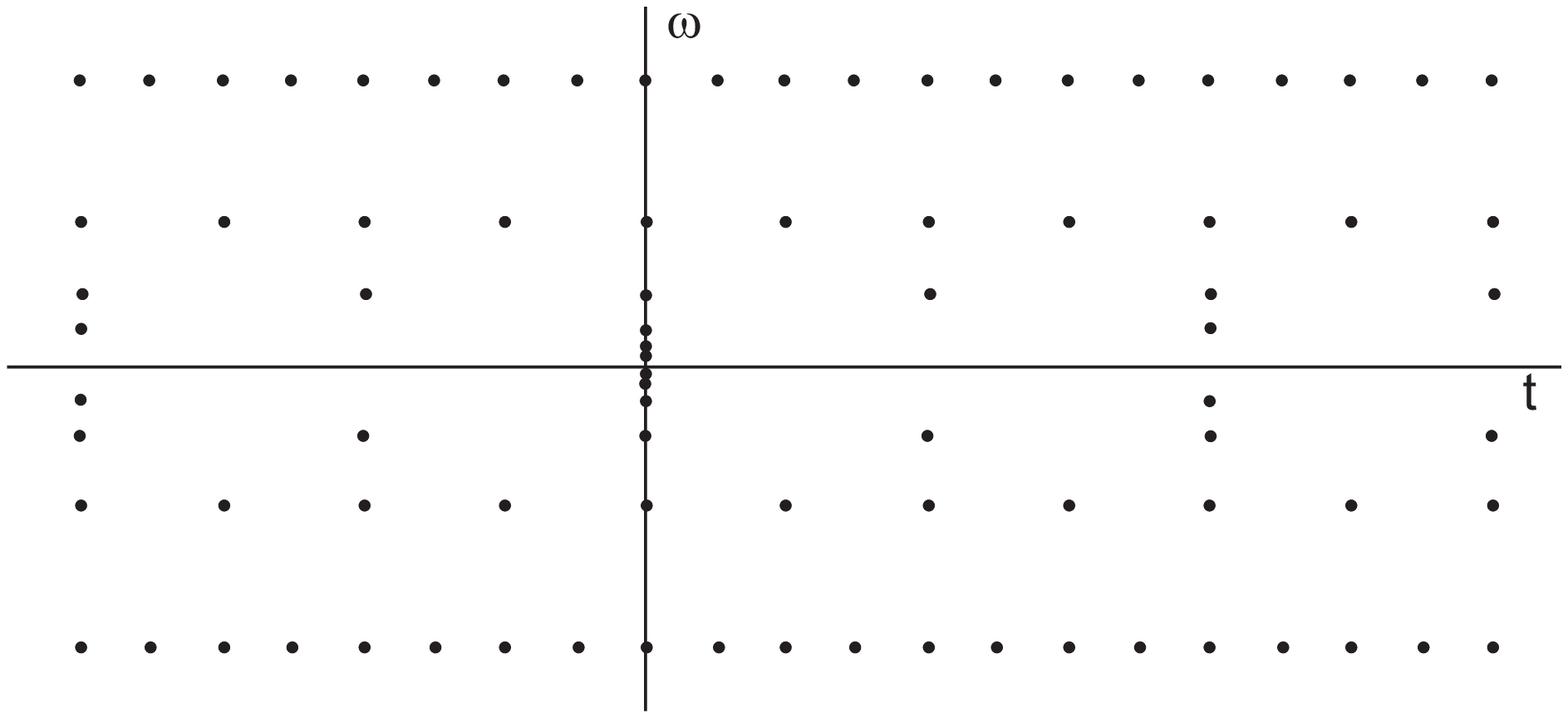,scale=0.5,clip=}&
\epsfig{file=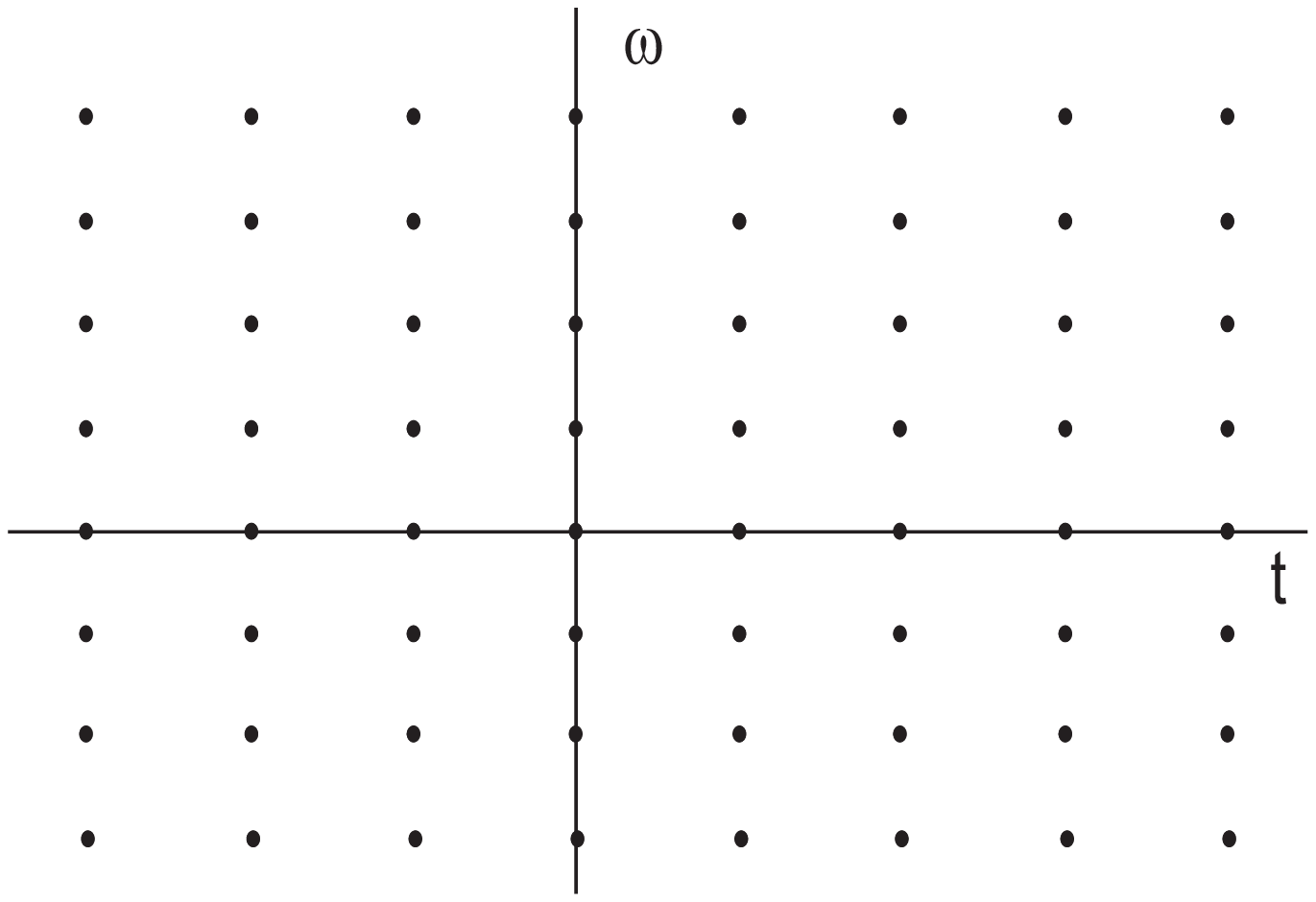,scale=0.5,clip=}\
\end{tabular}

Fig.10

\noindent\epsfig{file=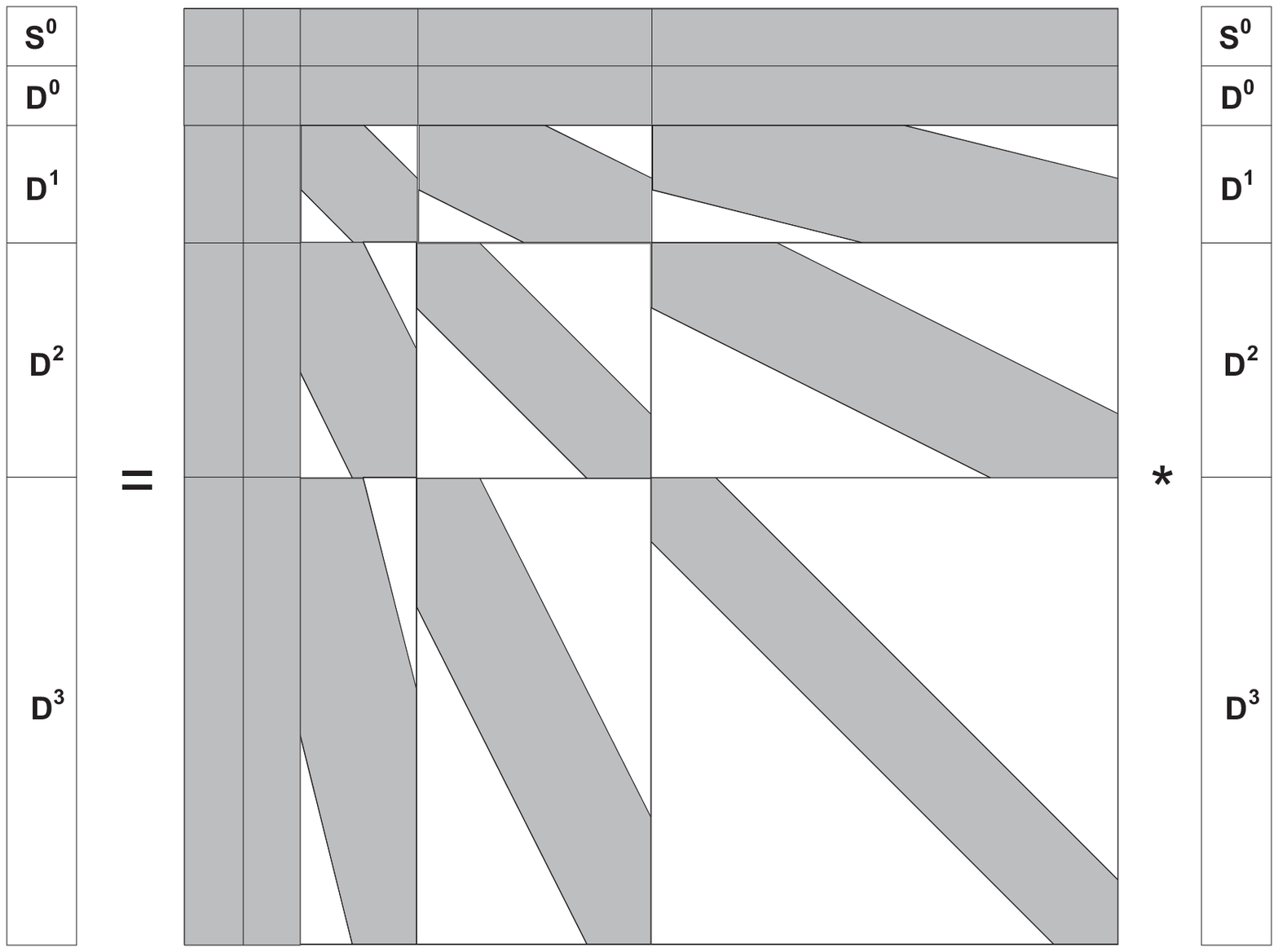,scale=0.5}

\vspace*{-1mm}
Fig.11

\bigskip

\noindent\epsfig{file=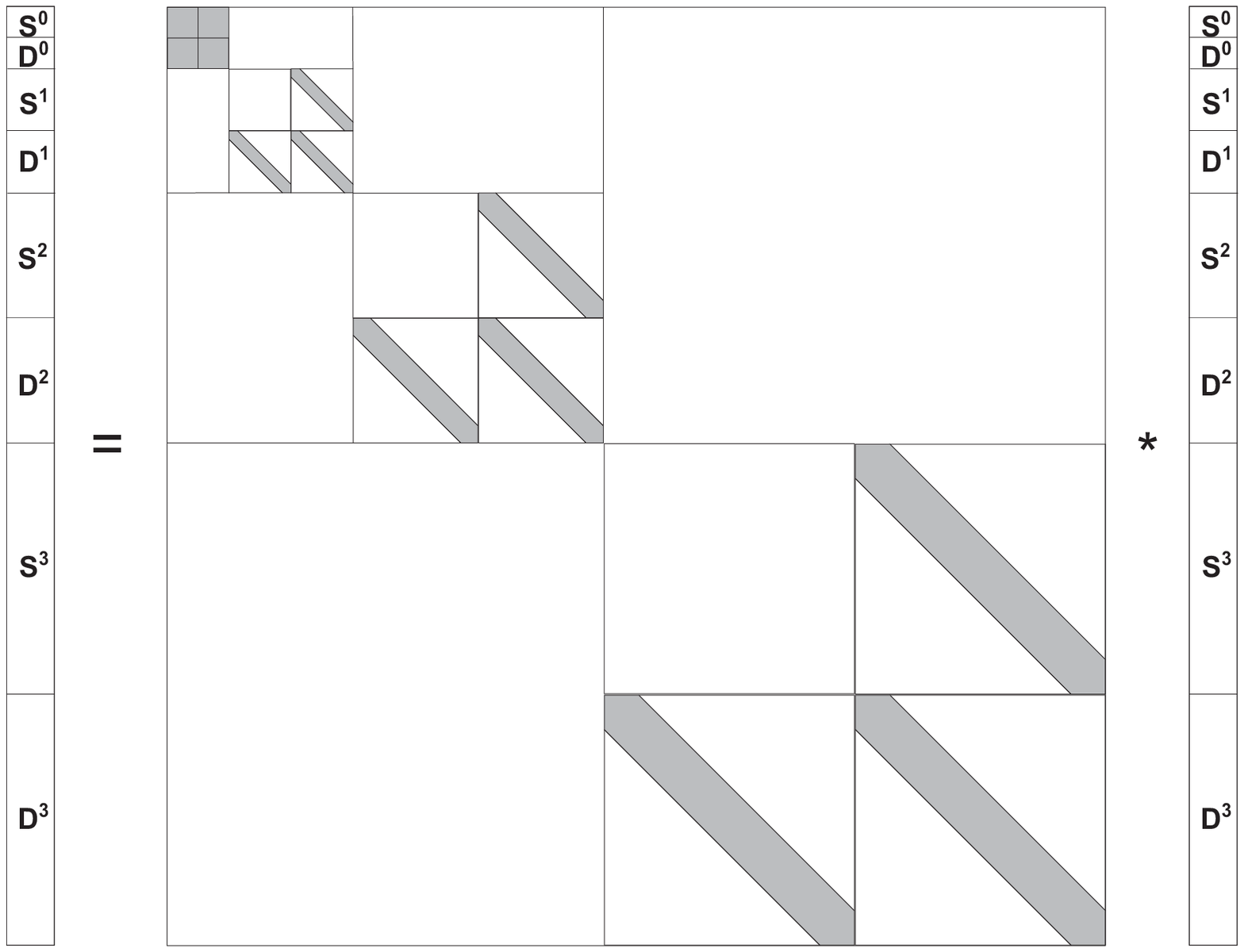,scale=0.5}

Fig.12

\epsfig{file=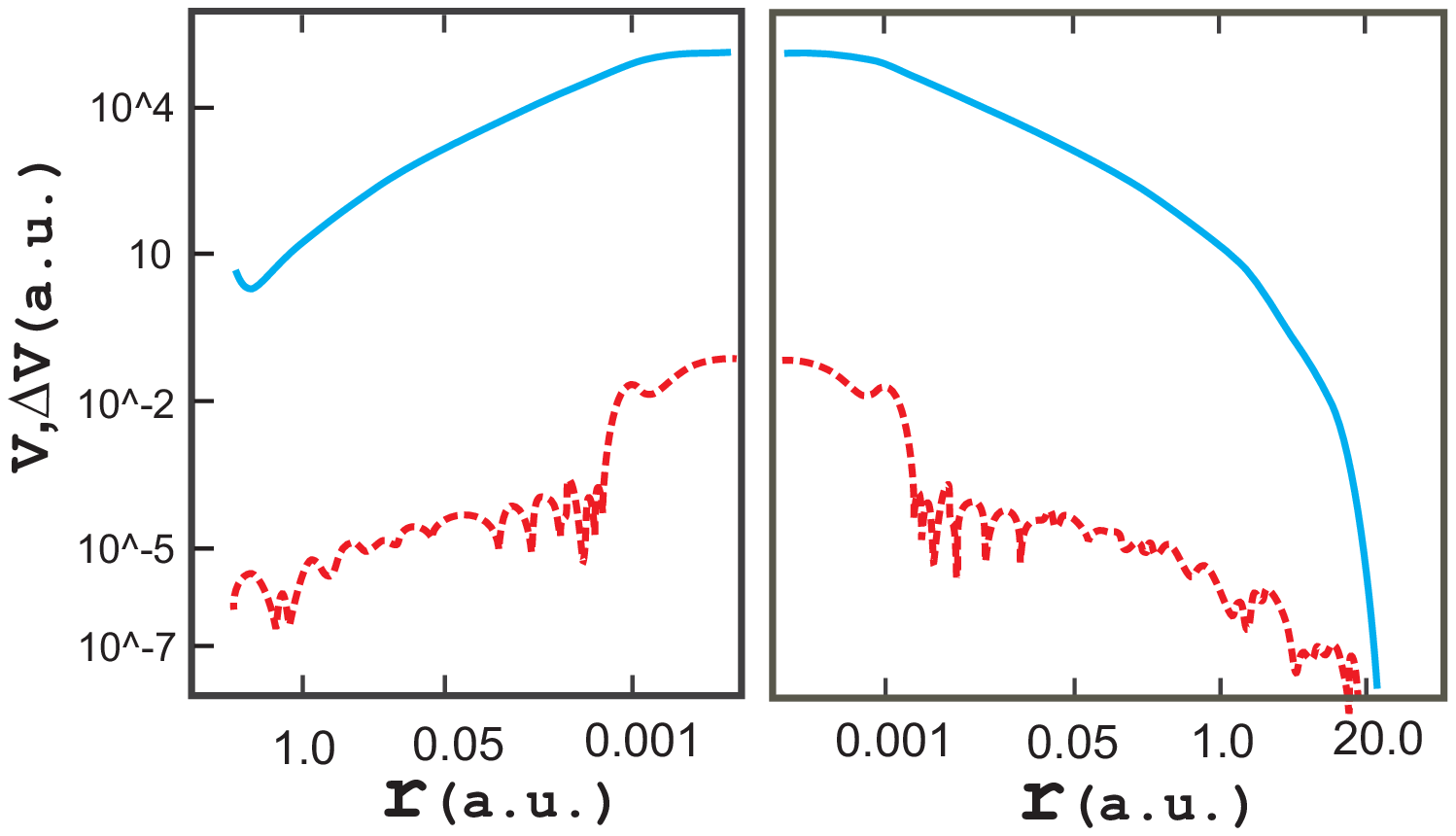,scale=0.7,clip=}

Fig.13

\epsfig{file=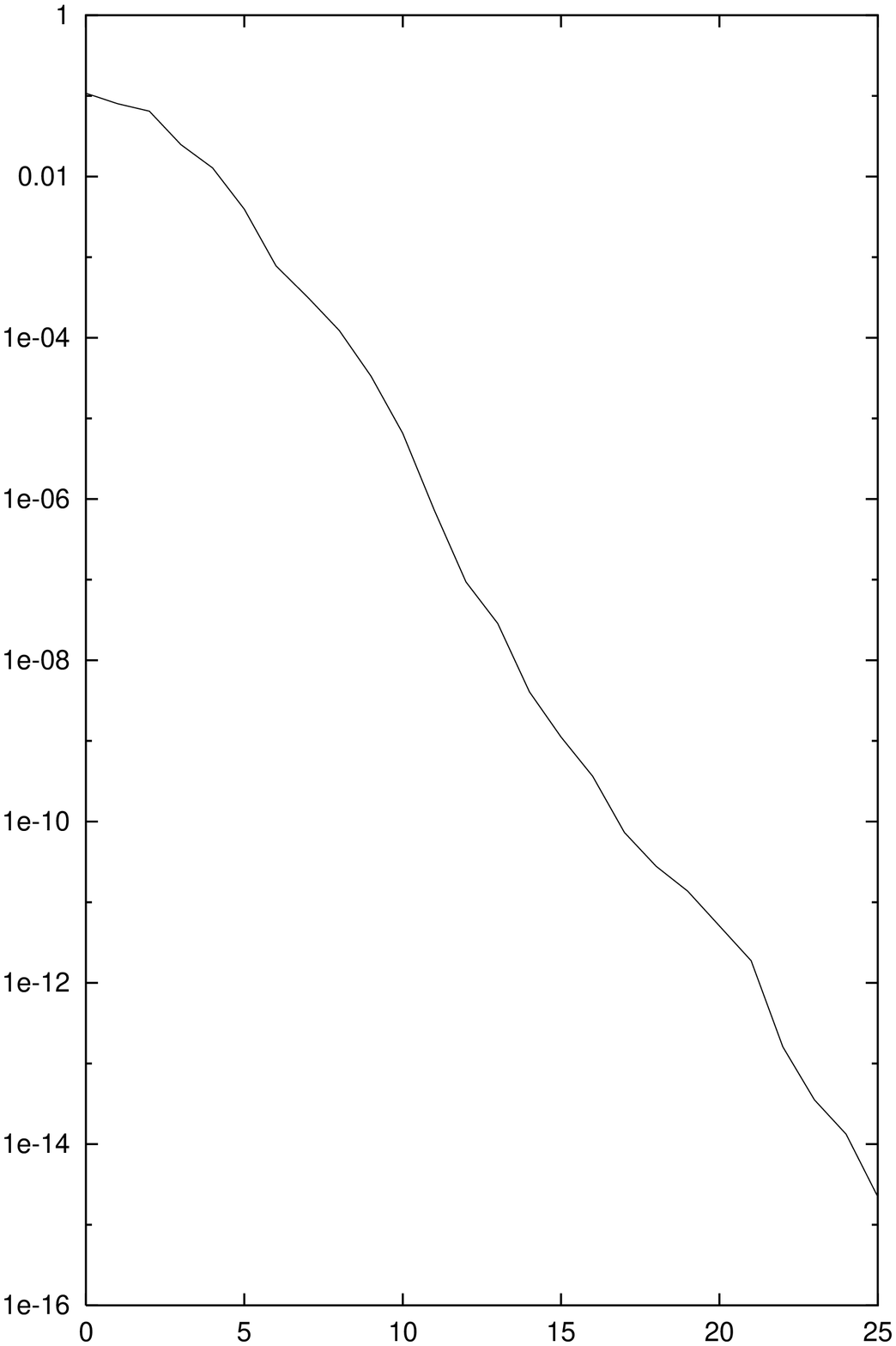,scale=0.5}

Fig.14

\epsfig{file=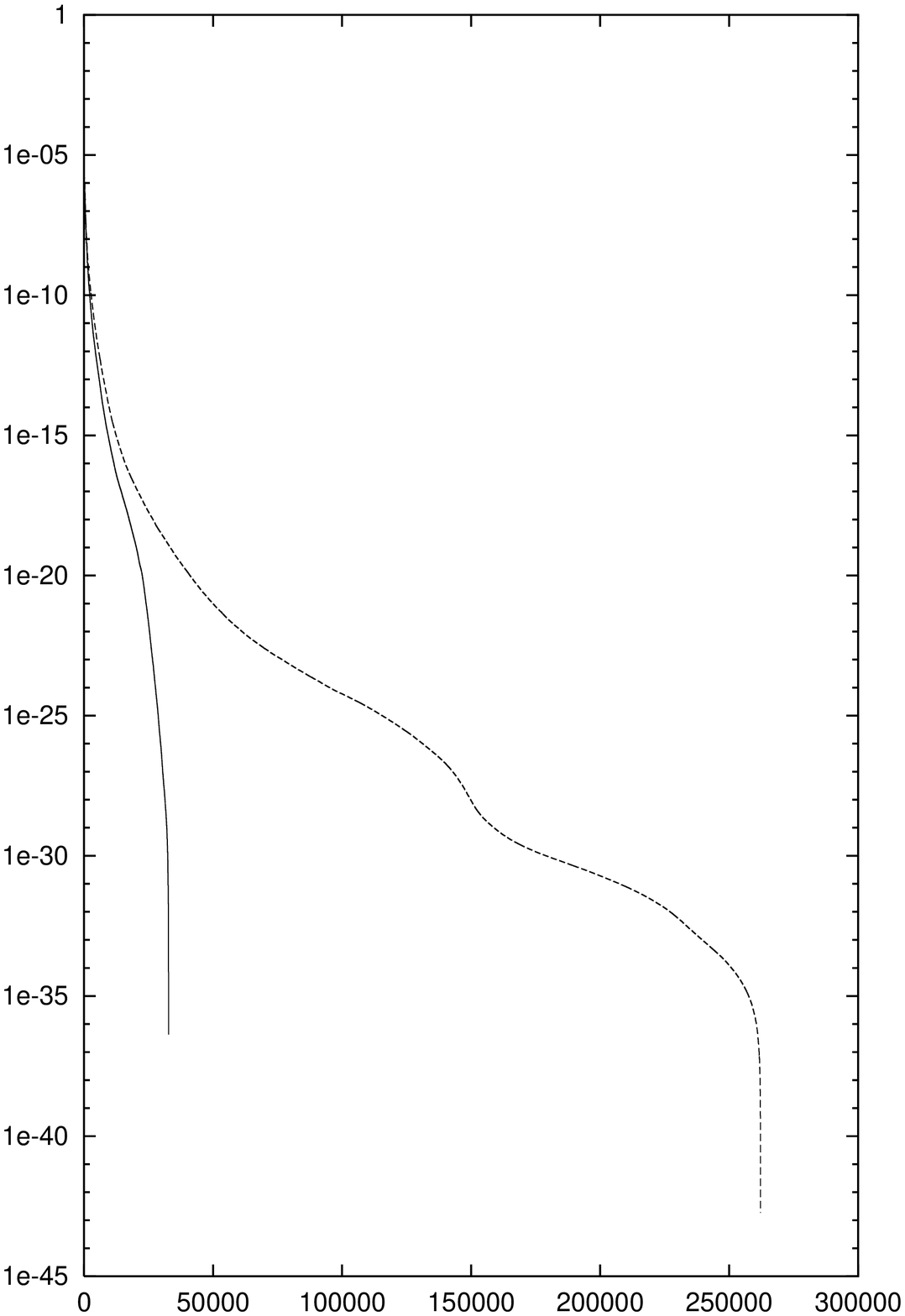,scale=0.5}

Fig.15

\begin{tabular}{cc}
\epsfig{file=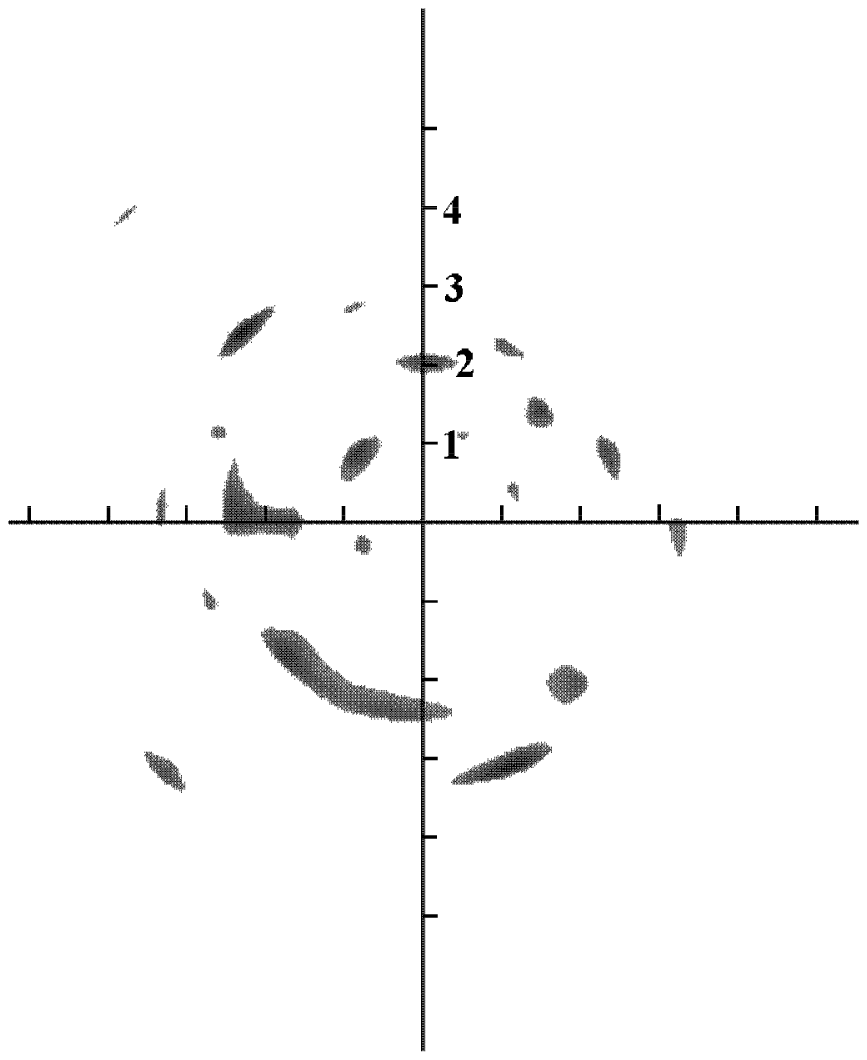,scale=0.7,clip=}&
\epsfig{file=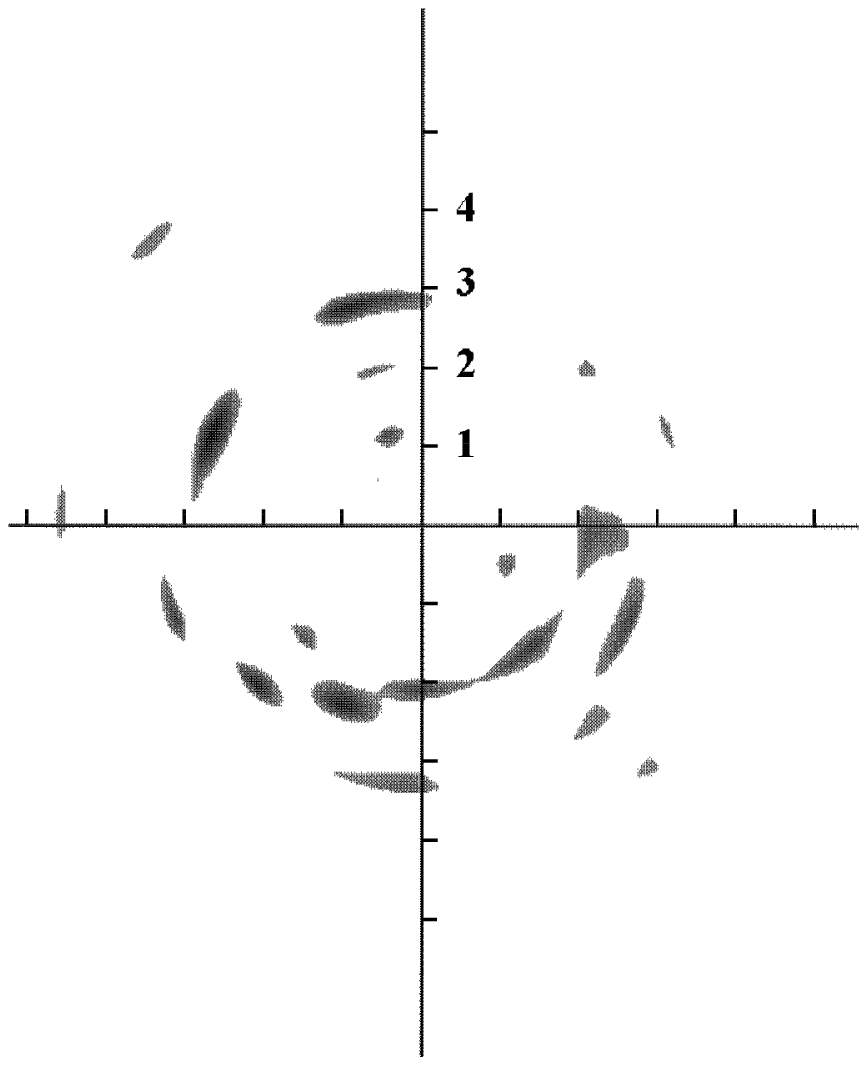,scale=0.7,clip=}\
\end{tabular}

Fig.16

\epsfig{file=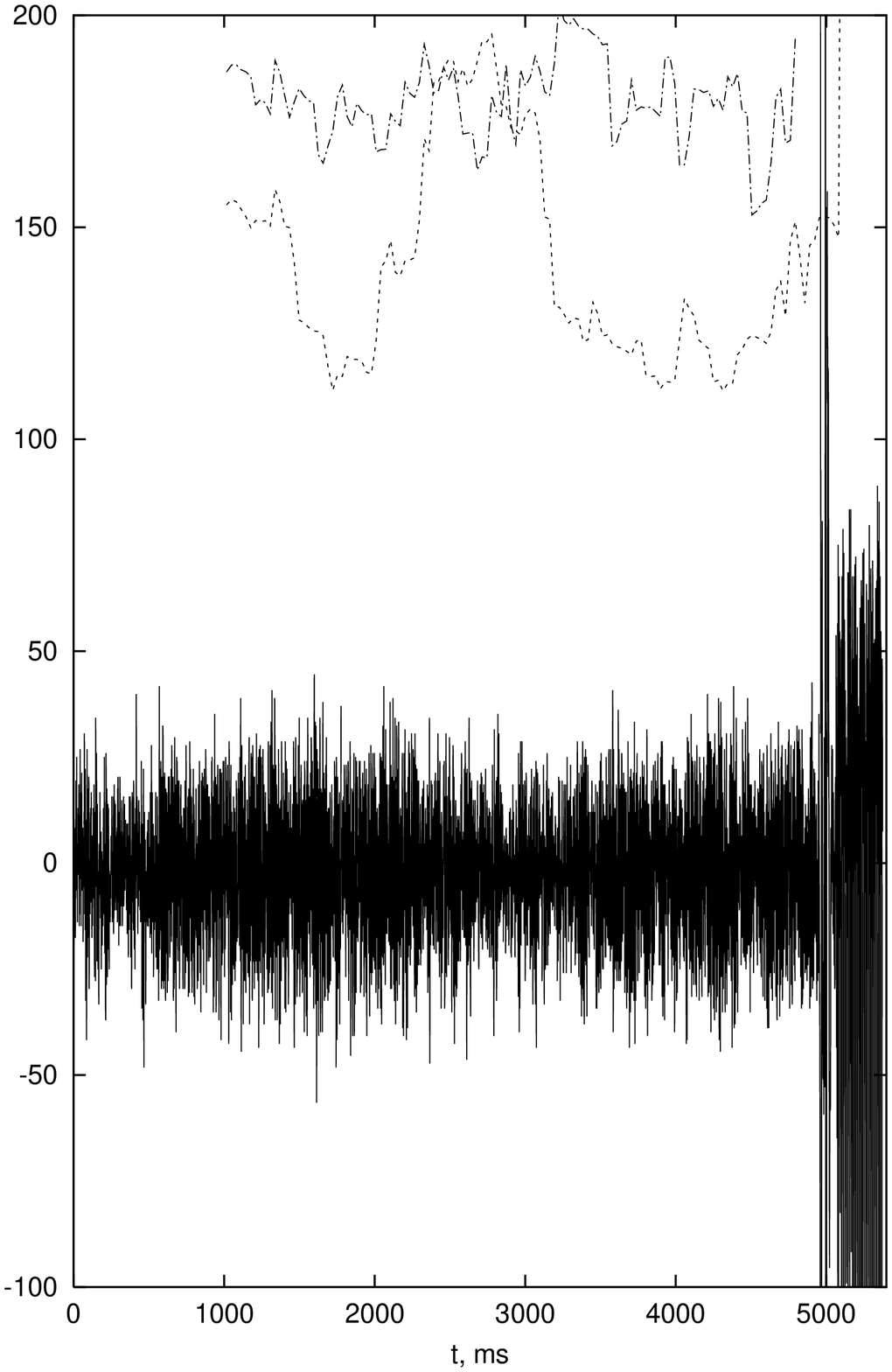,scale=0.5,clip=}

Fig.17

\epsfig{file=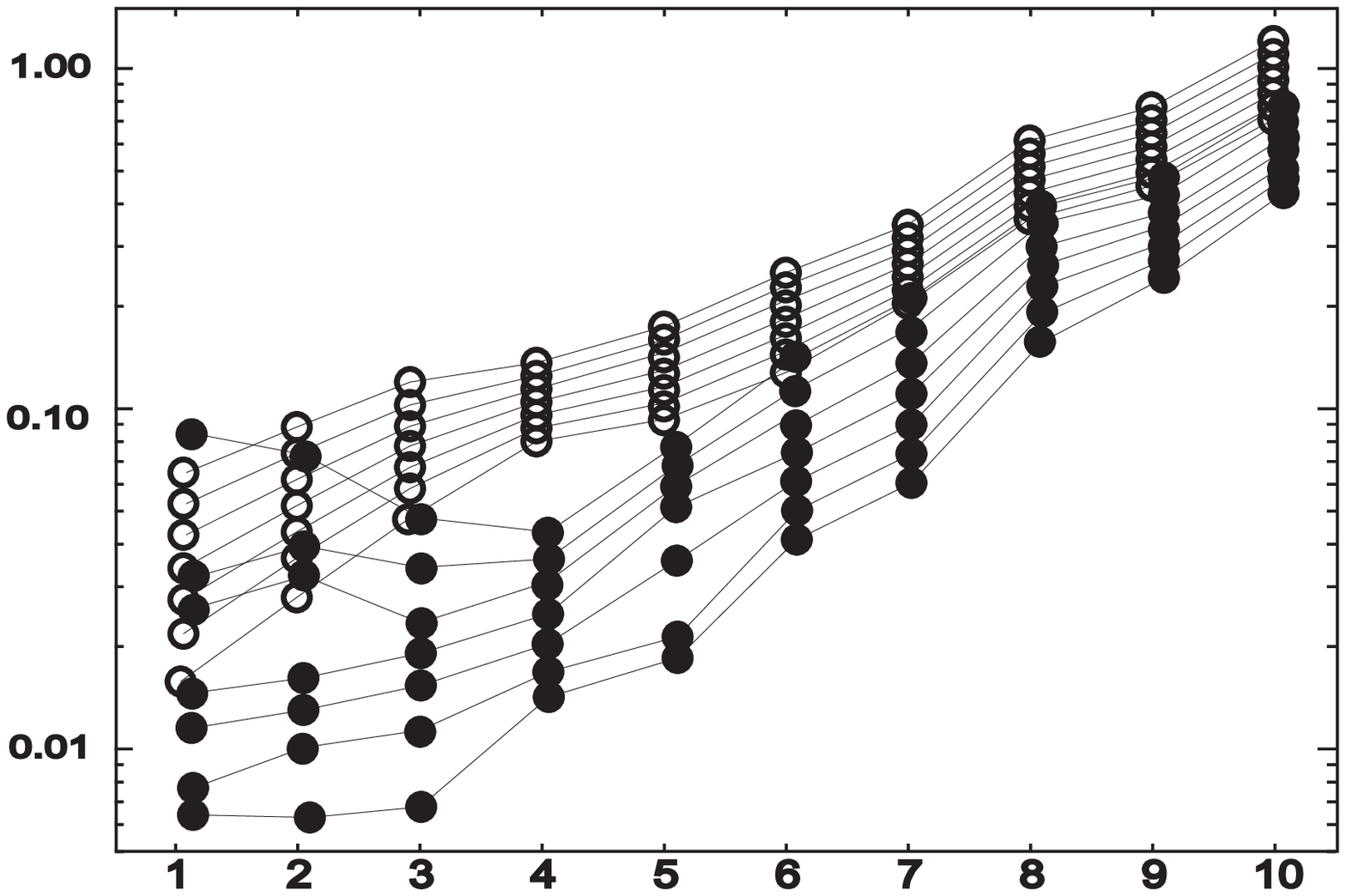,scale=0.7}

Fig.18

\begin{tabular}{cc}
\epsfig{file=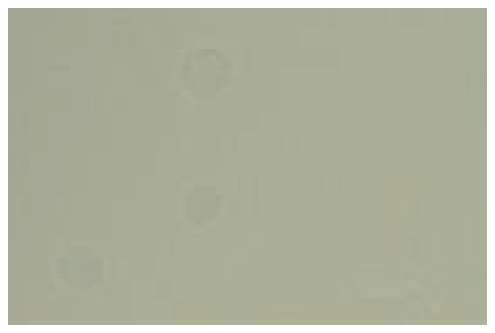,clip=}& \epsfig{file=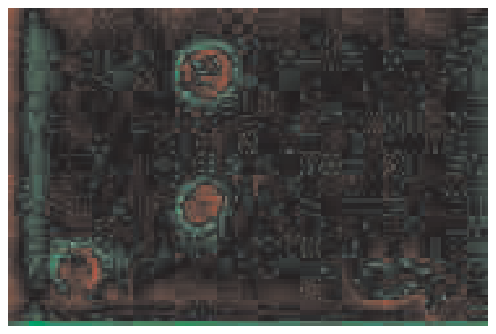,clip=}\
\end{tabular}

Fig.19

\epsfig{file=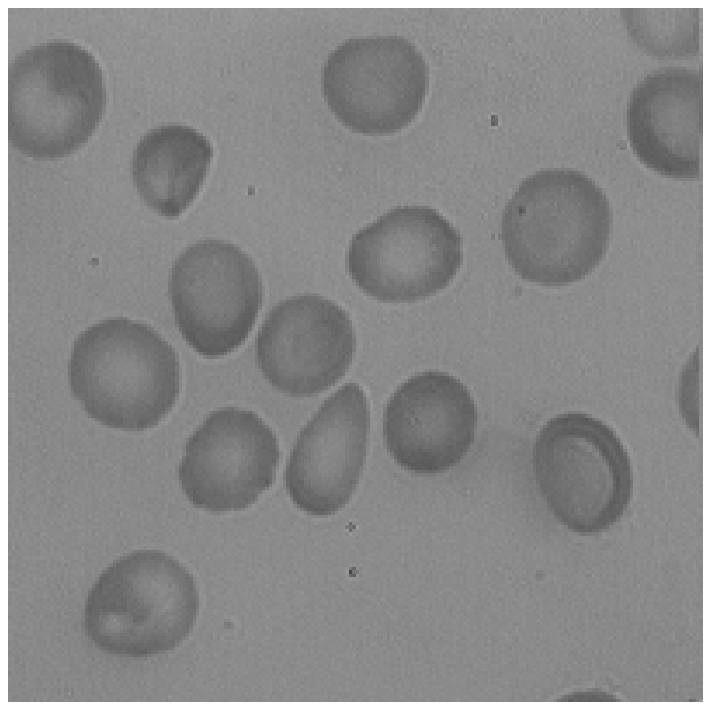}

Fig.20

\bigskip

\begin{tabular}{|c|c|c|}
\hline
\multicolumn{3}{|c|}{See JPEG-files:} \\
\hline
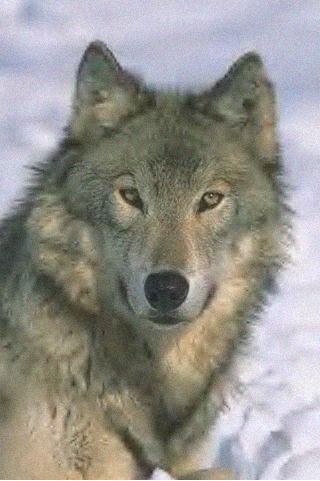& 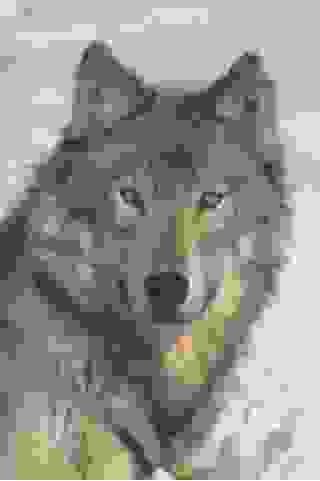&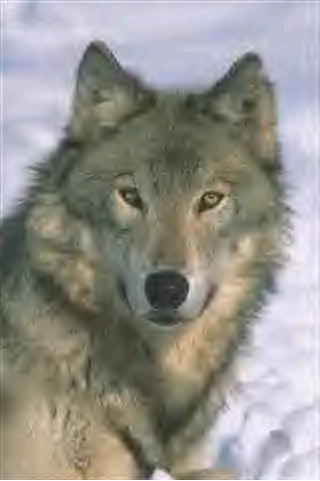 
\\
\hline
\end{tabular}

\bigskip

Fig.21

\epsfig{file=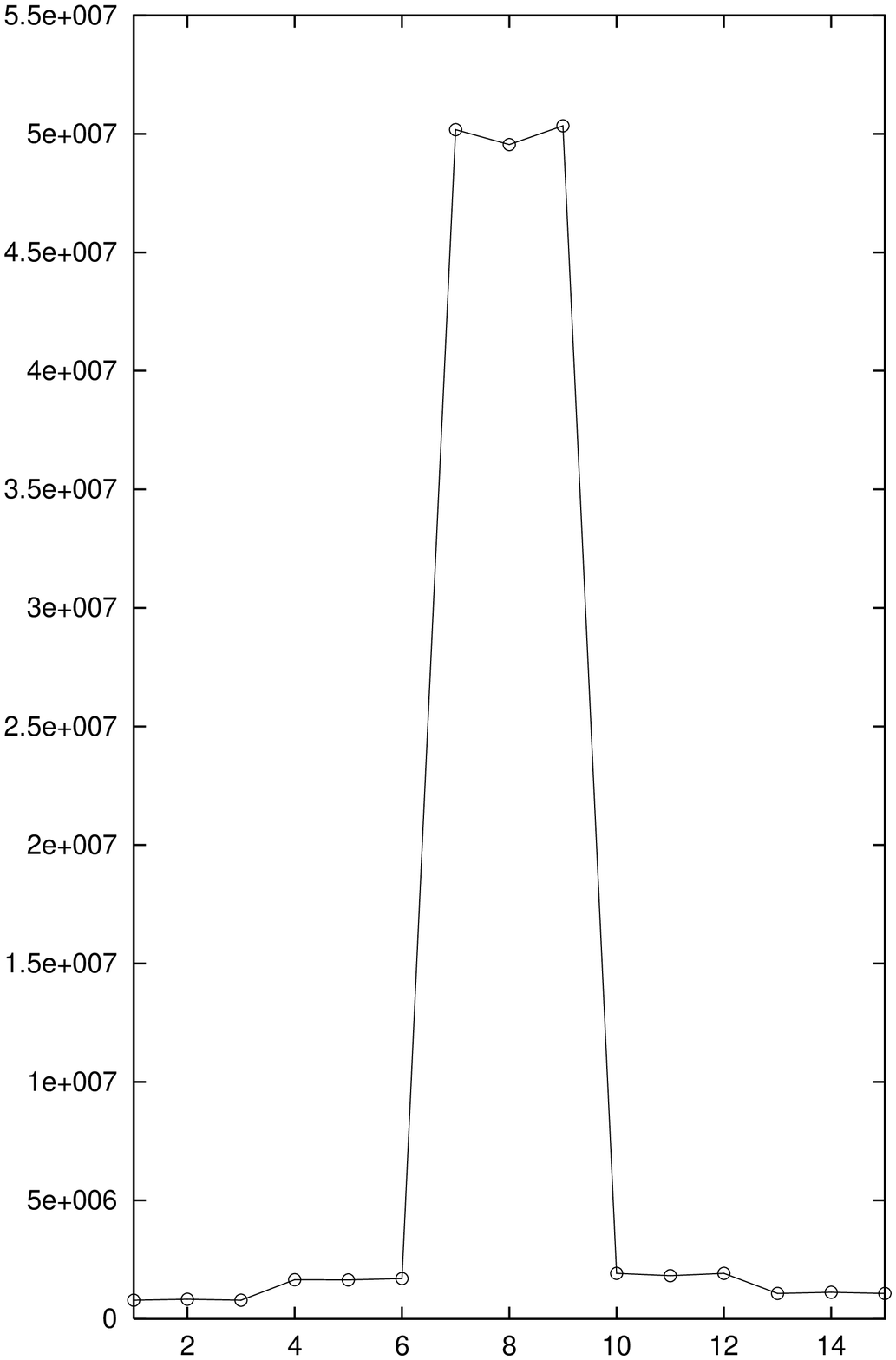,scale=0.5}

Fig.22

\end{document}